\documentclass[prx,aps,twocolumn,superscriptaddress]{revtex4-2}

\usepackage[pdftex]{graphicx}
\usepackage{amsbsy,amssymb,amsmath,bm,mathtools}

\usepackage{xspace}
\usepackage{bm}
\usepackage{color}
 
\usepackage{url}
\usepackage[section]{placeins}
\usepackage[colorlinks=true,linkcolor=blue,citecolor=blue]{hyperref}

\begin{document}


\title{Kinetics of orbital ordering in cooperative Jahn-Teller models: \\ Machine-learning enabled large-scale simulations}

\author{Supriyo Ghosh}
\affiliation{Department of Physics, University of Virginia, Charlottesville, VA 22904, USA}

\author{Sheng Zhang}
\affiliation{Department of Physics, University of Virginia, Charlottesville, VA 22904, USA}

\author{Chen Cheng}
\affiliation{Department of Physics, University of Virginia, Charlottesville, VA 22904, USA}

\author{Gia-Wei Chern}
\affiliation{Department of Physics, University of Virginia, Charlottesville, VA 22904, USA}

\date{\today}

\begin{abstract}
We present a scalable machine learning (ML) force-field model for the adiabatic dynamics of cooperative Jahn-Teller (JT) systems. Large scale dynamical simulations of the JT model also shed light on the orbital ordering dynamics in colossal magnetoresistance manganites. The JT effect in these materials describes the distortion of local oxygen octahedra driven by a coupling to the orbital degrees of freedom of $e_g$ electrons.  An effective electron-mediated interaction between the local JT modes leads to a structural transition and the emergence of long-range orbital order at low temperatures. Assuming the principle of locality, a deep-learning neural-network model is developed to accurately and efficiently predict the electron-induced forces that drive the dynamical evolution of JT phonons. A group-theoretical method is utilized to develop a descriptor that incorporates the combined orbital and lattice symmetry into the ML model. Large-scale Langevin dynamics simulations, enabled by the ML force-field models, are performed to investigate the coarsening dynamics of the composite JT distortion and orbital order after a thermal quench. The late-stage coarsening of orbital domains exhibits pronounced freezing behaviors which are likely related to the unusual morphology of the domain structures. Our work highlights a promising avenue for multi-scale dynamical modeling of correlated electron systems. 
\end{abstract}

\maketitle

\section{Introduction} 

\label{sec:intro}

Colossal magnetoresistance (CMR) refers to an extraordinary enhancement of the electrical conductivity induced by applying a moderate  magnetic field. Manganites of the formula ($R_{1-x} A_x$)MnO$_3$ where $R$ denotes rare earth ions and $A$ is a divalent alkaline ion are one of the canonical systems that exhibit CMR effect~\cite{tokura99,salamon01,dagotto01,dagotto-book}. In these compounds several degrees of freedom, including itinerant electrons, localized spins, orbitals, and lattice vibrations, are simultaneously important~\cite{dagotto05}. The complex interplay between these degrees of freedom leads to various symmetry-breaking phases with similar characteristic energy scales. Notable among them are ferromagnetic metallic phases and several antiferromagnetic insulating states accompanied by charge and/or orbital orders~\cite{schiffer95,chen96}. In particular, it has been proposed that nanoscale phase separation resulting from the various competing phases plays a crucial role in the emergence of CMR~\cite{dagotto01,dagotto-book,dagotto05,moreo99,uehara99,fath99}.
	
	
The metallic behaviors in manganites are mainly stabilized by the double-exchange (DE) mechanism~\cite{zener51,anderson55,degennes60}, which originates from a strong Hund's coupling between spins of itinerant carriers at the $e_g$ orbital and $S=3/2$ core spins from the localized $t_{2g}$ electrons. This charge-spin coupling in manganites and related compounds is described by a ferromagnetic Kondo-lattice model~\cite{yunoki98,dagotto98,chattopadhyay01}. Since parallel alignment between core spins facilitates the delocalization of $e_g$ electrons, DE mechanism tends to enhance ferromagnetism. Importantly, upon small hole doping, the Kondo-lattice model is shown to exhibit a mixed-phase regime where phase separation occurs between antiferromagnetic state, stabilized at half-filling, and hole-rich ferromagnetic regions~\cite{yunoki98}.  


Yet, early theoretical studies already showed that the DE mechanism alone is not sufficient to describe the complex physics of CMR materials. Indeed, the high-temperature insulating phases in manganites cannot be produced by DE models with realistic electron densities.  Strong electron correlation due to the on-site Coulomb repulsion generally could lead to electron localizations and the emergence of insulating phases in manganites~\cite{held00,maezono98a,maezono98b,yang10}. Several works have also shown that electron-phonon couplings via the Jahn-Teller (JT) mechanism are crucial to a complete description of the phase diagrams and transport behaviors of manganites~\cite{roder96,millis95,millis96,millis96b,nagaosa98,maezono03,hotta00}. 


The JT effect also highlights the importance of collective orbital behaviors and orbital ordering in manganites.   While long-range orbital order can be induced by super-exchange interactions as described by the Kugel-Khomskii Hamiltonians~\cite{kugel72,kugel73,kugel82,brink01,brink04}, the cooperative JT effect is another mechanism for collective orbital behaviors and orbital ordering.  Although the JT coupling describes a localized interaction between the orbital degrees of freedom of an electron and the vibronic normal modes of the surrounding MO$_6$ octahedron, an effective long-range interaction between local vibronic modes is generated by itinerant electrons. This mechanism is similar to the electron-mediated effective Ruderman-Kittel-Kasuya-Yosida (RKKY) interaction between localized magnetic moments in spin glasses~\cite{Ruderman1954,Kasuya1956,Yosida1957}. The long-range structural distortion induced by this effective interaction, in turn, could localize the electrons and establish a long-range orbital order~\cite{popovic00,hotta99}.

In particular, the orbital order in LaMnO$_3$, which is the parent compound of many CMR manganites, can be stabilized by a predominant cooperative JT effect, even in the absence of on-site Coulomb interaction~\cite{hotta99}. In real materials, it is most likely that both mechanisms contribute to the stabilization of the observed orbital order and structural distortions. An $A$-type antiferromagnetic order with the wave vector $\mathbf q = (0, 0, \pi)$ is found to accompany the observed orbital order. In this magnetic structure, the localized $t_{2g}$ spins are ferromagnetically ordered within a 2D plane by the DE mechanism, while the spin-polarized planes are antiferromagnetically aligned in the $c$-direction by a moderate super-exchange interaction. The orbital order, of the $C$-type, is characterized by an in-plane checkerboard arrangement of $d_{3x^2 - r^2}$ and $d_{3y^2 - r^2}$ orbitals and a concomitant $\mathbf K = (\pi, \pi)$ JT distortions.

Despite extensive studies on the orbital and JT transitions in manganites, the phase-ordering dynamics of this composite order has yet to be investigated. As the in-plane JT distortions of the $C$-type order break the $Z_2$ sublattice symmetry of the square lattice, the orbital/structural transition is expected to belong to the Ising universality class. While several universal coarsening behaviors of Ising-type order have been well established, it is unclear whether they can be applied to describe the kinetics of the $C$-type orbital order. First-principles simulations of orbital ordering in the cooperative JT model are difficult since an electronic structure problem has to be solved at every time-step in order to obtain the electronic forces that drive the JT distortions. For JT models without electron-electron interactions, the electronic structure can be solved by exact diagonalization, without resorting to more sophisticated many-body techniques. Yet, even repeated diagonalizations could be overwhelmingly expensive for large-scale simulations, which partly explains the lack of progress on the study of the dynamical aspects of JT phase transition.

In this paper, we develop a machine learning (ML) force-field model to enable large-scale dynamical simulations of orbital ordering in a representative cooperative JT model on a square lattice. The ML force-field framework, originally developed in the pioneering works of Behler and Parrinello~\cite{behler07}, and Bart\'ok {\em et al.}~\cite{bartok10}, has now been widely used for large-scale {\em ab inito} molecular dynamics with desired quantum accuracy. Here we generalized  this approach for the JT lattice models, in particular by  developing proper descriptors to account for lattice symmetries.  Central to our approach is a deep-learning neural network that could efficiently and accurately predict the electronic forces that drive the dynamics of JT modes. Moreover, the ML force-field approach is linearly scalable, which means the NN model trained by ED solutions from small systems can be directly applied to much larger lattices without rebuilding or retraining. 


We apply the ML force-field models to simulate thermal quenches of a cooperative JT system at half-filling and a slightly doped regime. The dynamical evolution of the local JT modes is described by an effective Langevin equation with driving forces obtained from ML models. In addition to the doublet JT distortions that break the cubic symmetry of a MO$_6$ octahedron, the dynamics of the $A_1$ breathing mode is also included. The relaxation process after a thermal quench is dominated by the formation and subsequent coarsening of the $C$-type JT/orbital order. Our simulations also find long-lasting localized meta-stable checkerboard density modulations. Intriguingly, the late-stage growth of the orbital-ordered domains does not follow the Allen-Cahn law expected for a non-conserved Ising type order. Instead, a freezing behavior is observed for the coarsening of orbital and JT domains.

The rest of the paper is organized as follows. We discuss the prototype cooperative JT model on a square lattice that is relevant for the orbital ordering in manganites in Sec.~\ref{sec:JT-model}. An effective Langevin equation is presented to describe the adiabatic dynamics of the JT distortions. In Sec.~\ref{sec:ML} we present a scalable ML force field model for cooperative JT systems. The ML model is benchmarked against the exact diagonalization for the force prediction and time-dependent distortion correlation functions. Large-scale thermal quench simulations of the JT model and detailed characterizations of the phase ordering dynamics and the growth law of orbital domains are presented in Sec.~\ref{sec:coarsening}. Finally, a summary and outlook is presented in Sec.~\ref{sec:conclusion}.

\section{Adiabatic dynamics of cooperative Jahn-Teller systems}

\label{sec:JT-model}

We consider a model of itinerant $e_g$ electrons interacting with the vibrational modes of MO$_6$ octahedra on a square lattice. Since our main interest here is the orbital ordering, we will neglect the electron spin degrees of freedom and the $t_{2g}$ core spins. In LaMnO$_3$, the orbital ordering (accompanied by a JT transition) takes place at $T_{\rm OO} = 750$~K, which is much higher than the magnetic transition at $T_M = 120$~K, below which an A-type antiferromagnetic order develops. In the paramagnetic phase above $T_M$, the spin degrees of freedom form a homogeneous, fluctuating background and we only consider the orbital degrees of freedom in this work. Although orbital orders in manganites are three-dimensional, as a first step towards large-scale simulations,  in this work we focus on a two-dimensional version of the cooperative JT systems, which exhibits the same $\mathbf K = (\pi, \pi)$ $C$-type order/JT order in the ground state. 

The Hamiltonian of the cooperative JT model on a square lattice is given by
 \begin{eqnarray}
 	\label{eq:H_total}
 	\hat{\mathcal{H}} = \hat{\mathcal{H}}_{\rm K} + \hat{\mathcal{H}}_{\rm JT} + {\mathcal{E}}_{\rm L}.
 \end{eqnarray}
The three terms correspond to the electron kinetic energy, the electron-phonon coupling, and the lattice elastic energy, respectively. The Hamiltonian $\hat{\mathcal{H}}_{\rm K}$ describes the nearest-neighbor hopping of $e_g$ electrons
 \begin{eqnarray}
 	\label{eq:H_kin}
 	 \hat{\mathcal{H}}_{\rm K} = \sum_{\gamma = x, y} \sum_{\langle ij \rangle \parallel \gamma} \sum_{\mu\nu = a, b} \left( t^\gamma_{\mu\nu} \hat{c}^\dagger_{i\mu} c^{\,}_{j\nu} + {\rm h.c.} \right),
 \end{eqnarray}
 where $\hat{c}^\dagger_{i, \mu}$ and $\hat{c}^{\,}_{i, \mu}$ represent the creation and annihilation operators, respectively, of an electron with orbital flavor $\mu$ at the $i$-th site, $a$ and $b$ denote the two basis  \(d_{x^2-y^2}\) and \(d_{3z^2-r^2}\) of $e_g$ orbitals, $t^\gamma_{\mu\nu}$ denotes the orbital-dependent anisotropic hopping between nearest-neighbor pairs $\langle ij \rangle$ parallel to the direction $\gamma = x, y$ on the square lattice.  The following hopping coefficients are used in our model calculations~\cite{yunoki98,hotta99}: \(t^{aa}_{ij}=-\sqrt{3}t^{ab}_{ij}=-\sqrt{3}t^{ba}_{ij}=3t^{bb}_{ij}=t_{\rm nn}\) for hopping along the \(x\)-direction, and \(t^{aa}_{ij}=\sqrt{3}t^{ab}_{ij}=\sqrt{3}t^{ba}_{ij}=3t^{bb}_{ij}=t_{\rm nn}\) along the \(y\)-direction. 

\begin{figure}[b]
\centering
\includegraphics[width=0.99\columnwidth]{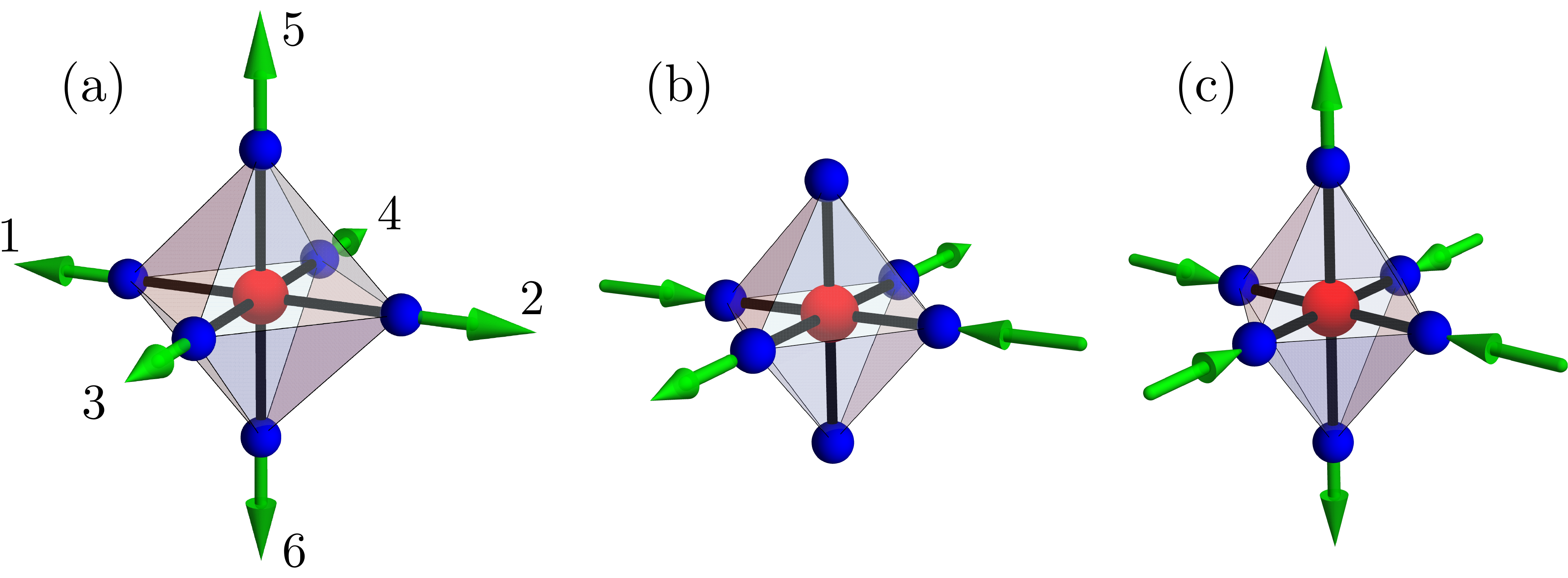}
\caption{Schematic diagram of the vibronic modes for the MnO6 octahedron: (a) the symmetry-preserving breathing mode, (b) and (c) the symmetry-breaking JT modes. In terms of oxygen displacements, the coordinates of these normal modes are: $Q^{A_1} = (-X_1 + X_2 - X_3 + X_4 - X_5 + X_6)/\sqrt{6}$, $Q^x = (-X_1+X_2 +X_3 - X_4)/2$, and $Q^z =  (-X_1 + X_2 - X_3 + X_4 + 2 X_5 - 2 X_6)/\sqrt{12}$, where $X_i$ denotes the $x$ coordinates of the $i$-th oxygen, and so on.  }
\label{fig:JT-modes}
\end{figure}

 The second term in $\hat{\mathcal{H}}$ describes the JT coupling:
 \begin{eqnarray}
 	\label{eq:JT}
 	\hat{\mathcal{H}}_{\rm JT} = -\lambda \sum_i \left( Q^{A_1}_i \hat{n}_i + \bm Q^E_i \cdot \hat{\bm \tau}  \right),
 \end{eqnarray}
 where $Q^{A_1}_i = Q^1_i$ denotes the breathing mode of the MO$_6$ octahedron at site-$i$, which couples to the electron number operator 
 \begin{eqnarray}
 	\hat{n}_i = \hat{c}^\dagger_{i a} \hat{c}^{\,}_{i a} + \hat{c}^\dagger_{i b} \hat{c}^{\,}_{i b}.
\end{eqnarray} 
The doublet $\bm Q^E_i = (Q^x_i, Q^z_i) = (Q^2_i, Q^3_i)$ describes the JT modes that break the cubic symmetry of the octahedron. Schematic diagrams of these three lattice modes are shown in Fig.~\ref{fig:JT-modes}. The JT doublet couples to the pseudo-spin operator $\hat{\bm\tau}_i = (\hat{\tau}^x_i, \hat{\tau}^z_i)$ representing the electron orbital degrees of freedom:
\begin{eqnarray}
 	\hat{\tau}^x_i = \hat c_{i a}^\dagger \hat c^{\,}_{i b}+ \hat c_{i b}^\dagger \hat c^{\,}_{i a}, \qquad 
	\hat\tau^z_i = \hat c_{i a}^\dagger \hat c^{\,}_{i a}- \hat c_{i b}^\dagger \hat c^{\,}_{i b}.
 \end{eqnarray}
The third term in Eq.~(\ref{eq:H_total}) is the classical elastic energy of the lattice distortions
 \begin{eqnarray}
 	\label{eq:E_elastic}
 	{\mathcal{E}}_{\rm L} = \frac{K}{2} \sum_i \left[ \beta \bigl(Q^{A_1}_i \bigr)^2 + \bigl|\bm Q^E_i \bigr|^2  \right],
 \end{eqnarray}
where $K$ denotes an effective elastic coefficient. The parameter $\beta$ is defined as $\beta = (\omega_{A_1} / \omega_{E})^2$, where $\omega_{A_1}$ and $\omega_{\rm E}$ are the vibration energies for the breathing and doublet JT modes, respectively, assuming that the reduced masses for these two modes are the same. Following previous works~\cite{hotta99,lliev98}, this parameter is set to $\beta = 2$ in the following calculations. 


It is worth noting that the breathing and JT modes of each octahedra are independent of each other in this elastic model. More realistic approach naturally needs to include couplings between neighboring octahedra, for example: $\mathcal{E}_L' = \sum_{ij} K^{mn}_{ij} Q^m_i Q^n_j$, where the indices $m, n = 1, x, z$. The equilibrium structural distortion is thus determined by both such direct elastic interactions as well as the effective electron-mediated interaction through the cooperative JT effect discussed above.  In general, as same oxygen atoms are shared by two neighboring octahedra, nearest-neighbor couplings between these $Q$ modes are antiferromagnetic, i.e. $K > 0$, which are compatible with both the $C$-type orbital order and the meta-stable CDW order to be discussed later. For simplicity, in this work we neglect the direct elastic couplings and focus on the intrinsic cooperative JT mechanism.

\begin{figure}
\centering
\includegraphics[width=0.99\columnwidth]{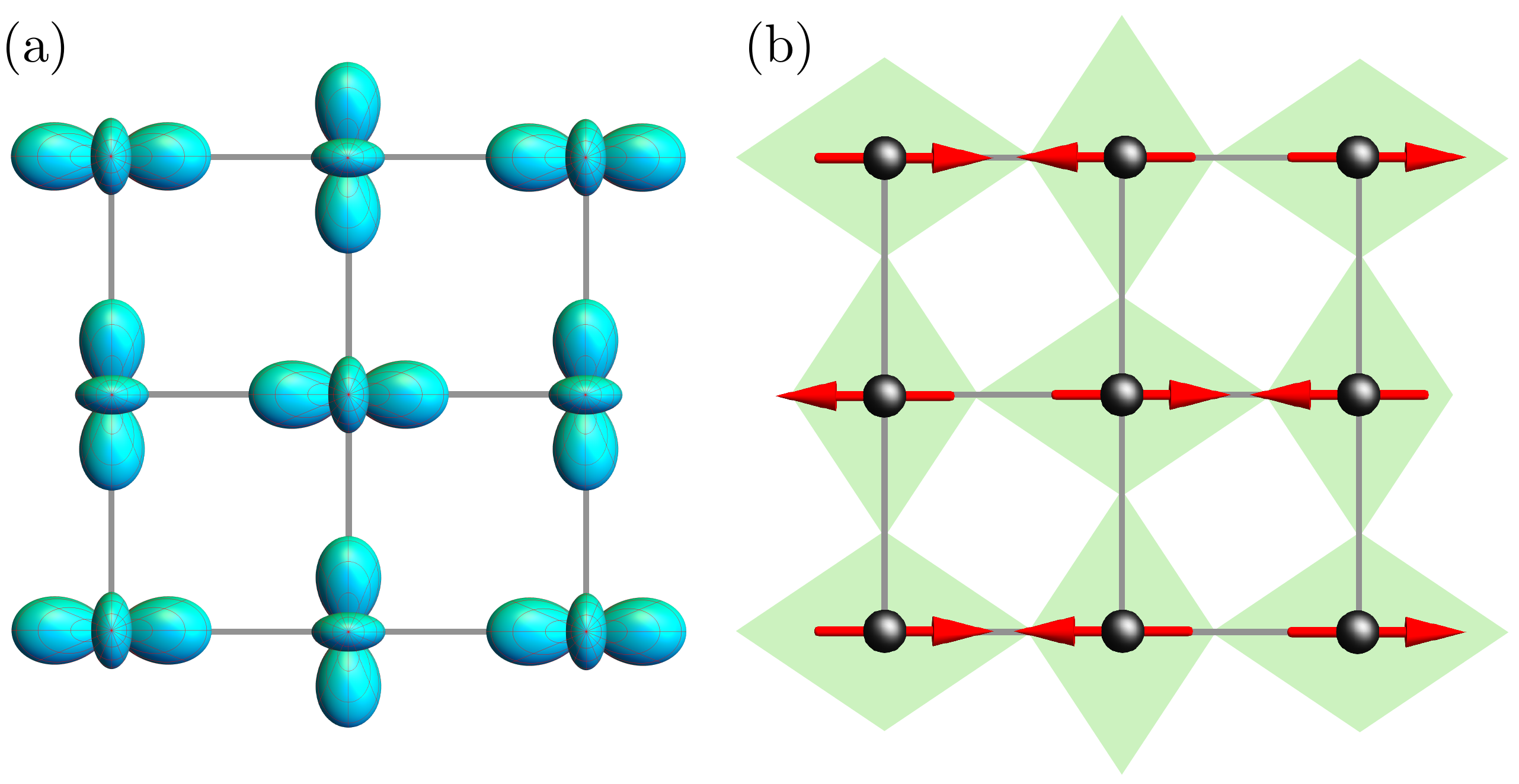}
\caption{Schematic diagram of (a) $C$-type orbital order, and (b) the concomitant antiferro-distortive JT order. Also shown in panel~(b) is the vector representation of the orbital/JT order. The arrows represent either the doublet vector $\bm Q^E$ or the expectation value of the orbital pseudo-spin $\langle \hat{\bm\tau} \rangle$. These two vectors are related to each other via Eq.~(\ref{eq:tau-Q}) in the ground state.}
\label{fig:C-type-OO}
\end{figure}
 
The ground state of the cooperative JT model in Eq.~(\ref{eq:H_total}) at half-filling exhibits a $C$-type orbital ordering, characterized by a wave vector $\mathbf K = (\pi, \pi)$, accompanied by a structural distortion of the antiferro-distortive order with a predominate $Q^x$ mode~\cite{yunoki98b,hotta99,hotta03,salafranca06,nanda10}. A schematic diagram of the orbital/JT order is shown in Fig.~\ref{fig:C-type-OO}. The orbital order can be described by the expectation values of the pseudo-spins. To this end, minimization of the total energy with respect to the JT distortions yields the relation
 \begin{eqnarray}
 	\label{eq:tau-Q}
 	\langle \hat{\bm \tau}_i \rangle = \frac{K}{\lambda} \bm Q^E_i,
 \end{eqnarray}
It is worth noting that the linear JT coupling in Eq.~(\ref{eq:JT}) of a single octahedron is given by the inner product $(\hat{\bm\tau} \cdot \bm Q^E)$, which implies that the interaction is invariant under simultaneous rotation of the orbital pseudo-spin and JT doublet  vector. Inclusion of quadratic JT couplings of the form $QQ\tau$  reduces this O(2) symmetry to a 3-fold degeneracy already at the single-octahedron level. On the other hand, in the cooperative JT scenario, the determination of the lattice distortions needs to include the kinetic energy of itinerant electrons on a lattice. Since the electron hopping in Eq.~(\ref{eq:H_kin}) is anisotropic with a strong orbital-dependence, the O(2) symmetry of the doublet vector is reduced to a two-fold mirror symmetry $Q^x \to -Q^x$ for the half-filled cooperative JT system~\cite{popovic00}. The resultant minima are found to be along the $Q^x$ direction $\bm Q^E_* = \pm( Q_*, 0)$, and the checkerboard arrangement of these two symmetry-related minima $\bm Q^E_*$ in the $C$-type order comes from a dominant electron-mediated nearest-neighbor interaction of antiferro-distortive sign~\cite{hotta99}.

The $Z_2$ mirror symmetry discussed above corresponds to a tetragonal lattice symmetry. As shown in Fig.~\ref{fig:JT-modes}(b), changing the sign of $Q^x$ sends a tetrahedron elongated in the $x$-direction to one along the $y$-direction. The eigenstates of $\hat{\tau}_x$ with eigenvalue $+1$ and $-1$ are dominated by $d_{3x^2 - r^2}$ and $d_{3y^2-r^2}$, respectively. The breaking of the global $Z_2$ symmetry of the  cooperative JT system leads to the $C$-type orbital order which can be described viewed as a N\'eel type order of Ising variable $Q^x$. On symmetry ground, the orbital JT phase transition is expected to belong to the Ising universality class. A detailed thermodynamic study of the orbital/JT phase transition by, e.g. Monte Carlo simulations, remain to be done. 

Also of interest is the transition dynamics of the simultaneous JT distortion and orbital ordering. Of particular interest is the coarsening behaviors of orbital domains and whether the resultant domain-growth law falls into well-established universality classes of phase-ordering dynamics.  To this end, we consider the dynamical evolution of the JT systems based on the approximation that electron relaxations are much faster than the dynamics of JT distortions. This adiabatic approximation is similar to the Born-Oppenheimer approximation widely used in the {\em ab initio} molecular dynamics methods~\cite{marx09}. In particular, this means that the electronic contribution to the driving forces are computed from an equilibrium Fermi liquid of the instantaneous lattice distortion. As both breathing and JT distortions are vibronic normal modes, their dynamics are governed by an effective Newton equation of motion 
\begin{eqnarray}
	\label{eq:langevin_eq}
	& & \mu_{A_1} \frac{d^2 Q^{A_1}_i}{dt^2} + \gamma_{A_1} \frac{d Q^{A_1}_i}{dt} 
	= - \frac{\partial \mathcal{E}_L}{\partial Q^{A_1}_i} 
	- \frac{\partial \langle \mathcal{H}_{\rm JT} \rangle}{\partial Q^{A_1}_i} +  \eta^{A_1}_i(t), \nonumber \\
	& & \mu_E \frac{d^2 \bm Q^{E}_i}{dt^2} + \gamma_{E} \frac{d \bm Q^{E}_i}{dt} 
	= -\frac{\partial \mathcal{E}_L}{\partial \bm Q^E_i} 
	- \frac{\partial \langle \mathcal{H}_{\rm JT} \rangle}{\partial \bm Q^{E}_i} +  \bm\eta^{E}_i(t).
\end{eqnarray}
Here $\mu_\alpha$ is the effective mass of the octahedral normal modes $\alpha = A_1$ and $E$, the corresponding damping coefficients and Langevin noises are denoted by $\gamma_\alpha$ and $\eta_\alpha$, respectively. For simplicity, we shall assume the same parameters for the breathing and JT modes. In particular, as discussed above, the difference in effective mass can be accounted for by the ratio $\beta$ of the elastic constants. As in standard Langevin method, the thermal forces are Gaussian random variables with zero mean and variance consistently related to the damping coefficients $\gamma$ through the dissipation-fluctuation theorem. 

The deterministic forces have two contributions, corresponding to the first two terms on the right-hand side: the  elastic restoring forces and the electronic forces via JT coupling. The calculation of forces can be simplified using the Hellmann-Feynman theorem: $\partial \langle \hat{\mathcal{H}} \rangle / \partial Q = \langle \partial \hat{\mathcal{H}} / \partial Q \rangle$. Using Eq.~(\ref{eq:JT}) and (\ref{eq:E_elastic}), one obtains the following expressions for the driving forces
\begin{eqnarray}
	\label{eq:forces}
	& & F^{A_1}_i = -\beta K Q^{A_1}_i + \lambda \langle \hat{n}^{\,}_i \rangle, \nonumber \\
	& & \bm F^E_i = - K \bm Q^E_i + \lambda \langle \hat{\bm \tau}^{\,}_i \rangle.
\end{eqnarray}
The equilibrium condition $\bm F^E_i = 0$ gives the relation in Eq.~(\ref{eq:tau-Q}). As discussed above, expectation values $\langle \cdots \rangle$ are computed based on an equilibrium electron liquid corresponding to the instantaneous lattice configuration $\{Q^{A_1}_i, \bm Q^E_i\}$. Explicitly, for example, the expectation values of orbital pseudo-spins are given by
\begin{eqnarray}
	\langle \hat{\bm \tau}_i \rangle = \frac{1}{Z_e} {\rm Tr}\!\left[\hat{\bm\tau}_i \, e^{-\beta \hat{\mathcal{H}}_e\left(Q^{A_1}_i, \bm Q^E_i \right)} \right]
\end{eqnarray}
where $\hat{\mathcal{H}}_e = \hat{\mathcal{H}}_{\rm K} + \hat{\mathcal{H}}_{\rm JT}$ is the electron Hamiltonian, $\beta = 1/k_B T$ is the inverse temperature, and $Z_e = {\rm Tr}e^{-\beta \hat{\mathcal{H}}_e}$ is the electron partition function. As the electron Hamiltonian $\mathcal{H}_e$ is quadratic in the fermion creation/annihilation operators, it can be solved by the exact diagonalization (ED) in real space. Yet, since the electronic forces have to be computed at every time-step of the Langevin dynamics simulation, the $\mathcal{O}(N^3)$ time complexity of  ED can be overwhelmingly time-consuming for large-scale simulations.

 \section{Machine learning force-field models}
 
 \label{sec:ML}
 
 \begin{figure*}
\centering
\includegraphics[width=2.0\columnwidth]{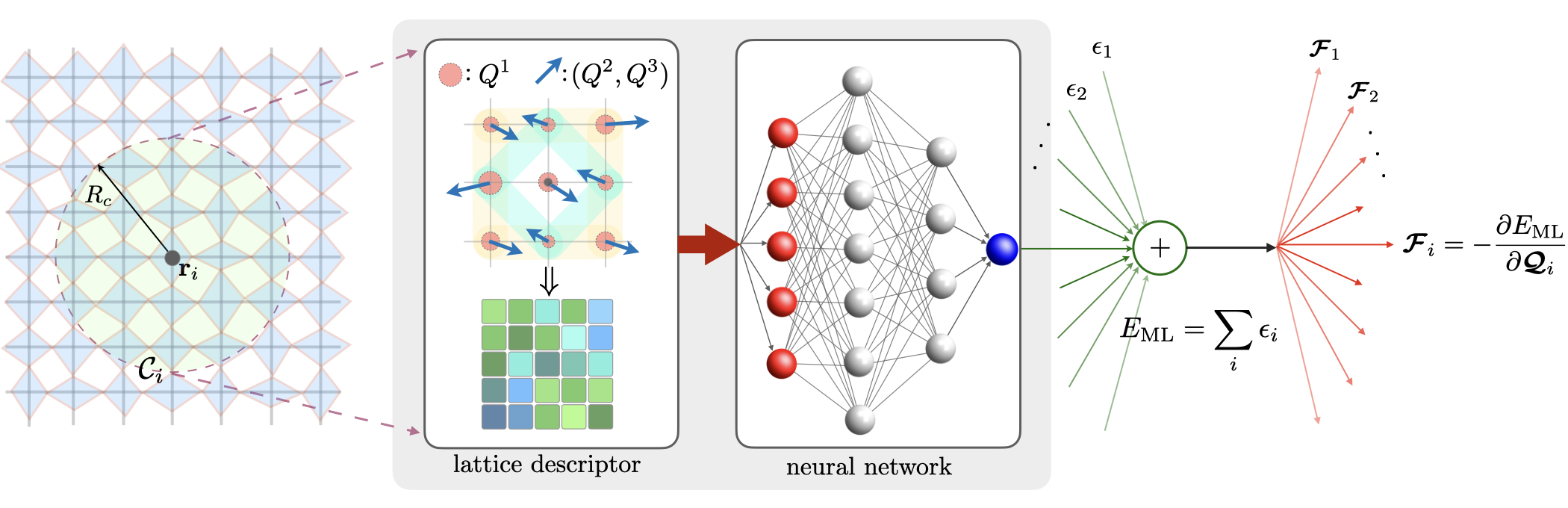}
\caption{Schematic diagram of the ML force-field model for the cooperative JT models. A lattice descriptor transforms the neighborhood distortion configuration $\mathcal{C}_i$ into effective coordinates $\{ G_m \}$ which are then fed into a fully connected neural network (NN). The output node of the NN corresponds to the local site-energy $\epsilon_i$. The combination of the descriptor and the NN provides an approximation for the universal function $\varepsilon(\cdot)$.  The corresponding total potential energy $E$ is obtained from the summation of the local energies. Automatic differentiation is employed to compute the derivatives $\partial E_{\rm ML} / \partial \bm{\mathcal{Q}}_i$ for the effective forces acting on the breathing and JT modes. } 
\label{fig:ML-schematic}
\end{figure*}

Here we present a machine-learning (ML) framework for computing the electronic forces in cooperative JT systems with a linear-scaling complexity.  Fundamentally, as pointed out by W. Kohn, linear-scaling electronic structure methods are possible mainly because of the locality nature or ``nearsightedness'' principle of many-electron systems~\cite{kohn96,prodan05}. Importantly, modern ML techniques provide an explicit and efficient approach to incorporate the locality principle into the implementation of $O(N)$ methods. Perhaps the most prominent and successful demonstration of this approach is the ML force-field methods developed in quantum chemistry to enable large-scale {\em ab initio} molecular dynamics (MD) simulations~\cite{behler07,bartok10,li15,shapeev16,behler16,botu17,smith17,chmiela17,zhang18,chmiela18,deringer19,mcgibbon17,suwa19,sauceda20}. Similar to the Langevin dynamics for cooperative JT systems described above, the atomic forces in {\em ab initio} MD are computed by solving, for example, the Kohn-Sham equation, which has to be repeated at every time-step~\cite{marx09}. The central idea behind the linear scalability of ML force-field methods is the divide-and-conquer approach proposed in
the pioneering works of Behler and Parrinello~\cite{behler07} and Bart\'ok {\em et al.}~\cite{bartok10}. The ML model is trained to produce a {\em local} atomic energy from a finite neighborhood. The atomic forces are obtained indirectly from the total energy, which is the sum of all atomic energies.

Similar ML frameworks have recently been developed to enable large-scale dynamical simulations in several condensed-matter lattice systems~\cite{zhang20,zhang21,zhang22,zhang22b,zhang23,cheng23,cheng23b}. In particular, the ML force-field approach was applied to the ferromagnetic Kondo-lattice or s-d models~\cite{zhang20,zhang21}. The strong-coupling regime of this model corresponds to the DE mechanism which is another important component of CMR physics as discussed in Sec.~\ref{sec:intro}.  The spin dynamics in the DE system is driven by itinerant electrons, similar to the cooperative JT models. Based on a generalized Behler-Parrinello scheme, a neural network model was trained to predict local effective fields induced by propagating electrons. The ML force-field methods have also been applied to the semiclassical Holstein model on a square lattice~\cite{cheng23}, a canonical system for studying the physics of electron-phonon coupling and phonon-assisted charge-density wave orders.  The scalar Einstein phonons in the standard Holstein model can be viewed as a simplified model for the breathing mode $Q^{A_1}$. Large-scale dynamical simulations enabled by the ML methods unveiled intriguing anomalous coarsening behavior of charge-density waves in the Holstein model~\cite{cheng23}.

\subsection{Behler-Parrinello machine-learning framework}

The Langevin dynamics discussed in Sec.~\ref{sec:JT-model} can be viewed as a MD method for the octahedral normal modes. This prompts a generalization of the Behler-Parrinello (BP) approach for the adiabatic dynamics of JT systems. However, there are two major differences between the two systems. First, both breathing and JT modes are defined on a lattice, in contrast to coordinates of atoms in free space. Second, the symmetry of the JT modes is tied to the symmetry of the underlying lattice, while MD systems are characterized by continuous translational and rotational symmetries. The BP scheme is modified to account for these two important issues; see FIG.~\ref{fig:ML-schematic} for a schematic diagram of the ML force-field model for the JT systems.  First, the total energy of the system in the adiabatic approximation is given by the expectation value of the Hamiltonian in Eq.~(\ref{eq:H_total}), which is to be approximated by a ML energy $E_{\rm ML}$. As in the BP scheme, this system energy is partitioned into local energies, each associated with a lattice site:
\begin{eqnarray}
	\langle \hat{\mathcal{H}} \rangle \approx E_{\rm ML} = \sum_i \epsilon_i = \sum_i \varepsilon(\mathcal{C}_i).
\end{eqnarray}
 Here we have invoked the locality assumption and express the site-energy $\epsilon_i$ as a function of lattice distortions in a local neighborhood, denoted as $\mathcal{C}_i$. Explicitly, we define the local distortion configuration as
 \begin{eqnarray}
 	\mathcal{C}_i = \left\{ \bm{\mathcal{Q}}_j \,\, \big| \,\, |\mathbf r_j - \mathbf r_i| < r_c \right\},
 \end{eqnarray}
 where $r_c$ is a cutoff distance which is determined by the locality of the forces, and we have grouped the breathing and JT modes into a three-component vector $\bm{\mathcal{Q}} =  (Q^1, Q^2, Q^3) = (Q^{A_1}, Q^x, Q^z)$ for convenience. It should be noted that the vector notation here does not imply an underlying O(3) rotation symmetry. The complex dependence of the site-energy on local environment is encoded in the function $\varepsilon(\cdot)$ which is universal for a given JT electronic Hamiltonian and electron filling fraction. Importantly, this universal function is to be approximated by a ML model in the BP approach. 
 
 As shown in FIG.~\ref{fig:ML-schematic}, there are two central components in the BP-type ML model: a descriptor and a learning model. The former is to transform the neighborhood configuration into a proper feature variables, while the latter is to approximate the universal function $\varepsilon(\cdot)$. In this work a feedforward neural network (NN) is employed  as the learning model which, according to universal approximation theorem~\cite{cybenko89,hornik89}, offers the capability of accurately representing complex functions to the desired accuracy. 
The total energy is given by the sum of all site-energies in the system, which are obtained by applying the same ML model to all lattice sites. The fact that the same ML model is used for all lattice sites simply reflects the translational symmetry of the original Hamiltonian. The effective forces acting on the octahedral normal modes are given by the derivatives of the total energy 
\begin{eqnarray}
	\label{eq:f_ML}
	\bm{\mathcal{F}}_i = -\frac{\partial E_{\rm ML}} { \partial \bm{\mathcal{Q}}_i}, 
\end{eqnarray}
which can be efficiently computed using automatic differentiation~\cite{paszke17,baydin18}. This expression also indicates that the ML predicted forces are conservative, which is an appealing feature of the PB-type ML models.


\subsection{Descriptors for lattice distortions}

\label{sec:lattice-descriptor}

The BP scheme also allows for a systematic approach to incorporate symmetry requirements into ML models through feature engineering. Since the local energy $\epsilon_i$ at the output of the NN is a scalar, it is invariant under symmetry operations of the system. However, despite the powerful approximation capability of NNs, such symmetry constraints can only be learnt statistically, sometimes with the help of techniques such as data augmentation. However, the symmetry constraints cannot be exactly implemented based on deep-learning alone. A descriptor is introduced here to provide a proper representation of the neighborhood configuration in such a way that the representation is itself invariant under transformations of the relevant symmetry group. The resultant feature variables are input to the NN model. As a result, symmetry-related configurations $\mathcal{C}_i$ are described by exactly the same feature variables, which in turn produces exactly the same local energy at the output. 

The importance of descriptors in the implementation of ML force field models for quantum MD was also emphasized in the original works of Behler and Parrinello~\cite{behler07}. A set of feature variables called the atom-centered symmetry functions (ACSFs) are introduced to represent local atomic configurations such that the rotation and reflection symmetries are exactly incorporated into the atomic energy function~\cite{behler07,behler16}. The building blocks of ACSFs are relative distances and angles of atomic position vectors, which are manifestly invariant under rotations of the SO(3) symmetry group~\cite{behler16}. The ACSF descriptor is physically intuitive, yet to some extent {\em ad hoc}, approach to ML force-field models for MD systems. Since then many atomic descriptors have been proposed and implemented~\cite{bartok10,li15,shapeev16,behler11,ghiringhelli15,bartok13,drautz19,himanen20,huo22}.

Since the JT models are defined on a lattice, the relevant symmetry group is reduced from the SO(3) rotation group of free space to the point group associated with the underlying lattice. Moreover, the JT modes are also characterized by well-defined transformation rules of the same point group. To describe the combined symmetry transformations, consider the neighborhood $\mathcal{C}_i$ centered at site-$i$, and a discrete rotation or reflection $\hat{g}$ of the point group that sends site-$j$ to $k$, i.e. $\mathbf R_{ki} = O(\hat{g}) \cdot \mathbf R_{ji}$, where $O(\hat{g})$ is the orthogonal matrix representation of~$\hat{g}$. The transformation of the octahedral distortions is described by
\begin{eqnarray}
	\tilde Q^{(\gamma,m)}_k = M^{(\gamma)}_{mn}(\hat{g}) \, Q^{(\gamma, n)}_j.
\end{eqnarray}
Here $\gamma$ indicates the irreducible representation (IR) of the vibronic modes, and $M^{(\gamma)}$ is the orthogonal transformation matrix of IR-$\gamma$, the indices $m, n = 1,2, \cdots n_{\gamma}$ label the different components in this IR. For example, for the double $\bm Q^{E}$, a matrix corresponding to a 90$^\circ$-rotation is $M^E(C_{\pi/4}) = {\rm diag}(-1, 1)$. 

Importantly, a proper representation of the neighborhood $\mathcal{C}_i$ needs to be invariant under these coupled symmetry transformations. A systematic approach to derive invariants of a symmetry group is the group-theoretical bispectrum method~\cite{kondor07}. Atomic descriptors based on bispectrum coefficients have been used in conjunction with Gaussian processing learning models for quantum MD simulations~\cite{bartok10,bartok13}. The group-theoretical method has also been employed to develop a general theory of descriptors for electronic lattice models in condensed-matter systems~\cite{zhang22,Ma19,Liu22,Tian23}. In particular, a descriptor based on the idea of reference IRs was developed for the Holstein model~\cite{cheng23}, which is essentially an electron-phonon model with the breathing modes. 

Here we outline the group-theoretical derivation of invariant feature variables for the JT models; more details can be found in Appendix~\ref{app:descriptor}. First, the octahedral distortions as represented by the set $\mathcal{C}_i$ essentially form a high-dimensional representation of the point group. They can then be decomposed into fundamental IRs of the point group. This decomposition can be highly simplified as the original representation matrix is automatically block-diagonalized, with each block corresponding to a fixed distance from the center-site. Standard methods can then be applied to the decomposition of each block~\cite{hamermesh62}. 

For the square-lattice JT system described by the $D_4$ point-group symmetry, there are three types of blocks with a dimension of either 4 or 8. The derivation of the relevant IRs for each block can be further simplified. This is because while the JT modes $\bm Q^E$ transform as a doublet in the octahedral group $O_h$, they are reduced to the direct sum of two 1D IRs when restricted to the $D_4$ group. Explicitly, under symmetry operations of $D_4$, both $Q^{A_1}$ and $Q^z$ transform as $A_1$ IR, while $Q^x$ transforms according to IR $B_1$, acquiring a $-1$ under reflections about the $y= \pm x$ diagonals. Take the vibronic modes $\{ \bm{\mathcal{Q}}_A, \bm{\mathcal{Q}}_B, \bm{\mathcal{Q}}_C, \bm{\mathcal{Q}}_D \}$ at the four nearest-neighbor sites as an example, the four $Q^z$ modes can be decomposed as $4Q^z = 1A_1 + 1B_1 + 1E$, with the following IR coefficients:
\begin{eqnarray}
\label{eq:type1-nb-z}
 f^{A_1} &= Q^z_{A}+Q^z_{B}+Q^z_{C}+Q^z_{D}, \nonumber \\
 f^{B_1} &= Q^z_{A}-Q^z_{B}+Q^z_{C}-Q^z_{D}, \\
 \bm f^E &= (Q^z_{A}-Q^z_{C}, Q^z_{B}-Q^z_{D}). \nonumber
\end{eqnarray}
The decomposition of the breathing $Q^{A_1}$ modes is described by the same formulas. On the other hand, while the four JT $Q^x$ modes are also decomposed as $4Q^x = 1A_1 + 1B_1 + 1E$, the $B_1$ symmetry of $Q^x$ gives rise to different IR coefficients: 
\begin{align}
\label{eq:type1-nb-x}
 f^{A_1} &= Q^x_{A}-Q^x_{B}+Q^x_{C}-Q^x_{D}, \nonumber \\
 f^{B_1} &= Q^x_{A}+Q^x_{B}+Q^x_{C}+Q^x_{D}, \\
 \bm f^E &= (Q^x_{A}-Q^x_{C}, -Q^x_{B}+Q^x_{D}). \nonumber
\end{align}
The calculation of IR coefficients for other types of blocks is presented in Appendix~\ref{app:descriptor}.

For convenience, we arrange the coefficients of the $r$-th IR of type $\Gamma$ in the overall decomposition of $\mathcal{C}_i$ into a vector $\bm f^{\Gamma}_r = (f^{(\Gamma, r)}_1, f^{(\Gamma, r)}_2, \cdots, f^{(\Gamma, r)}_{n_\Gamma} )$. As the sum of the squared coefficients of an IR, $p^{\Gamma}_r = \left| \bm f^{\Gamma}_r \right|^2$, is manifestly invariant under the point group, the set of all amplitudes $\{ p^\Gamma_r \}$, known as the power spectrum, offers a set of invariant feature variables. The power spectrum is a subset of more general invariant variables called the bispectrum coefficients~\cite{kondor07,bartok13}. A bispectrum coefficient is a product of three IR coefficients and the Clebsch-Gordon coefficients which account for the different transformation properties of the three IRs. Importantly, the relative phases between different IRs are encoded in these coefficients, which provide a faithful invariant representation of the neighborhood. 

For most point groups, the dimensions $n_\Gamma$ of individual IRs are small, which means there is a large multiplicity (indexed by $r$) for each IR. This in turn results in a large number of possible bispectrum coefficients, often with considerable redundancy. A more economic approach to encode the phase information is the idea of reference IR coefficients $\bm f^\Gamma_\text{ref}$, one for each IR type of the point group~\cite{zhang21}. These reference IR coefficients are derived using the same decomposition formulas, but based on distortions $\overline{\bm{\mathcal{Q}}}$ obtained by averaging large blocks of bond and chirality variables in the local neighborhood in order to reduce sensitivity to small variations.   Importantly, the reference IR allows one to introduce a ``phase" variable for each IR in the decomposition $\exp({\phi^\Gamma_r}) \equiv \bm f^\Gamma_r \cdot \bm f^\Gamma_{\rm ref} / |\bm f^\Gamma_r |\, |\bm f^\Gamma_{\rm ref}| = \pm 1$. The relative phase between IRs of the same type can then be inferred from their respective phases relative to the reference. Finally, the relative phases between IRs of different types are provided by the bispectrum coefficients from reference IR alone. 

The power spectrum can be combined with the phases to form invariant feature variables $G^\Gamma_r = p^\Gamma_r \,\exp(i \phi^\Gamma_r)$. These are to be supplemented by the bispectrum coefficients $B^{\Gamma1, \Gamma_2, \Gamma_3}_{\rm ref}$ obtained from the reference IR; see Appendix~\ref{app:descriptor} for details. The descriptor can be summarized by the following sequence of representations of the neighborhood
\begin{eqnarray}
	\label{eq:feature-variables}
	\bigl\{ \bm{\mathcal{Q}}_j \bigr\}  \,\, \to \,\, \bigl\{ \bm f^\Gamma_r \bigr\} \,\, \to \,\, \bigl\{ G^\Gamma_r, B^{\Gamma_1, \Gamma_2, \Gamma_3}_{\rm ref} \bigr\}.
\end{eqnarray}
As symmetry-related configurations are represented by exactly the same feature variables, the site-energy $\epsilon_i$ at the output of the NN is guaranteed to be the same, thus ensuring that the symmetry is preserved in the ML model.

\subsection{Implementation details and benchmarks}

Here we used PyTorch~\cite{paszke19} to construct fully connected neural networks with six hidden layers. The number of neurons in successive hidden layers are: $1024 \times 512 \times 256 \times 128 \times 64 \times 64$. With a cutoff radius of $r_{c}=7a$, where $a$ is the lattice constant, for defining the size of the neighborhood, the number of neurons at the input layer is determined by the number of feature variables and is fixed at 450.  The ReLU function is used as the activation function between layers~\cite{nair10,barron17}. The NN model is trained based on a loss function including the mean square error (MSE) of both the effective field and total energy:
\begin{eqnarray}
	L = \mu_F \frac{1}{N}\sum_{i=1}^N \Bigl| \bm{\mathcal{F}}^{\rm ED}_i - \bm{\mathcal{F}}^{\rm ML}_i  \Bigr|^2 
	+ \mu_E \Bigl| E_{\rm ED} - E_{\rm ML} \Bigr|^2, \quad
\end{eqnarray}
where $\mu_H$ and $\mu_E$ determine the relative weights of the force and energy constraints in the loss function. Different combinations of these two weights have been experimented. Overall, a better performance was obtained by putting more emphasis on the force accuracy.   As shown in Eq.~(\ref{eq:f_ML}), the effective forces are obtained from the derivative of the sum of local energies. This can be efficiently done using automatic differentiation in PyTorch~\cite{paszke17}. Trainable parameters of the NN are optimized by the Adam stochastic optimizer~\cite{kingma17} with a learning rate of 0.001. A 5-fold cross-validation and early stopping regularization are performed to prevent overfitting.

The above ML framework was applied to the JT model at exactly half-filling and for a filling fraction of $f = 0.49$. The nearest-neighbor hopping parameter $t_{\rm nn}$ served as the energy unit. The competition between JT coupling and elastic energy gives a distortion scale of $Q_0 \sim \lambda / K$, indicating an energy scale $E_{\rm JT} \sim \lambda^2/K$. Comparing this with the electron bandwidth, this suggests a dimensionless coupling constant $\tilde\lambda = E_{\rm JT}/W \sim \lambda^2 / K t_{\rm nn}$, which is set to 1.35 throughout the simulations. The effective masses $\mu$ and the elastic constant determine a the time-scale for the classical vibronic dynamics: $\tau_0 = \Omega^{-1}$, where the characteristic frequency $\Omega = \sqrt{K/\mu}$. Given this time-scale, an effective damping coefficient $\gamma = 0.25 \tau_0^{-1}$ and a time-step $\Delta t = 0.05 \tau_0$ were used in our Langevin simulations. 
The training dataset was obtained from ED calculations of random distortions as well as ED-based Langevin dynamics simulations, both on a $30\times 30$ lattice.  It included $1700$ random configurations, $1300$ intermediate states during the relaxations, and $800$ nearly equilibrium states, a total of $\sim 3.4\times 10^6$ force data.

\begin{figure}
\centering
\includegraphics[width=0.99\columnwidth]{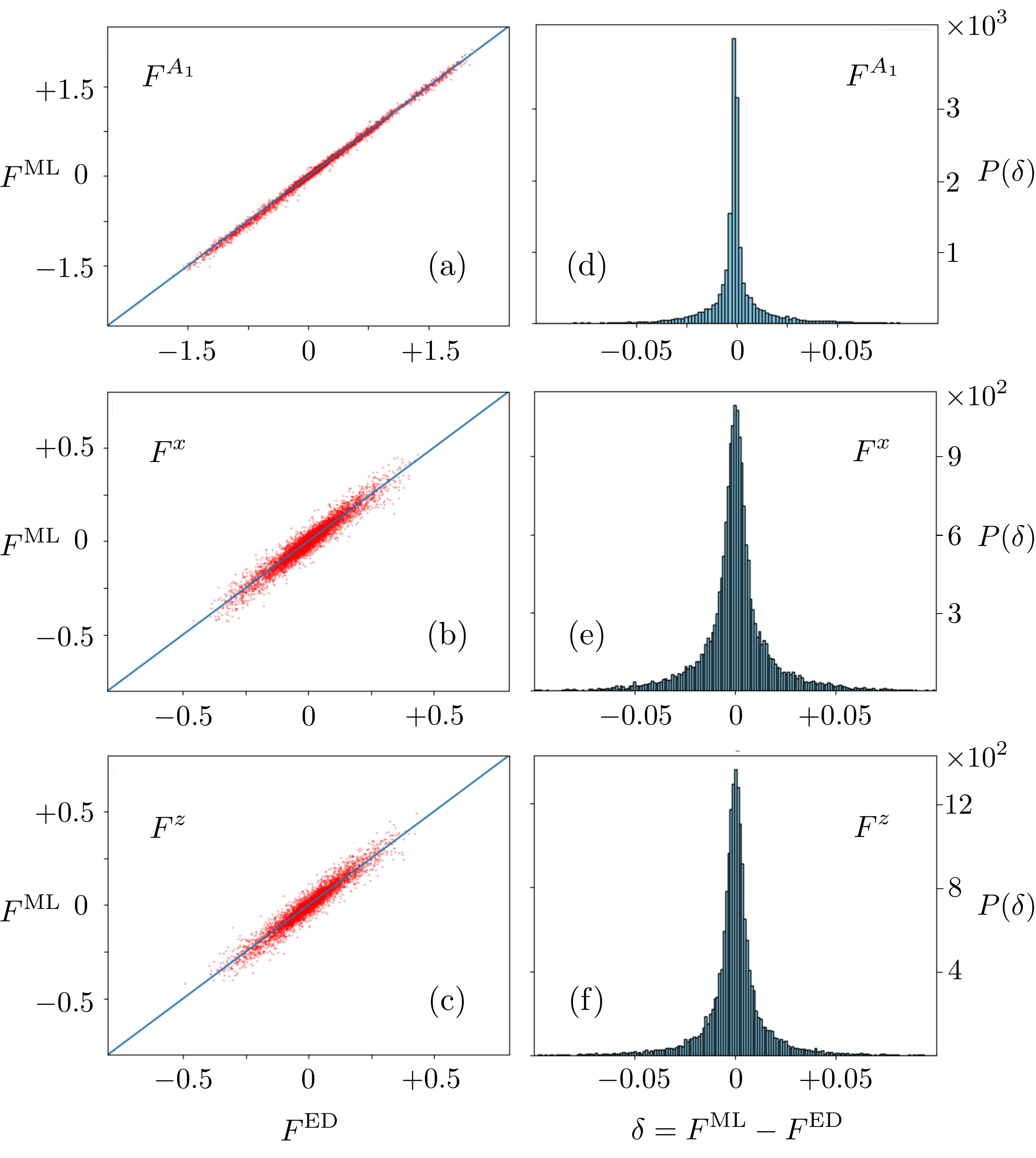}
\caption{Benchmark of the ML force-field models for JT model with a filling fraction $f = 0.49$. Panels~(a)--(c) on the left show the ML predicted forces $F^{\rm ML}$ versus the ground truth $F^{\rm ED}$ obtained from exact diagonalization method for the three vibronic modes of the octahedron. These forces are normalized by the electron-phonon coupling constant, which is set to $\lambda = 1.25$. The corresponding histograms of the prediction errors are shown in panels~(d)--(f). Similar results are also obtained for the ML models for the half-filling case.}
\label{fig:forces}
\end{figure}

Although the ML method is originally designed to model the time-consuming electronic structure calculations, the electronic contribution to the forces, i.e. the~$\langle \hat{\bm \tau}_i \rangle$ term in Eq.~(\ref{eq:forces}), exhibits a strong bimodal distribution as the system approaches the ground-state $C$-type orbital order. The two centers of the bimodal distribution correspond to the $d_{3x^2-r^2}$ and $d_{3y^2-r^2}$ orbitals in the checkerboard pattern. Since the net forces approach zero in such quasi-equilibrium states, the electronic force nearly cancels the classical part. The bimodal distribution can thus be attributed to a dominant leading-order dependence of the electronic force on the on-site JT distortion through the classical force, i.e. $\langle \hat{\bm\tau}_i \rangle = (K/\lambda) \bm Q^E_i + \bm h( \mathcal{C}_i)$; the second term here encodes the weaker yet subtle dependence on the neighborhood distortions. 
To avoid difficulty due to this bimodal distribution, it is more efficient to remove the dominant on-site classical term and focus the ML training on the second term. On the other hand, Eq.~(\ref{eq:forces}) shows that this function is simply proportional to the total force: $\bm h(\mathcal{C}_i) = \bm F^{E}_i / \lambda$. Indeed, our ED-Langevin simulations show that the total forces on the JT modes are well described by a Gaussian-like distribution.

\begin{figure}[t]
\centering
\includegraphics[width=0.99\columnwidth]{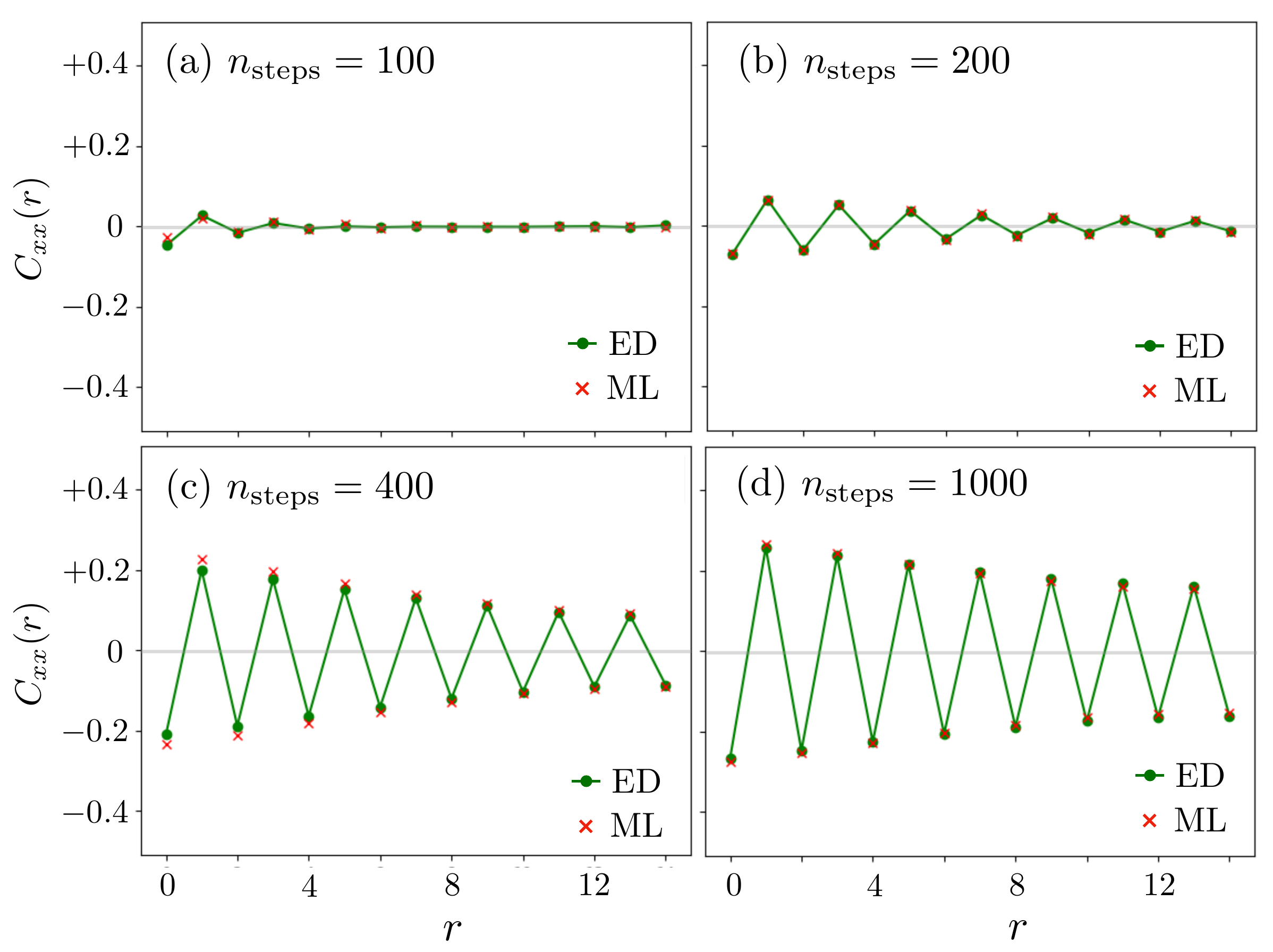}
\caption{Comparison of time-dependent correlation functions $C^{xx}(r, t)$ defined in Eq.~(\ref{eq:Cxx}) obtained from Langevin simulations with the ML force-field model and the ED method. The correlation functions were obtained from 40 independent thermal quench simulations on a $30\times 30$ lattice with a filling fraction $f = 0.49$. }	
\label{fig:corr}
\end{figure}

In addition to the generalized BP approach described above, we also implemented a modified ML scheme in which a separate NN is used to predict the forces acting on the breathing modes~$Q^{A_1}$. From the symmetry viewpoint, this is feasible since the breathing mode as well as its corresponding force $F^{A_1}$ are scalars, They are invariant under the point-group symmetry transformations, and the same descriptor can be also used for the corresponding NN model. The fact that no automatic differentiation is required for the breathing modes also enhances the efficiency. But more importantly, we found that the prediction accuracy of the $F^{A_1}$ forces is significantly improved. With this hybrid approach, rather accurate force predictions were achieved for all three vibronic modes, as shown in FIG.~\ref{fig:forces}. The histogram of prediction errors $( \delta = F^{ML} - F^{ED})$, shown on the right panels of FIG.~\ref{fig:forces}, are characterized by a small MSE of~$\sigma_F \sim 0.005$ for the force predictions.

It is worth noting that both the electron density and the local orbital configuration can be directly obtained from the ML model, thanks to the local and linear coupling between electrons and octahedral modes. As shown in Eq.~(\ref{eq:forces}), for example, the expectation value of orbital pseudo-spin can be obtained as: $\langle \hat{\bm\tau}_i \rangle = ( \hat{\bm F}^{E}_i + K \bm Q^E_i) / \lambda$, where $\hat{\bm F}^E_i = \bm h(\mathcal{C}_i)$ is the total doublet force predicted by the ML model.  Similarly, the on-site electron number can also be obtained from the predicted forces $\hat{F}^{A_1}_i$ for the breathing mode: $\langle \hat{n}_i \rangle = (\hat{F}^{A_1}_i + \beta K Q^{A_1}_i) / \lambda$.

\begin{figure*}[t]
\centering
\includegraphics[width=1.99\columnwidth]{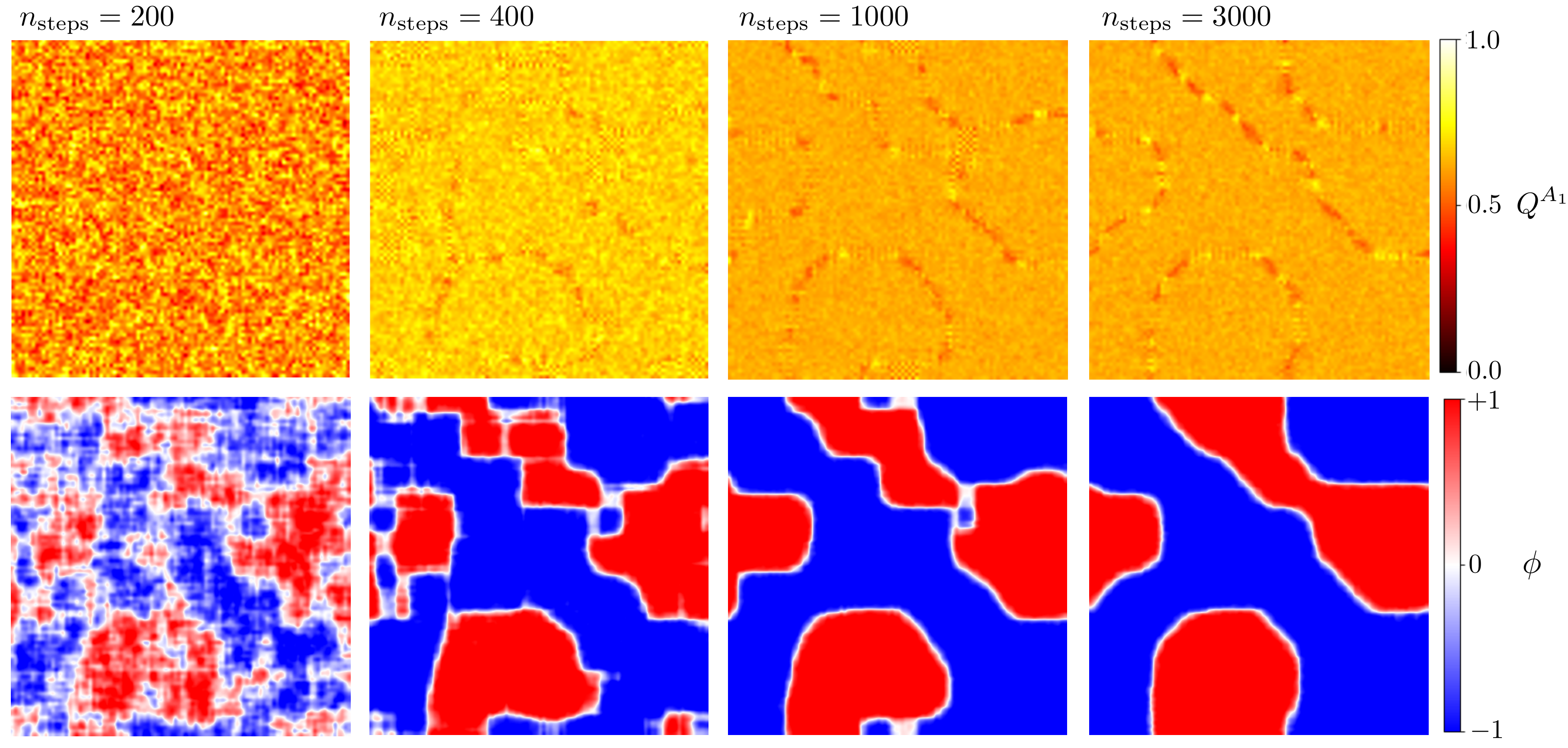}
\caption{Snapshots of local breathing mode $Q^{A_1}$ and orbital Ising order parameter $\phi_i$ at various time steps after a thermal quench of a half-filled $f = 0.5$ JT model.  An initially random configuration is suddenly quenched to a temperature $T = 0.001$ at time $t = 0$ ($n_{\rm step} = 0$). The ML-Langevin dynamics was used to simulate the relaxation of the system toward equilibrium. The red and blue regions correspond to orbital domains with order parameters $\phi = +1$ and $-1$, respectively. A time step $\Delta t = 0.05 \tau_0$ is used in the simulations.   }	
\label{fig:snapshots}
\end{figure*}

Next we integrated the trained ML models into the Langevin dynamics method and conducted thermal quench simulations of the JT model. We initiated the simulations with a random initial state, which was suddenly quenched to a temperature of $T = 0.01$ at $ t = 0$. The results from the ML-Langevin simulations were then compared with those from ED simulations. As the ground-state is characterized by a checkerboard pattern of the $Q^x$ JT distortions, we computed the correlation function 
\begin{eqnarray}
	\label{eq:Cxx}
	C^{xx}(r_{ij}) = \langle Q^x_{i}Q^x_{j}\rangle 
\end{eqnarray}
between the $Q^x$ modes at two octahedra separated by a distance $r_{ij}$ along either the $x$- or the $y$-direction. FIG.~\ref{fig:corr} displays the time-dependent correlation functions obtained from both ML and ED Langevin simulations, demonstrating excellent agreement. The relaxation process is characterized by a gradual built-up of staggered correlations of the $Q^x$ distortion, characteristic of the $C$-type orbital/JT oder. This dynamical benchmark shows that the ML model not only accurately predicts the forces, but also captures the dynamical evolution of the JT system.


\section{Coarsening of orbital order in cooperative Jahn-Teller model}
 
\label{sec:coarsening}

The ML force-field model is applied to study the large-scale coarsening dynamics of orbital order in the JT model. In addition to the linear scalability of ML method, the efficiency is further enhanced by running the simulations on GPU machines. We performed the thermal quench simulations on a $100 \times 100$ lattice where an initial state with random local distortions was suddenly cooled to a low temperature $T = 0.01$ at time $t = 0$. As discussed above, the low-temperature $C$-type orbital/JT order exhibits a broken $Z_2$ symmetry, which is physically related to both the sublattice and mirror transformation about the diagonals. For the case of half-filling, this ground state is also characterized by a uniform electron density of $\langle \hat{n}_i \rangle = 1$. The lattice distortions in the ground state can be described by two parameters:
\begin{eqnarray}
	\label{eq:Q-gs}
	Q^{A_1}_i = \eta, \qquad \bm Q^E_i = (\delta, 0) \exp( i \mathbf K \cdot \mathbf r_i).
\end{eqnarray}
where $\bold{{K}}=(\pi,\pi)$ is the ordering wave vector, and the phase factor $\exp(i \mathbf K \cdot \mathbf r_i) = \pm 1$ for lattice sites at the two sublattices. The parameter $\eta = \lambda / \beta K$ is the amplitude of the uniform expansion of octahedra, while $\delta = \lambda \langle \hat{\tau}^x\rangle / K$ can be viewed as a global $Z_2$ order parameter for the staggered JT distortions.  To characterize the inhomogeneous states with multiple orbital domains after a thermal quench, we define a scalar order parameter that measures the local staggered JT distortion
\begin{eqnarray}
	\phi_i = \Bigl( Q^x_i - \langle Q^x_{\rm nn} \rangle_i \Bigr) \exp(i \mathbf K \cdot \mathbf r_i).
\end{eqnarray}
where $\langle Q^x_{\rm nn} \rangle_i \equiv (Q^x_{i + \hat{\mathbf x}} + Q^x_{i - \hat{\mathbf x}} + Q^x_{i + \hat{\mathbf y}} + Q^x_{i - \hat{\mathbf y}})/4$ denotes the average JT distortion at the four nearest neighbors of site-$i$. A nonzero $\phi_i$ thus indicates the presence of a local difference in JT distortion. Indeed, this Ising order parameter is $\phi_i = \delta$ in the long-range staggered JT order described in Eq.~(\ref{eq:Q-gs}).

\begin{figure*}[t]
\centering
\includegraphics[width=1.99\columnwidth]{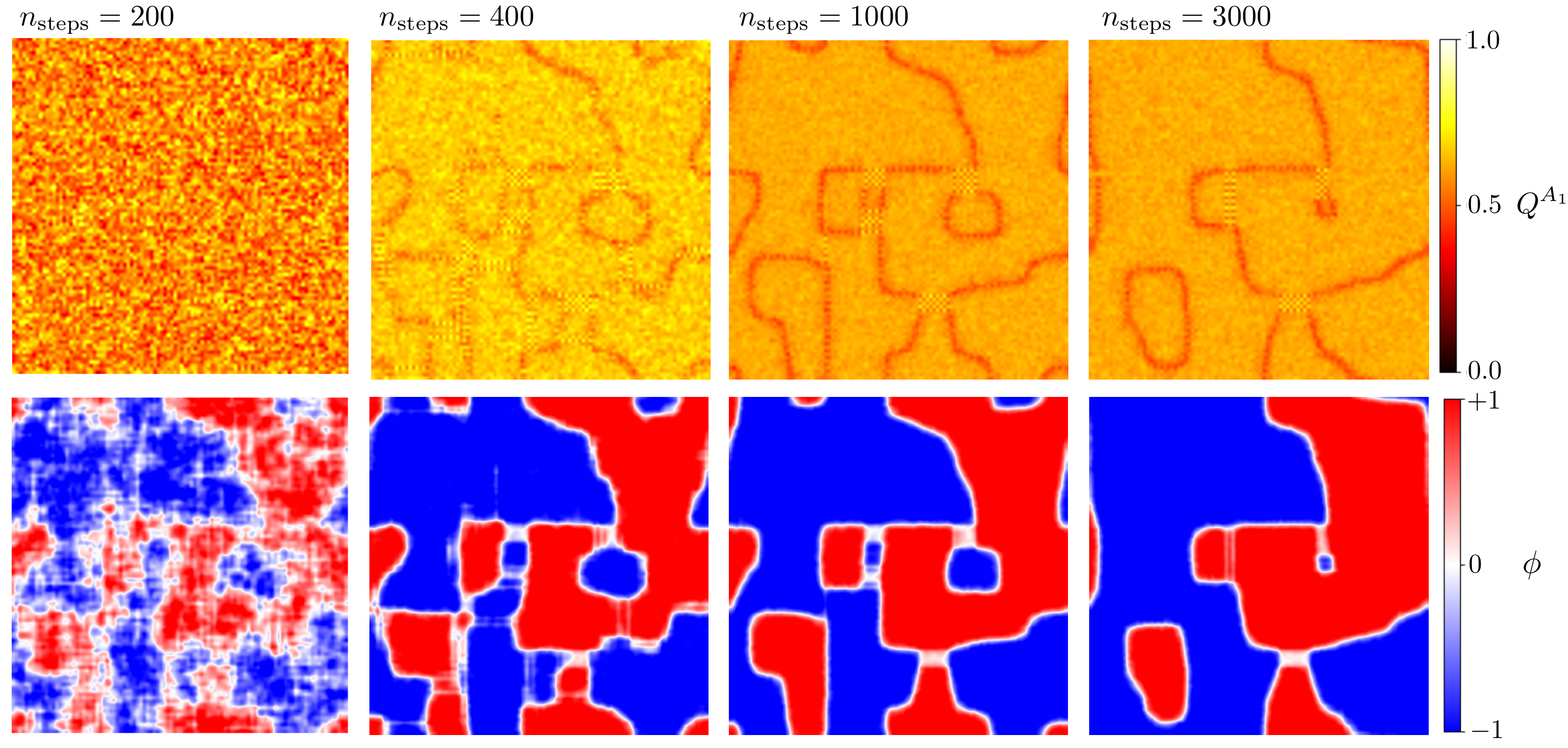}
\caption{Snapshots of local breathing mode $Q^{A_1}$ and orbital Ising order parameter $\phi_i$ at various time steps after a thermal quench for a JT model with $f = 0.49$ electron filling.   }	
\label{fig:snapshots2}
\end{figure*}

 

FIG.~\ref{fig:snapshots} show the snapshots of the breathing-mode amplitude~$Q^{A_1}_i$ and the local Ising order parameter $\phi_i$ at different times after a thermal quench of a half-filled JT model. The red and blue regions, corresponding to $\phi_i =+1$ and $-1$, respectively, are $C$-type orbital orders related by the $Z_2$ sublattice symmetry. The two types of orbital domains are separated by interfaces of vanishing $\phi_i$, corresponding to the white regions. On the other hand, a rather uniform ordering of $Q^{A_1}$ is quickly developed throughout the system, except at the interfaces of the two orbital domains. The nearly constant amplitude $Q^{A_1}$ corresponds to the uniform charge distribution $\langle \hat{n}_i \rangle \approx 1$ in the ground state as discussed above. At the interfaces that separate Ising domains of opposite signs, there are segments with an electron density both above and below the average value of one electron per site. 

The snapshots during the relaxation of a JT system with an electron filling of $f = 0.49$ are shown in FIG.~\ref{fig:snapshots2}. Overall, a relaxation behavior similar to that of the half-filled case was obtained. However, there is a major difference regarding the breathing mode and the electron density. In contrast to the milder inhomogeneity for the half-filling case, the amplitude of the $Q^{A_1}$ mode as well as the electron density are significantly reduced at the interfaces, as shown in the top panels of FIG.~\ref{fig:snapshots2}. As the $C$-type orbital/JT order is stabilized in the half-filling limit, this result is consistent with the phase separation scenario where nearly all of the doped holes go into the interfaces between orbital domains. 

\begin{figure}[b]
\centering
\includegraphics[width=0.99\columnwidth]{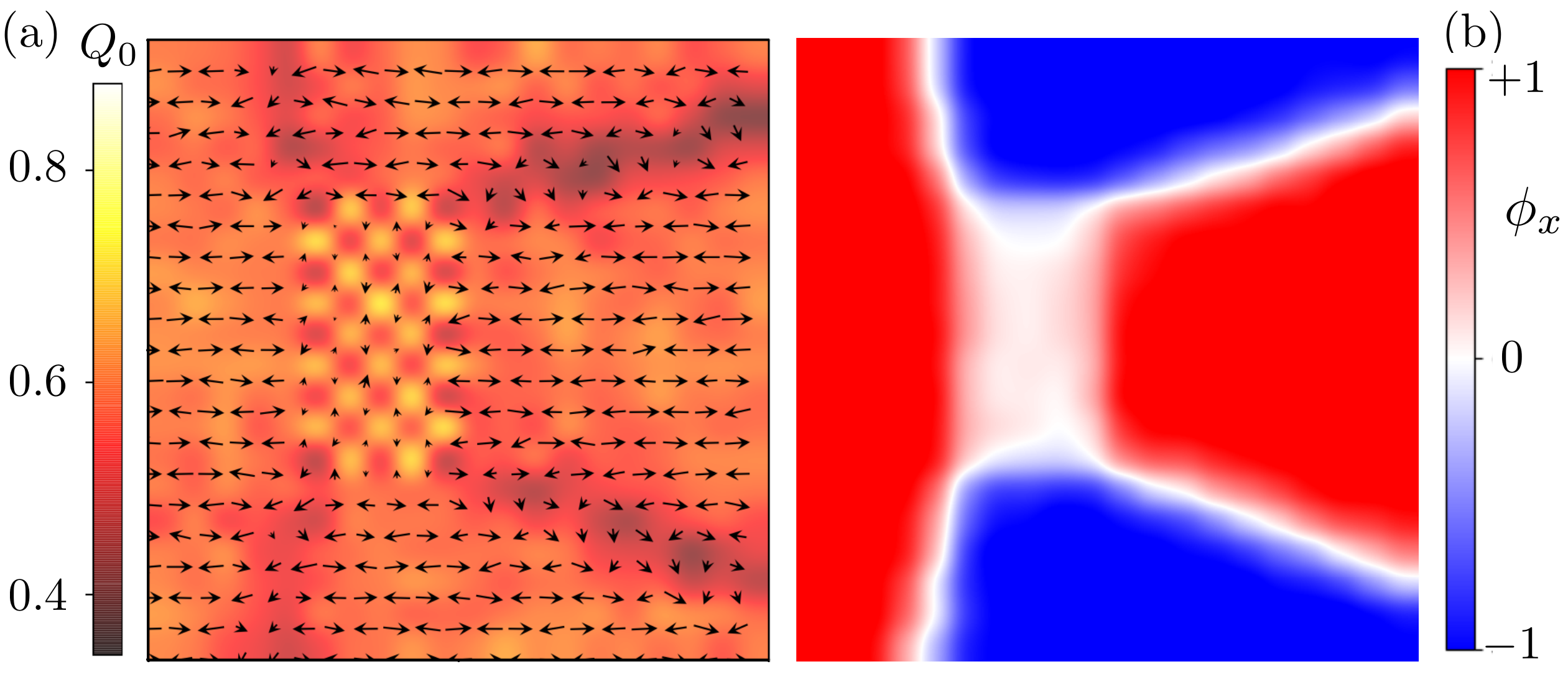}
\caption{Localized checkerboard modulation of breathing mode~$Q^{A_1}$ and electron density. An example of such local structures is displayed in terms of (a) the amplitude of breathing mode as well as the $\bm Q^E$ vector and (b) the orbital Ising order parameter.  }	
\label{fig:Q1_CDW}
\end{figure}

Another intriguing feature of the $Q^{A_1}$ configurations is the localized checkerboard modulations of the breathing-mode amplitudes. Such localized checkerboard modulations can be seen in both filling fractions, but more pronounced in the $f = 0.49$ case. A close-up view of this localized structure is shown in FIG.~\ref{fig:Q1_CDW}(a). A region of checkerboard modulation of the $Q^{A_1}$ amplitudes is enclosed in orbital domains represented by a N\'eel-type order of the doublet $\bm Q^E$ vectors predominantly in the $Q^x$ direction. 
As the electron number $\hat{n}_i$ directly couples to the breathing mode, this checkerboard structure also indicates a concomitant charge density wave (CDW) characterized by the wave vector $\mathbf K = (\pi, \pi)$. It is worth noting that these CDW states are essentially the same as the CDW state of the Holstein model at half-filling~\cite{noack91,hohenadler19,esterlis19}. 
In terms of the orbital Ising parameter $\phi$, shown in FIG.~\ref{fig:Q1_CDW}(b), the checkerboard pattern is accompanied by a vanishing orbital order, similar to the interface that separates different orbital domains. However, the average electron density of the CDW is found to be exactly one electron per site, which means that the doped holes are not accommodated in such local structures.

It is worth noting that these localized CDW structures are also obtained in the ED-based Langevin simulations on smaller systems, indicating that they are not an artifact of ML models. However, such charge modulation patterns are mostly meta-stable during the relaxation process. Our ED simulations on a $30\times 30$ lattice found that a state initialized to a homogeneous CDW order is unstable and will decay into the $C$-type order. Yet, as shown in the snapshots from large-scale ML-Langevin simulations, e.g. FIG.~\ref{fig:snapshots2}, such local CDW order persists even at late stage of the phase-ordering where large domains of orbital order have been established. It is likely that the local CDW order is stabilized as an intermediate structure in a multi-domain state during relaxation.

Next we discuss the coarsening of the orbital domains. As discussed in Sec.~\ref{sec:intro}, the $C$-type orbital order is described by an Ising order parameter. And since the orbital order and JT distortion are not subject to a conservation law, the coarsening of the orbital/JT order is expected to be described by the universality class of the phase ordering of a non-conserved Ising order. 
It is worth noting that the coarsening dynamics of Ising order has been thoroughly characterized and classified into several super-universal classes which depend on whether the Ising order is conserved and the presence of quenched disorder. For non-conserved Ising order, which is also the case for the $C$-type orbital order here, the coarsening dynamics is described by a curvature-driven mechanism summarized in the Allen-Cahn equation~\cite{Allen1972}. This results in a specific power-law growth $L \sim t^{1/2}$ of Ising domains applicable to two and higher dimensions~\cite{Bray1994,Onuki2002,Puri2009}. A natural question is to see whether the coarsening of the orbital domains falls into this Allen-Cahn universality class.

\begin{figure}
\centering
\includegraphics[width=0.99\columnwidth]{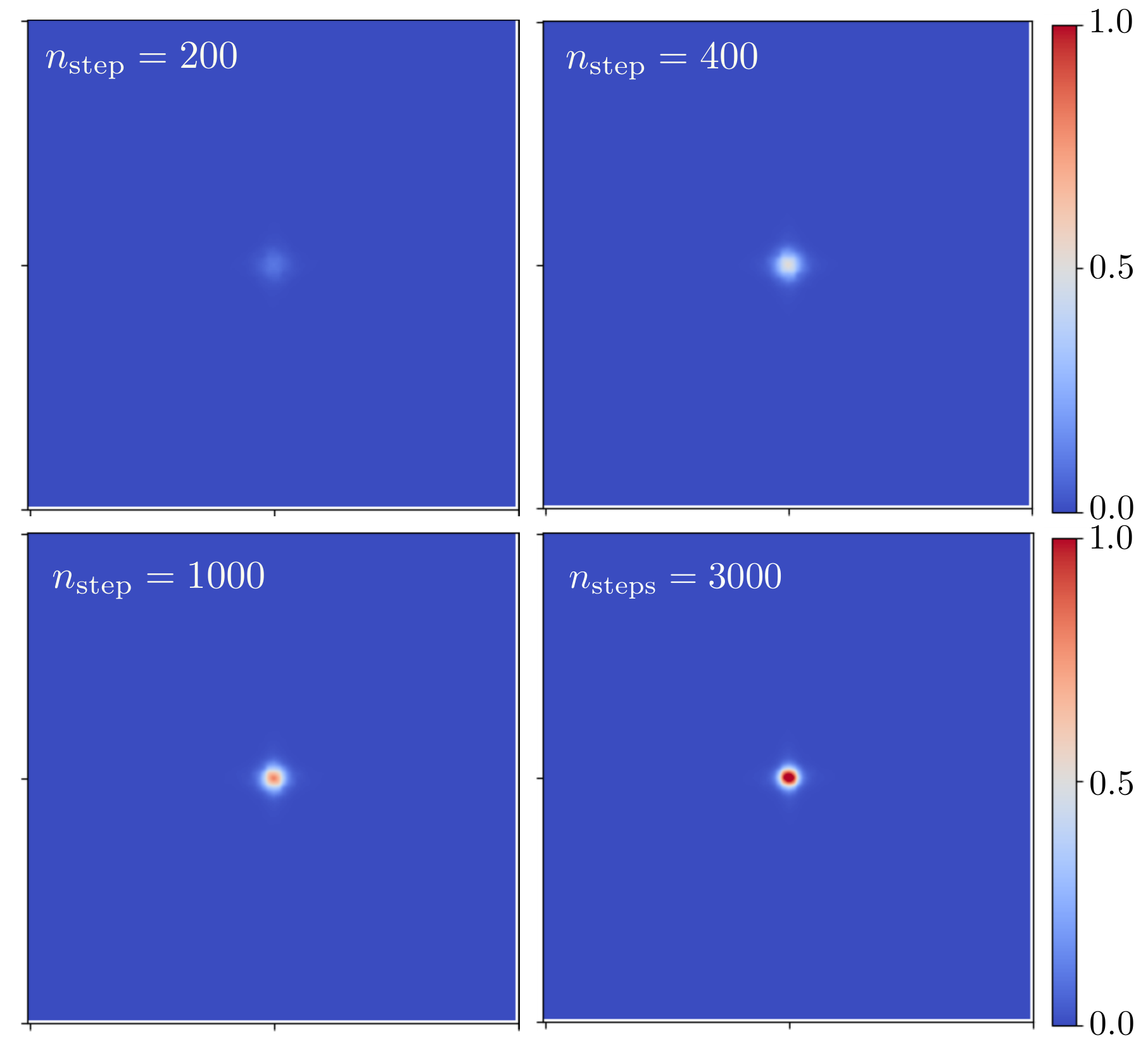}
\caption{Structure factor $S(\bold{k}, t)$ of the JT $Q^x$ distortion mode at different time steps after  suddenly cooled to a temperature $T = 0.001$, starting from a random state. The system size is $100\times 100$ with filling fraction $f=0.49$, and averaged over 40 independent configurations.}	
\label{fig:SF}
\end{figure}

To quantify the coarsening dynamics, we consider the time-dependent correlation length of the orbital/JT order. To this end, we first compute the time-dependent structural factor of the staggered distortion defined as $S(\mathbf k, t) = \left| Q^x(\mathbf k, t) \right|^2$, where $Q^x(\bold{k},t)$ is the Fourier transform the JT $Q^x$ configuration
\begin{eqnarray}
    Q^x(\bold{k},t)=\frac{1}{N}\sum_i Q^x_i \, e^{i \mathbf k \cdot \mathbf r_i }.
\end{eqnarray}
In FIG.~\ref{fig:SF} the structure factor is plotted at various time steps after a sudden quench to a low temperature $T = 0.01$. The emergence of the $C$-type order corresponds to a structure-factor peak at the wave vector $\bold{K} = (\pi, \pi)$. However, instead of a delta peak which is characteristic of a long-range order, a diffusive peak was observed even at late times of the phase ordering. The finite width of the diffusive peak is due to the presence of multiple orbital domains of opposite Ising orders. As the system progresses towards equilibrium, the coarsening of these ordered domains results in a stronger and sharper peak at $\bold{K}$, as can be seen in FIG.~\ref{fig:SF}.  The inverse of the width can thus provide a quantitative estimate for the characteristic length scale of ordered domains
\begin{eqnarray}
	L^{-1}(t) = \Delta k = {\sum_{\bold{k}}  S(\bold{k},t) |\bold{k}-\bold{K}|} \Big/ {\sum_{\bold{k}} S(\bold{k},t) },
\end{eqnarray}
This characteristic size of orbital domains also provides a measure of the orbital correlation, i.e. 
\begin{eqnarray}
	C^{xx}(\mathbf r, t) \sim e^{-|\mathbf r|/\xi(t)} e^{i \mathbf K \cdot \mathbf r},
\end{eqnarray}
where the correlation length is proportional to the characteristic domain size, $\xi \sim L$.

\begin{figure}[t]
\centering
\includegraphics[width=0.95\columnwidth]{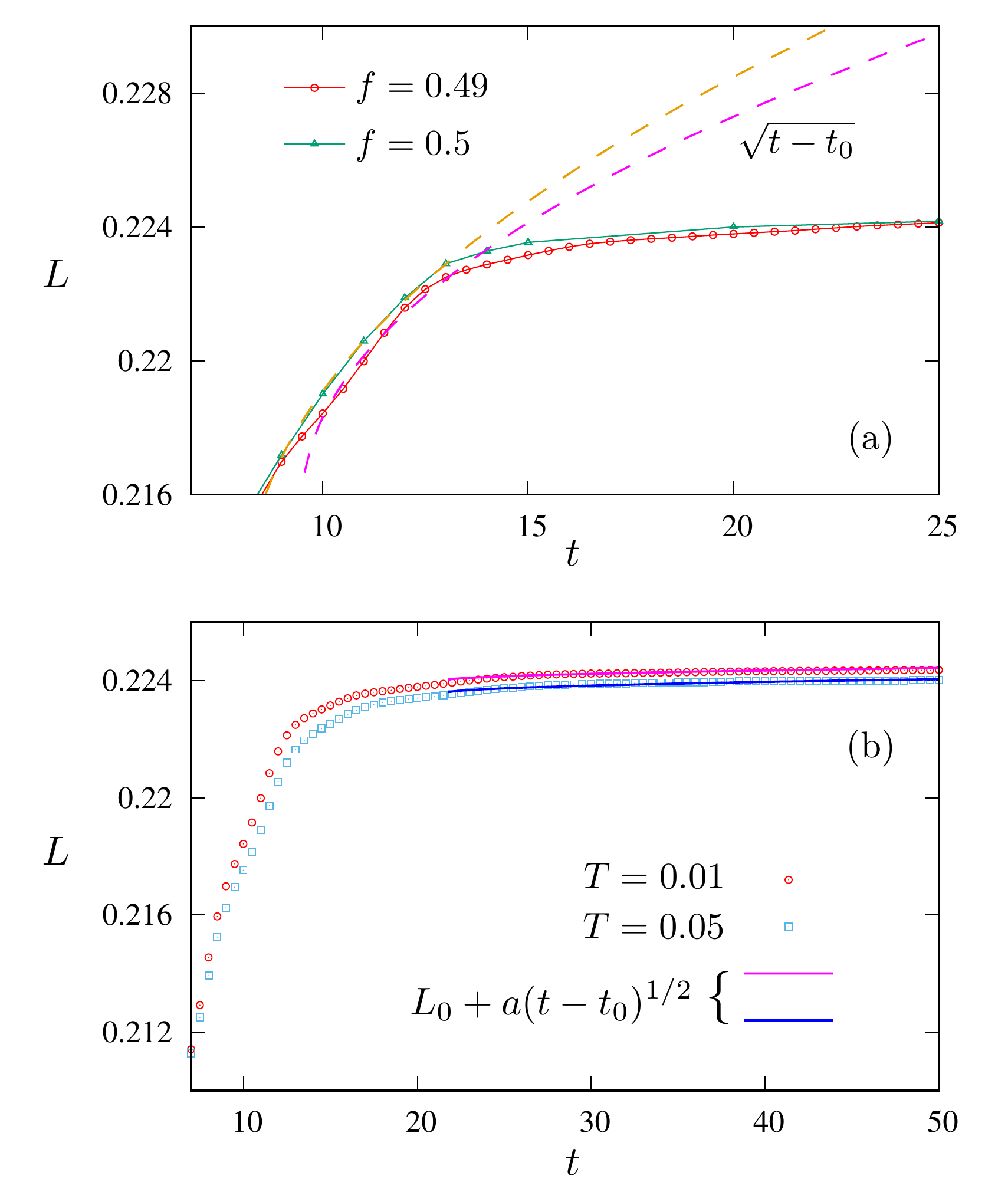}
\caption{The characteristic length $L(t)$ of orbital domains as a function of time, averaging over $40$ independent ML-Langevin simulations. Panel (a) shows $L(t)$ curves for the case of half-filling and a filling fraction $f = 0.49$. Panel~(b) shows the $L(t)$ curves at two different temperatures for the case of $f = 0.49$ electron filling. The two dashed lines in (a) and the two solid lines in (b) are fit using the Allen-Cahn growth law with a shifted time $t_0$ and an initial length $L_0$. The fitted expansion coefficients are $a = 8.1\times 10^{-5}$ and $a = 9.1\times 10^{-5}$ for temperature $T = 0.001$ and $T=0.005$, respectively. }	
\label{fig:L_t}
\end{figure}

FIG.~\ref{fig:L_t} shows the characteristic length as a function of time for two filling fractions, $f = 0.5$ (one electron per site) and $f = 0.49$, both obtained by averaging over 40 independent ML-Langevin simulations on a $100\times 100$ system. The $L(t)$ curves of the $f = 0.49$ case are also shown for two different temperatures. For all cases, the coarsening of the orbital domains shows a clear two-stage behavior: there is an initial rapid domain growth, which is followed by a much slower coarsening dynamics. This is also consistent with the snapshots shown in both FIG.~\ref{fig:snapshots} and~\ref{fig:snapshots2}. The initial fast stage, as represented by the two $n_{\rm steps} = 200$ configurations, is characterized by emerging orbital domains accompanied by a featureless random distribution of $Q^{A_1}$ and electron density. However, once the orbital orders are well developed in each domain at $n_{\rm steps} \gtrsim 400$, or for times $t \gtrsim 20 \tau_0$, the coarsening rate is significantly decreased.  

By fitting the initial relatively fast growth with the Allen-Cahn law, as shown as dashed lines in FIG.~\ref{fig:L_t}(a), one can see that the $L(t)$ from simulations starts to deviate from the $\sqrt{t}$ behavior for $t \gtrsim 15$. On the other hand, the snapshots shown in FIG.~\ref{fig:snapshots} and~\ref{fig:snapshots2} clearly show that there is still coarsening going on even at late times of the relaxation. If we attempt to fit this late-stage coarsening using the equation $L(t) = L_0 + a (t - t_0)^{1/2}$, the extracted expansion coefficient $a \sim 10^{-4}$ is extremely small. This corresponds to a relative domain growth of the order of $\Delta L / L_0 \sim 10^{-3}$ over a time span of 30 oscillation cycles $t_0$ of the JT modes. This indicates a freezing behavior at the late stage of the phase ordering regardless of whether the phase-ordering can be described by the Allen-Cahn domain growth.

It is worth noting that the Allen-Cahn domain growth law is intimately related to the domain morphologies of the inhomogeneous states. Fundamentally, the Allen-Cahn equation for the domain-wall motion describes a curvature-driven coarsening process. Mathematically, the domain-wall velocity is given by $v = -r \kappa$, where  $\kappa$ is the curvature, and $r$ is a proportional constant. Indeed, this equation implies a power-law domain growth. For inhomogeneous states characterized by a characteristic length $L$, the domain-wall velocity can be approximated by $v \sim dL/dt$, while the curvature is on average given by $\kappa \sim 1/L$. The Allen-Cahn equation indicates a differential equation $dL/dt \sim -1/L$, which can be readily integrated to give the growth behavior $L \sim t^{1/2}$.

Within the framework of curvature-driven domain growth, the late-stage freezing observed in our simulations could be attributed to the rather straight interfaces that separate orbital Ising domains of opposite signs; see FIG.~\ref{fig:snapshots} and \ref{fig:snapshots2}. Moreover, the nearly straight interfaces tend to run parallel along the $x$ or $y$-directions. As discussed in Sec.~\ref{sec:intro}, since the effective orbital interactions originate from the electrons through a mechanism similar to RKKY interactions, this interfacial anisotropy can be attributed to the highly directional hopping of $e_g$ electrons. The nearly zero curvature of such straight domain boundaries indicate a vanishing velocity for the domain-wall motion. Consequently, the coarsening is mostly driven by the corner regions of orbital domains, where a finite interfacial curvature remains. Let $\kappa_c$ be a characteristic curvature of the corner regions, the corresponding length scales of curved interfaces is $\ell \sim 1/\kappa_c$. It should be noted that this curvature $\kappa_c$ is determined by the competition of the electron and elastic energies of the JT model.  When typical domain size $L$ is greater than this length scale, curvature driven domain-wall motion is suppressed by the interfacial stiffness. This indicates a threshold length scale $L_{\rm th} \sim \ell \sim 1/\kappa_c$ such that orbital domains with a linear size $L \gtrsim L_{\rm th}$ start to show freezing behavior. The $L(t)$ curves shown in FIG.~\ref{fig:L_t} seem to be consistent with this threshold scenario.


\section{Conclusion and outlook}

\label{sec:conclusion}

To summarize, we have presented a comprehensive study on the coarsening dynamics of orbital order in a cooperative JT model using a scalable ML force-field approach. Within the adiabatic approximation, the dynamical evolution of the orbital order is governed by the dynamics of the MO$_6$ octahedra with a driving force computed from a quasi-equilibrium electron liquid. We have generalized the Behler-Parrinello ML force-field method, which was originally proposed for ML-based quantum MD simulations, to model the vibronic dynamics of cooperative JT systems. Assuming the locality principle for the electronic forces, a deep-learning neural network was developed to accurately encode the complex dependence of the effective forces on local distortion configurations. In order to preserve the symmetry of the original Hamiltonian, a lattice descriptor based on the group-theoretical bispectrum method was developed to incorporate of the original Hamiltonian.

The ML models  are  trained by datasets from exact diagonalization of $30\times 30$ systems. By integrating the ML force-field model with the Langevin dynamics simulation, our dynamical benchmarks showed that the ML model not only accurately predicts the local forces that drive the lattice dynamics, but also faithfully captures the dynamical evolution of the JT system. Importantly, compared with the exact diagonalization calculation for electronic structures, the ML algorithms significantly improve the time complexity for electronic force calculation which in turn reduces the time complexity for dynamical simulations by orders of magnitudes.

While numerous studies have devoted to the study of the rich low-temperature phases and the complex phase separation phenomena in CMR manganites, dynamical processes such as phase ordering, domain growth, and kinetics of phase separation, are much less explored theoretically. This is partly due to the technical difficulty of large-scale dynamical simulations for such complex systems. Our work on the dynamics of orbital ordering makes an important step towards the multi-scale dynamical modeling of manganites and other correlated electron systems. Our large-scale Langevin simulations enabled by the ML models uncovered an intriguing freezing phenomenon at the late stage of the phase ordering. The freezing dynamics is likely caused by the unique geometry of orbital domains which exhibit rather straight domain boundaries along the $x$ or $y$-directions. Within the framework of curvature-driven domain growth, the vanishing curvature of such straight interfaces implies a stagnated domain-wall motion.

The successful demonstration of the ML force-field approach for the adiabatic dynamics of orbital/JT systems also laid the groundwork for applying similar ML methods to more complicated electronic models of CMR manganites that include spin degrees of freedom. It is worth noting that a BP-type ML model has been developed for the adiabatic spin dynamics of the single-band ferromagnetic Kondo-lattice model~\cite{zhang20,zhang21}. As discussed in Sec.~\ref{sec:intro}, the DE mechanism in this model gives rise to a phase separation regime upon hole doping, which is an important ingredient for CMR effect.  It is envisioned a BP type neural network which predicts a local energy based on both magnetic and JT distortion configurations in an immediate neighborhood could provide a unified ML force-field model for such CMR systems.

It is worth noting that other linear-scaling numerical techniques, notably the kernel polynomial methods~\cite{weisse06,barros13,wang18}, can also be to solve quadratic fermion Hamiltonians, which is the case of the electron Hamiltonian in the cooperative JT model considered here. However, KPM and other similar $\mathcal{O}(N)$ methods cannot be directly generalized to fermionic models with electron-electron interactions such as the on-site Coulomb repulsion which has been shown to also play an important role in CMR manganites~\cite{held00,maezono98a,maezono98b,yang10}. On the other hand, the ML framework presented in this work provides a general linear-scaling approach to achieve linear scalability even for JT systems that include on-site Hubbard repulsion or other electron-electron interactions~\cite{roder96,millis95,millis96,millis96b,nagaosa98,maezono03,hotta00}. This is because, as discussed in Sec.~\ref{sec:ML}, linear-scaling electronic structure methods are fundamentally based on Kohn's locality principle. The divide-and-conquer approach intrinsic to the ML force-field models naturally take advantage of the locality property to achieve linear scalability. On the other hand, the ML models for systems with electron-electron interactions have to be trained by datasets from more sophisticated many-body techniques ranging from Hartree-Fock, Gutzwiller mean-field type methods to quantum Monte Carlo or dynamical mean-field theory.

\appendix

\section{Descriptor for JT systems}

\label{app:descriptor}

\begin{figure*}[t]
\centering
\includegraphics[width=1.98\columnwidth]{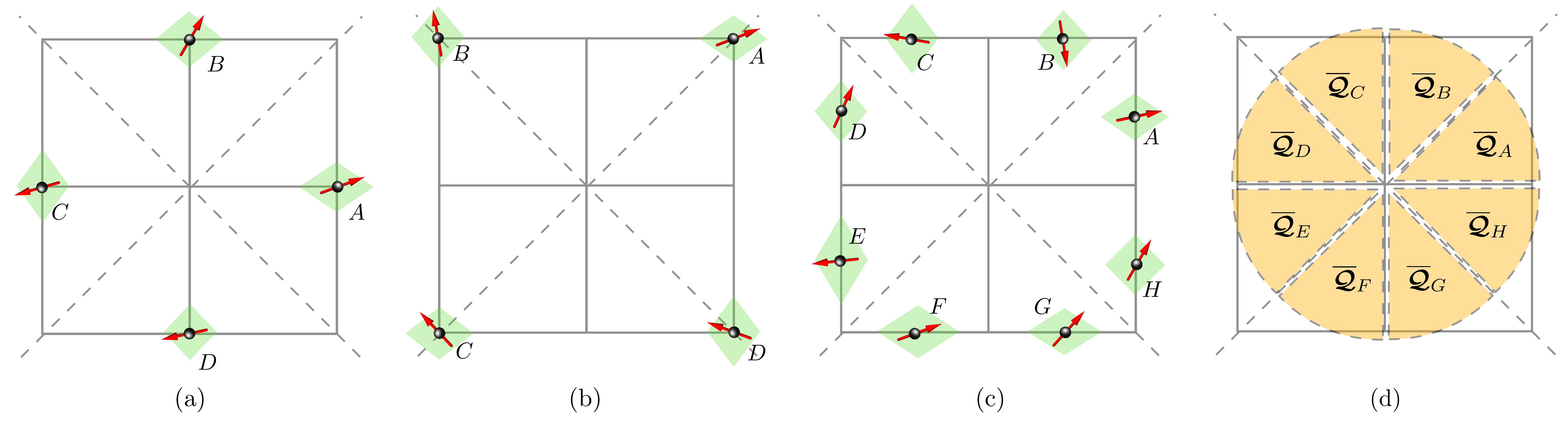}
\caption{Schematic diagrams of the three types of neighbor blocks: (a) type-I with four neighbor sites on the $x$ or $y$ axes, (b) type-II with four neighbors on the $y = \pm x$ diagonals, and (c) type-III with eight neighbors. The arrows represent the doublet vectors $\bm Q^E = (Q^x, Q^z)$. (d) An example of partitioning the neighborhood into eight symmetry-related regions. Average distortion vectors $\overline{\bm{\mathcal{Q}}}_K$ obtained from these regions are used to compute the reference IR coefficients $\bm f^{\Gamma}_{\rm ref}$ for the descriptor.   }	
\label{fig:neighbor-Q}
\end{figure*}

Here we present details of the calculations of IR coefficients. As mentioned in the main text, each block of the representation matrices corresponds to a group of neighbors at the same distance from the center site-$i$. For the square lattice, these invariant blocks can be classified into three types illustrated in FIG.~\ref{fig:neighbor-Q}. We first consider type-I neighbors labeled as $A$, $B$, $C$, and $D$ in panel~(a). Although the octahedral modes $\bm{\mathcal{Q}}_m$ at these four sites form a 12-dimensional representation of the $D_4$ group, it is automatically decomposed into three 4-dimensional blocks since the three components of $\bm{\mathcal{Q}}$ transform independent of each other. As discussed in Sec.~\ref{sec:lattice-descriptor}, both $Q^{A_1}$ and $Q^z$ behave as the singlet $A_1$ representation of $D_4$, while the $Q^x$ mode transforms according to the $B_1$ symmetry. Both are decomposed as $4 = A_1 + B_1 + E$, and the corresponding IR coefficients are listed in Eqs.~(\ref{eq:type1-nb-z}) and (\ref{eq:type1-nb-x}).

The type-II blocks correspond to neighbors sitting on the two 45$^\circ$ diagonals, as shown in FIG.~\ref{fig:neighbor-Q}(b). The decomposition of the breathing and $Q^z$ modes, both of $A_1$ symmetry, is given by $4 = A_1 + B_2 + E$. Take the $Q^z$ as an example, the IR coefficients are
\begin{align}
\nonumber f^{A_1} &= Q^z_{A}+Q^z_{B}+Q^z_{C}+Q^z_{D} \\
\nonumber f^{B_2} &= Q^z_{A}-Q^z_{B}+Q^z_{C}-Q^z_{D} \\
\nonumber f^E_x &= Q^z_{A}-Q^z_{B}-Q^z_{C}+Q^z_{D}, \\
\nonumber f^E_y & = Q^z_{A}+Q^z_{B}-Q^z_{C}-Q^z_{D}.
\end{align}
On the other hand, the four $Q^x$ modes are decomposed as $4 = A_2 + B_1 + E$, with the following linear combinations
\begin{align}
\nonumber f^{A_2} &= Q^x_{A}-Q^x_{B}+Q^x_{C}-Q^x_{D}, \\
\nonumber f^{B_1} &= Q^x_{A}+Q^x_{B}+Q^x_{C}+Q^x_{D},\\
\nonumber f^{E}_x &= Q^x_{A}-Q^x_{B}-Q^x_{C}+Q^x_{D},  \\
\nonumber f^E_y & = -Q^x_{A}-Q^x_{B}+Q^x_{C}+Q^x_{D}.
\end{align}
Finally, the type-III blocks correspond to a group of 8 neighbors shown in FIG.~\ref{fig:neighbor-Q}(c). Again, the 24-dimensional representation of the 8 $\bm{\mathcal{Q}}$ vectors is automatically decomposed into three 8-dimensional block. Both the $Q^{A_1}$ and $Q^z$ blocks are decomposed into six IRs: $8 = A_1 + A_2 + B_1 + B_2 + 2E$, with the following coefficients
\begin{align}
\label{eq:App_eq1}
\nonumber f^{A_1} &= Q^z_{A}+Q^z_{B}+Q^z_{C}+Q^z_{D}+ Q^z_{E}+Q^z_{F}+Q^z_{G}+Q^z_{H} \\
\nonumber f^{A_2} &= Q^z_{A}-Q^z_{B}+Q^z_{C}-Q^z_{D}+ Q^z_{E}-Q^z_{F}+Q^z_{G}-Q^z_{H}\\
\nonumber f^{B_1} &= Q^z_{A}-Q^z_{B}-Q^z_{C}+Q^z_{D}+ Q^z_{E}-Q^z_{F}-Q^z_{G}+Q^z_{H}\\
\nonumber f^{B_2} &= Q^z_{A}+Q^z_{B}-Q^z_{C}-Q^z_{D}+ Q^z_{E}+Q^z_{F}-Q^z_{G}-Q^z_{H}\\
\nonumber \bm f^{E} &= ( Q^z_{A}-Q^z_{D}-Q^z_{E}+Q^z_{H},  Q^z_{B}+Q^z_{C}-Q^z_{F}-Q^z_{G})\\
\nonumber \bm f^{E'} &= ( Q^z_{B}-Q^z_{C}-Q^z_{F}+Q^z_{G},  Q^z_{A}+Q^z_{D}-Q^z_{E}-Q^z_{H}) \\
\end{align}
The same decomposition also applies to the $Q^x$ blocks, but with different linear combinations
\begin{align}
\label{eq:App_eq2}
\nonumber f^{A_1} &= Q^x_{A}-Q^x_{B}-Q^x_{C}+Q^x_{D}+ Q^x_{E}-Q^x_{F}-Q^x_{G}+Q^x_{H}\\
\nonumber f^{A_2} &= Q^x_{A}+Q^x_{B}-Q^x_{C}-Q^x_{D}+ Q^x_{E}+Q^x_{F}-Q^x_{G}-Q^x_{H}\\
\nonumber f^{B_1} &= Q^x_{A}+Q^x_{B}+Q^x_{C}+Q^x_{D}+ Q^x_{E}+Q^x_{F}+Q^x_{G}+Q^x_{H} \\
\nonumber f^{B_2} &= Q^x_{A}-Q^x_{B}+Q^x_{C}-Q^x_{D}+ Q^x_{E}-Q^x_{F}+Q^x_{G}-Q^x_{H}\\
\nonumber \bm f^{E} &= ( Q^x_{A}-Q^x_{4}-Q^x_{E}+Q^x_{H},  -Q^x_{B}-Q^x_{C}+Q^x_{F}+Q^x_{G})\\
\nonumber \bm f^{E'} &= ( -Q^x_{B}+Q^x_{C}+Q^x_{F}-Q^x_{G},  Q^x_{A}+Q^x_{D}-Q^x_{E}-Q^x_{H}) \\
\end{align}
Using these formulas, one can systematically decompose the octahedral distortion modes $\bm{\mathcal{Q}}_j$ within the neighborhood into IRs. As discussed in Sec.~\ref{sec:lattice-descriptor}, the various coefficients of an IR in the decomposition is represented by a vector 
\[
	\bm f^{\Gamma}_r = (f^{(\Gamma, r)}_1, f^{(\Gamma, r)}_2, \cdots, f^{(\Gamma, r)}_{n_\Gamma} )
\]
where $\Gamma = A_1, A_2, B_1, B_2,$ or $E$ labels the IR of $D_4$ group, $n_\Gamma$ is the dimension of the IR, and $r$ enumerates the multiple occurrences of $\Gamma$ in the decomposition of $\mathcal{C}_i$. A set of invariants called the power spectrum can be readily obtained
\[
	p^{\Gamma}_r = \left| \bm f^{\Gamma}_r \right|^2.
\]
As discussed in the main text, a descriptor based only on power spectrum suffers from spurious symmetry since the relative phases between different IRs are not accounted for. A complete descriptor is given by the bispectrum coefficients~\cite{kondor07,bartok13}
\[
	B^{\Gamma, \Gamma_1, \Gamma_2}_{r, r_1, r_2} = \sum_{m, n, l} C^{\Gamma; \Gamma_1, \Gamma_2}_{m, n, l} 
	f^{\Gamma\, *}_{r, m} f^{\Gamma_1}_{r_1, n} f^{\Gamma_2}_{r_2, l},
\]
where $C^{\Gamma; \Gamma_1, \Gamma_2}_{m, n, l}$ are the Clebsch-Gordan coefficients of the point group, which are introduced to account for the different transformation properties of the three IRs. Although the bispectrum coefficients provide a faithful symmetry-invariant representation of the neighborhood distortion configurations, the rather large number of such coefficients combined with the fact that many of them are redundant render this an impractical approach for descriptors. 

A more efficient method for descriptor is based on the idea of reference IRs, which serve as an intermediate to store the phase information, as discussed in Sec.~\ref{sec:lattice-descriptor}. To obtain coefficients of these reference IRs, we first partition the neighborhood (defined as within a radius $r_c$ from the center site) into 8 symmetry-related regions, as shown in FIG.~\ref{fig:neighbor-Q}(d). An average distortion vector $\overline{\bm{\mathcal{Q}}}_K = \frac{1}{M} \sum_{j \in K} \bm{\mathcal{Q}}_j$ is obtained for each of the regions, where $M$ is the number of sites in each region. This construction has the advantage that the resultant reference IRs are less sensitive to small variations in the neighborhood.  The reference IR coefficients $\bm f^\Gamma_{\rm ref}$ are computed using Eqs.~(\ref{eq:App_eq1}) and (\ref{eq:App_eq2}). The reference bispectrum coefficients $B^{\Gamma, \Gamma_1, \Gamma_2}_{\rm ref}$ which are part of the feature variables in Eq.~(\ref{eq:feature-variables}) are obtained by substituting $\bm f^{\Gamma}_{\rm ref}$ into the above bispectrum equation. 

\bigskip 

\begin{acknowledgments}
The authors thank Puhan Zhang for useful discussions. The work was supported by the US Department of Energy Basic Energy Sciences under Contract No. DE-SC0020330. The authors also acknowledge the support of Research Computing at the University of Virginia.
\end{acknowledgments}

\bibliography{ref}

\begin{thebibliography}{95}%
\makeatletter
\providecommand \@ifxundefined [1]{%
 \@ifx{#1\undefined}
}%
\providecommand \@ifnum [1]{%
 \ifnum #1\expandafter \@firstoftwo
 \else \expandafter \@secondoftwo
 \fi
}%
\providecommand \@ifx [1]{%
 \ifx #1\expandafter \@firstoftwo
 \else \expandafter \@secondoftwo
 \fi
}%
\providecommand \natexlab [1]{#1}%
\providecommand \enquote  [1]{``#1''}%
\providecommand \bibnamefont  [1]{#1}%
\providecommand \bibfnamefont [1]{#1}%
\providecommand \citenamefont [1]{#1}%
\providecommand \href@noop [0]{\@secondoftwo}%
\providecommand \href [0]{\begingroup \@sanitize@url \@href}%
\providecommand \@href[1]{\@@startlink{#1}\@@href}%
\providecommand \@@href[1]{\endgroup#1\@@endlink}%
\providecommand \@sanitize@url [0]{\catcode `\\12\catcode `\$12\catcode
  `\&12\catcode `\#12\catcode `\^12\catcode `\_12\catcode `\%12\relax}%
\providecommand \@@startlink[1]{}%
\providecommand \@@endlink[0]{}%
\providecommand \url  [0]{\begingroup\@sanitize@url \@url }%
\providecommand \@url [1]{\endgroup\@href {#1}{\urlprefix }}%
\providecommand \urlprefix  [0]{URL }%
\providecommand \Eprint [0]{\href }%
\providecommand \doibase [0]{https://doi.org/}%
\providecommand \selectlanguage [0]{\@gobble}%
\providecommand \bibinfo  [0]{\@secondoftwo}%
\providecommand \bibfield  [0]{\@secondoftwo}%
\providecommand \translation [1]{[#1]}%
\providecommand \BibitemOpen [0]{}%
\providecommand \bibitemStop [0]{}%
\providecommand \bibitemNoStop [0]{.\EOS\space}%
\providecommand \EOS [0]{\spacefactor3000\relax}%
\providecommand \BibitemShut  [1]{\csname bibitem#1\endcsname}%
\let\auto@bib@innerbib\@empty
\bibitem [{\citenamefont {Tokura}\ and\ \citenamefont
  {Tomioka}(1999)}]{tokura99}%
  \BibitemOpen
  \bibfield  {author} {\bibinfo {author} {\bibfnamefont {Y.}~\bibnamefont
  {Tokura}}\ and\ \bibinfo {author} {\bibfnamefont {Y.}~\bibnamefont
  {Tomioka}},\ }\bibfield  {title} {\bibinfo {title} {Colossal magnetoresistive
  manganites},\ }\href
  {https://doi.org/https://doi.org/10.1016/S0304-8853(99)00352-2} {\bibfield
  {journal} {\bibinfo  {journal} {Journal of Magnetism and Magnetic Materials}\
  }\textbf {\bibinfo {volume} {200}},\ \bibinfo {pages} {1} (\bibinfo {year}
  {1999})}\BibitemShut {NoStop}%
\bibitem [{\citenamefont {Salamon}\ and\ \citenamefont
  {Jaime}(2001)}]{salamon01}%
  \BibitemOpen
  \bibfield  {author} {\bibinfo {author} {\bibfnamefont {M.~B.}\ \bibnamefont
  {Salamon}}\ and\ \bibinfo {author} {\bibfnamefont {M.}~\bibnamefont
  {Jaime}},\ }\bibfield  {title} {\bibinfo {title} {The physics of manganites:
  Structure and transport},\ }\href {https://doi.org/10.1103/RevModPhys.73.583}
  {\bibfield  {journal} {\bibinfo  {journal} {Rev. Mod. Phys.}\ }\textbf
  {\bibinfo {volume} {73}},\ \bibinfo {pages} {583} (\bibinfo {year}
  {2001})}\BibitemShut {NoStop}%
\bibitem [{\citenamefont {Dagotto}\ \emph {et~al.}(2001)\citenamefont
  {Dagotto}, \citenamefont {Hotta},\ and\ \citenamefont {Moreo}}]{dagotto01}%
  \BibitemOpen
  \bibfield  {author} {\bibinfo {author} {\bibfnamefont {E.}~\bibnamefont
  {Dagotto}}, \bibinfo {author} {\bibfnamefont {T.}~\bibnamefont {Hotta}},\
  and\ \bibinfo {author} {\bibfnamefont {A.}~\bibnamefont {Moreo}},\ }\bibfield
   {title} {\bibinfo {title} {Colossal magnetoresistant materials: the key role
  of phase separation},\ }\href
  {https://doi.org/https://doi.org/10.1016/S0370-1573(00)00121-6} {\bibfield
  {journal} {\bibinfo  {journal} {Physics Reports}\ }\textbf {\bibinfo {volume}
  {344}},\ \bibinfo {pages} {1} (\bibinfo {year} {2001})}\BibitemShut {NoStop}%
\bibitem [{\citenamefont {Dagotto}(2002)}]{dagotto-book}%
  \BibitemOpen
  \bibfield  {author} {\bibinfo {author} {\bibfnamefont {E.}~\bibnamefont
  {Dagotto}},\ }\href@noop {} {\emph {\bibinfo {title} {Nanoscale Phase
  Separation and Colossal Magnetoresistance}}}\ (\bibinfo  {publisher}
  {Springer Berlin},\ \bibinfo {year} {2002})\BibitemShut {NoStop}%
\bibitem [{\citenamefont {Dagotto}(2005)}]{dagotto05}%
  \BibitemOpen
  \bibfield  {author} {\bibinfo {author} {\bibfnamefont {E.}~\bibnamefont
  {Dagotto}},\ }\bibfield  {title} {\bibinfo {title} {Complexity in strongly
  correlated electronic systems},\ }\href
  {https://doi.org/10.1126/science.1107559} {\bibfield  {journal} {\bibinfo
  {journal} {Science}\ }\textbf {\bibinfo {volume} {309}},\ \bibinfo {pages}
  {257} (\bibinfo {year} {2005})}\BibitemShut {NoStop}%
\bibitem [{\citenamefont {Schiffer}\ \emph {et~al.}(1995)\citenamefont
  {Schiffer}, \citenamefont {Ramirez}, \citenamefont {Bao},\ and\ \citenamefont
  {Cheong}}]{schiffer95}%
  \BibitemOpen
  \bibfield  {author} {\bibinfo {author} {\bibfnamefont {P.}~\bibnamefont
  {Schiffer}}, \bibinfo {author} {\bibfnamefont {A.~P.}\ \bibnamefont
  {Ramirez}}, \bibinfo {author} {\bibfnamefont {W.}~\bibnamefont {Bao}},\ and\
  \bibinfo {author} {\bibfnamefont {S.-W.}\ \bibnamefont {Cheong}},\ }\bibfield
   {title} {\bibinfo {title} {Low temperature magnetoresistance and the
  magnetic phase diagram of
  {${\mathrm{La}}_{1\ensuremath{-}\mathit{x}}{\mathrm{Ca}}_{\mathit{x}}{\mathrm{MnO}}_{3}$}},\
  }\href {https://doi.org/10.1103/PhysRevLett.75.3336} {\bibfield  {journal}
  {\bibinfo  {journal} {Phys. Rev. Lett.}\ }\textbf {\bibinfo {volume} {75}},\
  \bibinfo {pages} {3336} (\bibinfo {year} {1995})}\BibitemShut {NoStop}%
\bibitem [{\citenamefont {Chen}\ and\ \citenamefont {Cheong}(1996)}]{chen96}%
  \BibitemOpen
  \bibfield  {author} {\bibinfo {author} {\bibfnamefont {C.~H.}\ \bibnamefont
  {Chen}}\ and\ \bibinfo {author} {\bibfnamefont {S.-W.}\ \bibnamefont
  {Cheong}},\ }\bibfield  {title} {\bibinfo {title} {Commensurate to
  incommensurate charge ordering and its real-space images in
  {L${\mathrm{a}}_{0.5}$C${\mathrm{a}}_{0.5}$Mn${\mathrm{O}}_{3}$}},\ }\href
  {https://doi.org/10.1103/PhysRevLett.76.4042} {\bibfield  {journal} {\bibinfo
   {journal} {Phys. Rev. Lett.}\ }\textbf {\bibinfo {volume} {76}},\ \bibinfo
  {pages} {4042} (\bibinfo {year} {1996})}\BibitemShut {NoStop}%
\bibitem [{\citenamefont {Moreo}\ \emph {et~al.}(1999)\citenamefont {Moreo},
  \citenamefont {Yunoki},\ and\ \citenamefont {Dagotto}}]{moreo99}%
  \BibitemOpen
  \bibfield  {author} {\bibinfo {author} {\bibfnamefont {A.}~\bibnamefont
  {Moreo}}, \bibinfo {author} {\bibfnamefont {S.}~\bibnamefont {Yunoki}},\ and\
  \bibinfo {author} {\bibfnamefont {E.}~\bibnamefont {Dagotto}},\ }\bibfield
  {title} {\bibinfo {title} {Phase separation scenario for manganese oxides and
  related materials},\ }\href {https://doi.org/10.1126/science.283.5410.2034}
  {\bibfield  {journal} {\bibinfo  {journal} {Science}\ }\textbf {\bibinfo
  {volume} {283}},\ \bibinfo {pages} {2034} (\bibinfo {year}
  {1999})}\BibitemShut {NoStop}%
\bibitem [{\citenamefont {Uehara}\ \emph {et~al.}(1999)\citenamefont {Uehara},
  \citenamefont {Mori}, \citenamefont {Chen},\ and\ \citenamefont
  {Cheong}}]{uehara99}%
  \BibitemOpen
  \bibfield  {author} {\bibinfo {author} {\bibfnamefont {M.}~\bibnamefont
  {Uehara}}, \bibinfo {author} {\bibfnamefont {S.}~\bibnamefont {Mori}},
  \bibinfo {author} {\bibfnamefont {C.~H.}\ \bibnamefont {Chen}},\ and\
  \bibinfo {author} {\bibfnamefont {S.-W.}\ \bibnamefont {Cheong}},\ }\bibfield
   {title} {\bibinfo {title} {Percolative phase separation underlies colossal
  magnetoresistance in mixed-valent manganites},\ }\href
  {https://doi.org/10.1038/21142} {\bibfield  {journal} {\bibinfo  {journal}
  {Nature}\ }\textbf {\bibinfo {volume} {399}},\ \bibinfo {pages} {560}
  (\bibinfo {year} {1999})}\BibitemShut {NoStop}%
\bibitem [{\citenamefont {F\"ath}\ \emph {et~al.}(1999)\citenamefont {F\"ath},
  \citenamefont {Freisem}, \citenamefont {Menovsky}, \citenamefont {Tomioka},
  \citenamefont {Aarts},\ and\ \citenamefont {Mydosh}}]{fath99}%
  \BibitemOpen
  \bibfield  {author} {\bibinfo {author} {\bibfnamefont {M.}~\bibnamefont
  {F\"ath}}, \bibinfo {author} {\bibfnamefont {S.}~\bibnamefont {Freisem}},
  \bibinfo {author} {\bibfnamefont {A.~A.}\ \bibnamefont {Menovsky}}, \bibinfo
  {author} {\bibfnamefont {Y.}~\bibnamefont {Tomioka}}, \bibinfo {author}
  {\bibfnamefont {J.}~\bibnamefont {Aarts}},\ and\ \bibinfo {author}
  {\bibfnamefont {J.~A.}\ \bibnamefont {Mydosh}},\ }\bibfield  {title}
  {\bibinfo {title} {Spatially inhomogeneous metal-insulator transition in
  doped manganites},\ }\href {https://doi.org/10.1126/science.285.5433.1540}
  {\bibfield  {journal} {\bibinfo  {journal} {Science}\ }\textbf {\bibinfo
  {volume} {285}},\ \bibinfo {pages} {1540} (\bibinfo {year}
  {1999})}\BibitemShut {NoStop}%
\bibitem [{\citenamefont {Zener}(1951)}]{zener51}%
  \BibitemOpen
  \bibfield  {author} {\bibinfo {author} {\bibfnamefont {C.}~\bibnamefont
  {Zener}},\ }\bibfield  {title} {\bibinfo {title} {Interaction between the
  $d$-shells in the transition metals. ii. ferromagnetic compounds of manganese
  with perovskite structure},\ }\href {https://doi.org/10.1103/PhysRev.82.403}
  {\bibfield  {journal} {\bibinfo  {journal} {Phys. Rev.}\ }\textbf {\bibinfo
  {volume} {82}},\ \bibinfo {pages} {403} (\bibinfo {year} {1951})}\BibitemShut
  {NoStop}%
\bibitem [{\citenamefont {Anderson}\ and\ \citenamefont
  {Hasegawa}(1955)}]{anderson55}%
  \BibitemOpen
  \bibfield  {author} {\bibinfo {author} {\bibfnamefont {P.~W.}\ \bibnamefont
  {Anderson}}\ and\ \bibinfo {author} {\bibfnamefont {H.}~\bibnamefont
  {Hasegawa}},\ }\bibfield  {title} {\bibinfo {title} {Considerations on double
  exchange},\ }\href {https://doi.org/10.1103/PhysRev.100.675} {\bibfield
  {journal} {\bibinfo  {journal} {Phys. Rev.}\ }\textbf {\bibinfo {volume}
  {100}},\ \bibinfo {pages} {675} (\bibinfo {year} {1955})}\BibitemShut
  {NoStop}%
\bibitem [{\citenamefont {de~Gennes}(1960)}]{degennes60}%
  \BibitemOpen
  \bibfield  {author} {\bibinfo {author} {\bibfnamefont {P.~G.}\ \bibnamefont
  {de~Gennes}},\ }\bibfield  {title} {\bibinfo {title} {Effects of double
  exchange in magnetic crystals},\ }\href
  {https://doi.org/10.1103/PhysRev.118.141} {\bibfield  {journal} {\bibinfo
  {journal} {Phys. Rev.}\ }\textbf {\bibinfo {volume} {118}},\ \bibinfo {pages}
  {141} (\bibinfo {year} {1960})}\BibitemShut {NoStop}%
\bibitem [{\citenamefont {Yunoki}\ \emph
  {et~al.}(1998{\natexlab{a}})\citenamefont {Yunoki}, \citenamefont {Hu},
  \citenamefont {Malvezzi}, \citenamefont {Moreo}, \citenamefont {Furukawa},\
  and\ \citenamefont {Dagotto}}]{yunoki98}%
  \BibitemOpen
  \bibfield  {author} {\bibinfo {author} {\bibfnamefont {S.}~\bibnamefont
  {Yunoki}}, \bibinfo {author} {\bibfnamefont {J.}~\bibnamefont {Hu}}, \bibinfo
  {author} {\bibfnamefont {A.~L.}\ \bibnamefont {Malvezzi}}, \bibinfo {author}
  {\bibfnamefont {A.}~\bibnamefont {Moreo}}, \bibinfo {author} {\bibfnamefont
  {N.}~\bibnamefont {Furukawa}},\ and\ \bibinfo {author} {\bibfnamefont
  {E.}~\bibnamefont {Dagotto}},\ }\bibfield  {title} {\bibinfo {title} {Phase
  separation in electronic models for manganites},\ }\href
  {https://doi.org/10.1103/PhysRevLett.80.845} {\bibfield  {journal} {\bibinfo
  {journal} {Phys. Rev. Lett.}\ }\textbf {\bibinfo {volume} {80}},\ \bibinfo
  {pages} {845} (\bibinfo {year} {1998}{\natexlab{a}})}\BibitemShut {NoStop}%
\bibitem [{\citenamefont {Dagotto}\ \emph {et~al.}(1998)\citenamefont
  {Dagotto}, \citenamefont {Yunoki}, \citenamefont {Malvezzi}, \citenamefont
  {Moreo}, \citenamefont {Hu}, \citenamefont {Capponi}, \citenamefont
  {Poilblanc},\ and\ \citenamefont {Furukawa}}]{dagotto98}%
  \BibitemOpen
  \bibfield  {author} {\bibinfo {author} {\bibfnamefont {E.}~\bibnamefont
  {Dagotto}}, \bibinfo {author} {\bibfnamefont {S.}~\bibnamefont {Yunoki}},
  \bibinfo {author} {\bibfnamefont {A.~L.}\ \bibnamefont {Malvezzi}}, \bibinfo
  {author} {\bibfnamefont {A.}~\bibnamefont {Moreo}}, \bibinfo {author}
  {\bibfnamefont {J.}~\bibnamefont {Hu}}, \bibinfo {author} {\bibfnamefont
  {S.}~\bibnamefont {Capponi}}, \bibinfo {author} {\bibfnamefont
  {D.}~\bibnamefont {Poilblanc}},\ and\ \bibinfo {author} {\bibfnamefont
  {N.}~\bibnamefont {Furukawa}},\ }\bibfield  {title} {\bibinfo {title}
  {Ferromagnetic {Kondo} model for manganites: Phase diagram, charge
  segregation, and influence of quantum localized spins},\ }\href
  {https://doi.org/10.1103/PhysRevB.58.6414} {\bibfield  {journal} {\bibinfo
  {journal} {Phys. Rev. B}\ }\textbf {\bibinfo {volume} {58}},\ \bibinfo
  {pages} {6414} (\bibinfo {year} {1998})}\BibitemShut {NoStop}%
\bibitem [{\citenamefont {Chattopadhyay}\ \emph {et~al.}(2001)\citenamefont
  {Chattopadhyay}, \citenamefont {Millis},\ and\ \citenamefont
  {Das~Sarma}}]{chattopadhyay01}%
  \BibitemOpen
  \bibfield  {author} {\bibinfo {author} {\bibfnamefont {A.}~\bibnamefont
  {Chattopadhyay}}, \bibinfo {author} {\bibfnamefont {A.~J.}\ \bibnamefont
  {Millis}},\ and\ \bibinfo {author} {\bibfnamefont {S.}~\bibnamefont
  {Das~Sarma}},\ }\bibfield  {title} {\bibinfo {title} {{$T=0$} phase diagram
  of the double-exchange model},\ }\href
  {https://doi.org/10.1103/PhysRevB.64.012416} {\bibfield  {journal} {\bibinfo
  {journal} {Phys. Rev. B}\ }\textbf {\bibinfo {volume} {64}},\ \bibinfo
  {pages} {012416} (\bibinfo {year} {2001})}\BibitemShut {NoStop}%
\bibitem [{\citenamefont {Held}\ and\ \citenamefont
  {Vollhardt}(2000)}]{held00}%
  \BibitemOpen
  \bibfield  {author} {\bibinfo {author} {\bibfnamefont {K.}~\bibnamefont
  {Held}}\ and\ \bibinfo {author} {\bibfnamefont {D.}~\bibnamefont
  {Vollhardt}},\ }\bibfield  {title} {\bibinfo {title} {Electronic correlations
  in manganites},\ }\href {https://doi.org/10.1103/PhysRevLett.84.5168}
  {\bibfield  {journal} {\bibinfo  {journal} {Phys. Rev. Lett.}\ }\textbf
  {\bibinfo {volume} {84}},\ \bibinfo {pages} {5168} (\bibinfo {year}
  {2000})}\BibitemShut {NoStop}%
\bibitem [{\citenamefont {Maezono}\ \emph
  {et~al.}(1998{\natexlab{a}})\citenamefont {Maezono}, \citenamefont
  {Ishihara},\ and\ \citenamefont {Nagaosa}}]{maezono98a}%
  \BibitemOpen
  \bibfield  {author} {\bibinfo {author} {\bibfnamefont {R.}~\bibnamefont
  {Maezono}}, \bibinfo {author} {\bibfnamefont {S.}~\bibnamefont {Ishihara}},\
  and\ \bibinfo {author} {\bibfnamefont {N.}~\bibnamefont {Nagaosa}},\
  }\bibfield  {title} {\bibinfo {title} {Orbital polarization in manganese
  oxides},\ }\href {https://doi.org/10.1103/PhysRevB.57.R13993} {\bibfield
  {journal} {\bibinfo  {journal} {Phys. Rev. B}\ }\textbf {\bibinfo {volume}
  {57}},\ \bibinfo {pages} {R13993} (\bibinfo {year}
  {1998}{\natexlab{a}})}\BibitemShut {NoStop}%
\bibitem [{\citenamefont {Maezono}\ \emph
  {et~al.}(1998{\natexlab{b}})\citenamefont {Maezono}, \citenamefont
  {Ishihara},\ and\ \citenamefont {Nagaosa}}]{maezono98b}%
  \BibitemOpen
  \bibfield  {author} {\bibinfo {author} {\bibfnamefont {R.}~\bibnamefont
  {Maezono}}, \bibinfo {author} {\bibfnamefont {S.}~\bibnamefont {Ishihara}},\
  and\ \bibinfo {author} {\bibfnamefont {N.}~\bibnamefont {Nagaosa}},\
  }\bibfield  {title} {\bibinfo {title} {Phase diagram of manganese oxides},\
  }\href {https://doi.org/10.1103/PhysRevB.58.11583} {\bibfield  {journal}
  {\bibinfo  {journal} {Phys. Rev. B}\ }\textbf {\bibinfo {volume} {58}},\
  \bibinfo {pages} {11583} (\bibinfo {year} {1998}{\natexlab{b}})}\BibitemShut
  {NoStop}%
\bibitem [{\citenamefont {Yang}\ and\ \citenamefont {Held}(2010)}]{yang10}%
  \BibitemOpen
  \bibfield  {author} {\bibinfo {author} {\bibfnamefont {Y.-F.}\ \bibnamefont
  {Yang}}\ and\ \bibinfo {author} {\bibfnamefont {K.}~\bibnamefont {Held}},\
  }\bibfield  {title} {\bibinfo {title} {Dynamical mean field theory for
  manganites},\ }\href {https://doi.org/10.1103/PhysRevB.82.195109} {\bibfield
  {journal} {\bibinfo  {journal} {Phys. Rev. B}\ }\textbf {\bibinfo {volume}
  {82}},\ \bibinfo {pages} {195109} (\bibinfo {year} {2010})}\BibitemShut
  {NoStop}%
\bibitem [{\citenamefont {R\"oder}\ \emph {et~al.}(1996)\citenamefont
  {R\"oder}, \citenamefont {Zang},\ and\ \citenamefont {Bishop}}]{roder96}%
  \BibitemOpen
  \bibfield  {author} {\bibinfo {author} {\bibfnamefont {H.}~\bibnamefont
  {R\"oder}}, \bibinfo {author} {\bibfnamefont {J.}~\bibnamefont {Zang}},\ and\
  \bibinfo {author} {\bibfnamefont {A.~R.}\ \bibnamefont {Bishop}},\ }\bibfield
   {title} {\bibinfo {title} {Lattice effects in the colossal-magnetoresistance
  manganites},\ }\href {https://doi.org/10.1103/PhysRevLett.76.1356} {\bibfield
   {journal} {\bibinfo  {journal} {Phys. Rev. Lett.}\ }\textbf {\bibinfo
  {volume} {76}},\ \bibinfo {pages} {1356} (\bibinfo {year}
  {1996})}\BibitemShut {NoStop}%
\bibitem [{\citenamefont {Millis}\ \emph {et~al.}(1995)\citenamefont {Millis},
  \citenamefont {Littlewood},\ and\ \citenamefont {Shraiman}}]{millis95}%
  \BibitemOpen
  \bibfield  {author} {\bibinfo {author} {\bibfnamefont {A.~J.}\ \bibnamefont
  {Millis}}, \bibinfo {author} {\bibfnamefont {P.~B.}\ \bibnamefont
  {Littlewood}},\ and\ \bibinfo {author} {\bibfnamefont {B.~I.}\ \bibnamefont
  {Shraiman}},\ }\bibfield  {title} {\bibinfo {title} {Double exchange alone
  does not explain the resistivity of
  {${{\mathrm{La}}_{1}}_{\ensuremath{-}\mathit{x}}{\mathrm{Sr}}_{\mathit{x}}{\mathrm{MnO}}_{3}$}},\
  }\href {https://doi.org/10.1103/PhysRevLett.74.5144} {\bibfield  {journal}
  {\bibinfo  {journal} {Phys. Rev. Lett.}\ }\textbf {\bibinfo {volume} {74}},\
  \bibinfo {pages} {5144} (\bibinfo {year} {1995})}\BibitemShut {NoStop}%
\bibitem [{\citenamefont {Millis}\ \emph {et~al.}(1996)\citenamefont {Millis},
  \citenamefont {Shraiman},\ and\ \citenamefont {Mueller}}]{millis96}%
  \BibitemOpen
  \bibfield  {author} {\bibinfo {author} {\bibfnamefont {A.~J.}\ \bibnamefont
  {Millis}}, \bibinfo {author} {\bibfnamefont {B.~I.}\ \bibnamefont
  {Shraiman}},\ and\ \bibinfo {author} {\bibfnamefont {R.}~\bibnamefont
  {Mueller}},\ }\bibfield  {title} {\bibinfo {title} {Dynamic {Jahn-Teller}
  effect and colossal magnetoresistance in
  {${\mathrm{La}}_{1\ensuremath{-}\mathit{x}}{\mathrm{Sr}}_{\mathit{x}}{\mathrm{MnO}}_{3}$}},\
  }\href {https://doi.org/10.1103/PhysRevLett.77.175} {\bibfield  {journal}
  {\bibinfo  {journal} {Phys. Rev. Lett.}\ }\textbf {\bibinfo {volume} {77}},\
  \bibinfo {pages} {175} (\bibinfo {year} {1996})}\BibitemShut {NoStop}%
\bibitem [{\citenamefont {Millis}(1996)}]{millis96b}%
  \BibitemOpen
  \bibfield  {author} {\bibinfo {author} {\bibfnamefont {A.~J.}\ \bibnamefont
  {Millis}},\ }\bibfield  {title} {\bibinfo {title} {Cooperative jahn-teller
  effect and electron-phonon coupling in
  {${\mathrm{La}}_{1\mathrm{\ensuremath{-}}\mathit{x}}$${\mathrm{A}}_{\mathit{x}}$${\mathrm{MnO}}_{3}$}},\
  }\href {https://doi.org/10.1103/PhysRevB.53.8434} {\bibfield  {journal}
  {\bibinfo  {journal} {Phys. Rev. B}\ }\textbf {\bibinfo {volume} {53}},\
  \bibinfo {pages} {8434} (\bibinfo {year} {1996})}\BibitemShut {NoStop}%
\bibitem [{\citenamefont {Nagaosa}\ \emph {et~al.}(1998)\citenamefont
  {Nagaosa}, \citenamefont {Murakami},\ and\ \citenamefont {Lee}}]{nagaosa98}%
  \BibitemOpen
  \bibfield  {author} {\bibinfo {author} {\bibfnamefont {N.}~\bibnamefont
  {Nagaosa}}, \bibinfo {author} {\bibfnamefont {S.}~\bibnamefont {Murakami}},\
  and\ \bibinfo {author} {\bibfnamefont {H.~C.}\ \bibnamefont {Lee}},\
  }\bibfield  {title} {\bibinfo {title} {Electron correlation and {Jahn-Teller}
  interaction in manganese oxides},\ }\href
  {https://doi.org/10.1103/PhysRevB.57.R6767} {\bibfield  {journal} {\bibinfo
  {journal} {Phys. Rev. B}\ }\textbf {\bibinfo {volume} {57}},\ \bibinfo
  {pages} {R6767} (\bibinfo {year} {1998})}\BibitemShut {NoStop}%
\bibitem [{\citenamefont {Maezono}\ and\ \citenamefont
  {Nagaosa}(2003)}]{maezono03}%
  \BibitemOpen
  \bibfield  {author} {\bibinfo {author} {\bibfnamefont {R.}~\bibnamefont
  {Maezono}}\ and\ \bibinfo {author} {\bibfnamefont {N.}~\bibnamefont
  {Nagaosa}},\ }\bibfield  {title} {\bibinfo {title} {{Jahn-Teller} effect and
  electron correlation in manganites},\ }\href
  {https://doi.org/10.1103/PhysRevB.67.064413} {\bibfield  {journal} {\bibinfo
  {journal} {Phys. Rev. B}\ }\textbf {\bibinfo {volume} {67}},\ \bibinfo
  {pages} {064413} (\bibinfo {year} {2003})}\BibitemShut {NoStop}%
\bibitem [{\citenamefont {Hotta}\ \emph {et~al.}(2000)\citenamefont {Hotta},
  \citenamefont {Malvezzi},\ and\ \citenamefont {Dagotto}}]{hotta00}%
  \BibitemOpen
  \bibfield  {author} {\bibinfo {author} {\bibfnamefont {T.}~\bibnamefont
  {Hotta}}, \bibinfo {author} {\bibfnamefont {A.~L.}\ \bibnamefont
  {Malvezzi}},\ and\ \bibinfo {author} {\bibfnamefont {E.}~\bibnamefont
  {Dagotto}},\ }\bibfield  {title} {\bibinfo {title} {Charge-orbital ordering
  and phase separation in the two-orbital model for manganites: Roles of
  {Jahn-Teller} phononic and {Coulombic} interactions},\ }\href
  {https://doi.org/10.1103/PhysRevB.62.9432} {\bibfield  {journal} {\bibinfo
  {journal} {Phys. Rev. B}\ }\textbf {\bibinfo {volume} {62}},\ \bibinfo
  {pages} {9432} (\bibinfo {year} {2000})}\BibitemShut {NoStop}%
\bibitem [{\citenamefont {Kugel'}\ and\ \citenamefont
  {Khomskii}(1972)}]{kugel72}%
  \BibitemOpen
  \bibfield  {author} {\bibinfo {author} {\bibfnamefont {K.~I.}\ \bibnamefont
  {Kugel'}}\ and\ \bibinfo {author} {\bibfnamefont {D.~I.}\ \bibnamefont
  {Khomskii}},\ }\bibfield  {title} {\bibinfo {title} {Superexchange ordering
  of degenerate orbitals and magnetic structure of dielectrics with {
  Jahn-Teller} ions},\ }\href {http://jetpletters.ru/ps/0/article_26676.shtml}
  {\bibfield  {journal} {\bibinfo  {journal} {JETP Letters}\ }\textbf {\bibinfo
  {volume} {15}},\ \bibinfo {pages} {446} (\bibinfo {year} {1972})}\BibitemShut
  {NoStop}%
\bibitem [{\citenamefont {Kugel}\ and\ \citenamefont
  {Khomskii}(1973)}]{kugel73}%
  \BibitemOpen
  \bibfield  {author} {\bibinfo {author} {\bibfnamefont {K.~I.}\ \bibnamefont
  {Kugel}}\ and\ \bibinfo {author} {\bibfnamefont {D.~I.}\ \bibnamefont
  {Khomskii}},\ }\bibfield  {title} {\bibinfo {title} {Crystal-structure and
  magnetic properties of substances with orbital degeneracy},\ }\href@noop {}
  {\bibfield  {journal} {\bibinfo  {journal} {Sov. Phys. JETP}\ }\textbf
  {\bibinfo {volume} {37}},\ \bibinfo {pages} {725} (\bibinfo {year}
  {1973})}\BibitemShut {NoStop}%
\bibitem [{\citenamefont {Kugel'}\ and\ \citenamefont
  {Khomskiĭ}(1982)}]{kugel82}%
  \BibitemOpen
  \bibfield  {author} {\bibinfo {author} {\bibfnamefont {K.~I.}\ \bibnamefont
  {Kugel'}}\ and\ \bibinfo {author} {\bibfnamefont {D.~I.}\ \bibnamefont
  {Khomskiĭ}},\ }\bibfield  {title} {\bibinfo {title} {The jahn-teller effect
  and magnetism: transition metal compounds},\ }\href
  {https://doi.org/10.1070/PU1982v025n04ABEH004537} {\bibfield  {journal}
  {\bibinfo  {journal} {Soviet Physics Uspekhi}\ }\textbf {\bibinfo {volume}
  {25}},\ \bibinfo {pages} {231} (\bibinfo {year} {1982})}\BibitemShut
  {NoStop}%
\bibitem [{\citenamefont {van~den Brink}\ and\ \citenamefont
  {Khomskii}(2001)}]{brink01}%
  \BibitemOpen
  \bibfield  {author} {\bibinfo {author} {\bibfnamefont {J.}~\bibnamefont
  {van~den Brink}}\ and\ \bibinfo {author} {\bibfnamefont {D.}~\bibnamefont
  {Khomskii}},\ }\bibfield  {title} {\bibinfo {title} {Orbital ordering of
  complex orbitals in doped mott insulators},\ }\href
  {https://doi.org/10.1103/PhysRevB.63.140416} {\bibfield  {journal} {\bibinfo
  {journal} {Phys. Rev. B}\ }\textbf {\bibinfo {volume} {63}},\ \bibinfo
  {pages} {140416} (\bibinfo {year} {2001})}\BibitemShut {NoStop}%
\bibitem [{\citenamefont {van~den Brink}(2004)}]{brink04}%
  \BibitemOpen
  \bibfield  {author} {\bibinfo {author} {\bibfnamefont {J.}~\bibnamefont
  {van~den Brink}},\ }\bibfield  {title} {\bibinfo {title} {Orbital-only
  models: ordering and excitations},\ }\href
  {https://doi.org/10.1088/1367-2630/6/1/201} {\bibfield  {journal} {\bibinfo
  {journal} {New Journal of Physics}\ }\textbf {\bibinfo {volume} {6}},\
  \bibinfo {pages} {201} (\bibinfo {year} {2004})}\BibitemShut {NoStop}%
\bibitem [{\citenamefont {Ruderman}\ and\ \citenamefont
  {Kittel}(1954)}]{Ruderman1954}%
  \BibitemOpen
  \bibfield  {author} {\bibinfo {author} {\bibfnamefont {M.~A.}\ \bibnamefont
  {Ruderman}}\ and\ \bibinfo {author} {\bibfnamefont {C.}~\bibnamefont
  {Kittel}},\ }\bibfield  {title} {\bibinfo {title} {{Indirect Exchange
  Coupling of Nuclear Magnetic Moments by Conduction Electrons}},\ }\href
  {https://doi.org/10.1103/PhysRev.96.99} {\bibfield  {journal} {\bibinfo
  {journal} {Phys. Rev.}\ }\textbf {\bibinfo {volume} {96}},\ \bibinfo {pages}
  {99} (\bibinfo {year} {1954})}\BibitemShut {NoStop}%
\bibitem [{\citenamefont {Kasuya}(1956)}]{Kasuya1956}%
  \BibitemOpen
  \bibfield  {author} {\bibinfo {author} {\bibfnamefont {T.}~\bibnamefont
  {Kasuya}},\ }\bibfield  {title} {\bibinfo {title} {{A Theory of Metallic
  Ferro- and Antiferromagnetism on Zener's Model}},\ }\href
  {https://doi.org/10.1143/PTP.16.45} {\bibfield  {journal} {\bibinfo
  {journal} {Prog. Theor. Phys.}\ }\textbf {\bibinfo {volume} {16}},\ \bibinfo
  {pages} {45} (\bibinfo {year} {1956})}\BibitemShut {NoStop}%
\bibitem [{\citenamefont {Yosida}(1957)}]{Yosida1957}%
  \BibitemOpen
  \bibfield  {author} {\bibinfo {author} {\bibfnamefont {K.}~\bibnamefont
  {Yosida}},\ }\bibfield  {title} {\bibinfo {title} {Magnetic properties of
  {Cu-Mn} alloys},\ }\href {https://doi.org/10.1103/PhysRev.106.893} {\bibfield
   {journal} {\bibinfo  {journal} {Phys. Rev.}\ }\textbf {\bibinfo {volume}
  {106}},\ \bibinfo {pages} {893} (\bibinfo {year} {1957})}\BibitemShut
  {NoStop}%
\bibitem [{\citenamefont {Popovic}\ and\ \citenamefont
  {Satpathy}(2000)}]{popovic00}%
  \BibitemOpen
  \bibfield  {author} {\bibinfo {author} {\bibfnamefont {Z.}~\bibnamefont
  {Popovic}}\ and\ \bibinfo {author} {\bibfnamefont {S.}~\bibnamefont
  {Satpathy}},\ }\bibfield  {title} {\bibinfo {title} {Cooperative jahn-teller
  coupling in the manganites},\ }\href
  {https://doi.org/10.1103/PhysRevLett.84.1603} {\bibfield  {journal} {\bibinfo
   {journal} {Phys. Rev. Lett.}\ }\textbf {\bibinfo {volume} {84}},\ \bibinfo
  {pages} {1603} (\bibinfo {year} {2000})}\BibitemShut {NoStop}%
\bibitem [{\citenamefont {Hotta}\ \emph {et~al.}(1999)\citenamefont {Hotta},
  \citenamefont {Yunoki}, \citenamefont {Mayr},\ and\ \citenamefont
  {Dagotto}}]{hotta99}%
  \BibitemOpen
  \bibfield  {author} {\bibinfo {author} {\bibfnamefont {T.}~\bibnamefont
  {Hotta}}, \bibinfo {author} {\bibfnamefont {S.}~\bibnamefont {Yunoki}},
  \bibinfo {author} {\bibfnamefont {M.}~\bibnamefont {Mayr}},\ and\ \bibinfo
  {author} {\bibfnamefont {E.}~\bibnamefont {Dagotto}},\ }\bibfield  {title}
  {\bibinfo {title} {A-type antiferromagnetic and {C}-type orbital-ordered
  states in ${\mathrm{lamno}}_{3}$ using cooperative jahn-teller phonons},\
  }\href {https://doi.org/10.1103/PhysRevB.60.R15009} {\bibfield  {journal}
  {\bibinfo  {journal} {Phys. Rev. B}\ }\textbf {\bibinfo {volume} {60}},\
  \bibinfo {pages} {R15009} (\bibinfo {year} {1999})}\BibitemShut {NoStop}%
\bibitem [{\citenamefont {Behler}\ and\ \citenamefont
  {Parrinello}(2007)}]{behler07}%
  \BibitemOpen
  \bibfield  {author} {\bibinfo {author} {\bibfnamefont {J.}~\bibnamefont
  {Behler}}\ and\ \bibinfo {author} {\bibfnamefont {M.}~\bibnamefont
  {Parrinello}},\ }\bibfield  {title} {\bibinfo {title} {Generalized
  neural-network representation of high-dimensional potential-energy
  surfaces},\ }\href {https://doi.org/10.1103/PhysRevLett.98.146401} {\bibfield
   {journal} {\bibinfo  {journal} {Phys. Rev. Lett.}\ }\textbf {\bibinfo
  {volume} {98}},\ \bibinfo {pages} {146401} (\bibinfo {year}
  {2007})}\BibitemShut {NoStop}%
\bibitem [{\citenamefont {Bart\'ok}\ \emph {et~al.}(2010)\citenamefont
  {Bart\'ok}, \citenamefont {Payne}, \citenamefont {Kondor},\ and\
  \citenamefont {Cs\'anyi}}]{bartok10}%
  \BibitemOpen
  \bibfield  {author} {\bibinfo {author} {\bibfnamefont {A.~P.}\ \bibnamefont
  {Bart\'ok}}, \bibinfo {author} {\bibfnamefont {M.~C.}\ \bibnamefont {Payne}},
  \bibinfo {author} {\bibfnamefont {R.}~\bibnamefont {Kondor}},\ and\ \bibinfo
  {author} {\bibfnamefont {G.}~\bibnamefont {Cs\'anyi}},\ }\bibfield  {title}
  {\bibinfo {title} {Gaussian approximation potentials: The accuracy of quantum
  mechanics, without the electrons},\ }\href
  {https://doi.org/10.1103/PhysRevLett.104.136403} {\bibfield  {journal}
  {\bibinfo  {journal} {Phys. Rev. Lett.}\ }\textbf {\bibinfo {volume} {104}},\
  \bibinfo {pages} {136403} (\bibinfo {year} {2010})}\BibitemShut {NoStop}%
\bibitem [{\citenamefont {Iliev}\ \emph {et~al.}(1998)\citenamefont {Iliev},
  \citenamefont {Abrashev}, \citenamefont {Lee}, \citenamefont {Popov},
  \citenamefont {Sun}, \citenamefont {Thomsen}, \citenamefont {Meng},\ and\
  \citenamefont {Chu}}]{lliev98}%
  \BibitemOpen
  \bibfield  {author} {\bibinfo {author} {\bibfnamefont {M.~N.}\ \bibnamefont
  {Iliev}}, \bibinfo {author} {\bibfnamefont {M.~V.}\ \bibnamefont {Abrashev}},
  \bibinfo {author} {\bibfnamefont {H.-G.}\ \bibnamefont {Lee}}, \bibinfo
  {author} {\bibfnamefont {V.~N.}\ \bibnamefont {Popov}}, \bibinfo {author}
  {\bibfnamefont {Y.~Y.}\ \bibnamefont {Sun}}, \bibinfo {author} {\bibfnamefont
  {C.}~\bibnamefont {Thomsen}}, \bibinfo {author} {\bibfnamefont {R.~L.}\
  \bibnamefont {Meng}},\ and\ \bibinfo {author} {\bibfnamefont {C.~W.}\
  \bibnamefont {Chu}},\ }\bibfield  {title} {\bibinfo {title} {Raman
  spectroscopy of orthorhombic perovskitelike ${\mathrm{ymno}}_{3}$ and
  ${\mathrm{lamno}}_{3}$},\ }\href {https://doi.org/10.1103/PhysRevB.57.2872}
  {\bibfield  {journal} {\bibinfo  {journal} {Phys. Rev. B}\ }\textbf {\bibinfo
  {volume} {57}},\ \bibinfo {pages} {2872} (\bibinfo {year}
  {1998})}\BibitemShut {NoStop}%
\bibitem [{\citenamefont {Yunoki}\ \emph
  {et~al.}(1998{\natexlab{b}})\citenamefont {Yunoki}, \citenamefont {Moreo},\
  and\ \citenamefont {Dagotto}}]{yunoki98b}%
  \BibitemOpen
  \bibfield  {author} {\bibinfo {author} {\bibfnamefont {S.}~\bibnamefont
  {Yunoki}}, \bibinfo {author} {\bibfnamefont {A.}~\bibnamefont {Moreo}},\ and\
  \bibinfo {author} {\bibfnamefont {E.}~\bibnamefont {Dagotto}},\ }\bibfield
  {title} {\bibinfo {title} {Phase separation induced by orbital degrees of
  freedom in models for manganites with jahn-teller phonons},\ }\href
  {https://doi.org/10.1103/PhysRevLett.81.5612} {\bibfield  {journal} {\bibinfo
   {journal} {Phys. Rev. Lett.}\ }\textbf {\bibinfo {volume} {81}},\ \bibinfo
  {pages} {5612} (\bibinfo {year} {1998}{\natexlab{b}})}\BibitemShut {NoStop}%
\bibitem [{\citenamefont {Hotta}\ \emph {et~al.}(2003)\citenamefont {Hotta},
  \citenamefont {Moraghebi}, \citenamefont {Feiguin}, \citenamefont {Moreo},
  \citenamefont {Yunoki},\ and\ \citenamefont {Dagotto}}]{hotta03}%
  \BibitemOpen
  \bibfield  {author} {\bibinfo {author} {\bibfnamefont {T.}~\bibnamefont
  {Hotta}}, \bibinfo {author} {\bibfnamefont {M.}~\bibnamefont {Moraghebi}},
  \bibinfo {author} {\bibfnamefont {A.}~\bibnamefont {Feiguin}}, \bibinfo
  {author} {\bibfnamefont {A.}~\bibnamefont {Moreo}}, \bibinfo {author}
  {\bibfnamefont {S.}~\bibnamefont {Yunoki}},\ and\ \bibinfo {author}
  {\bibfnamefont {E.}~\bibnamefont {Dagotto}},\ }\bibfield  {title} {\bibinfo
  {title} {Unveiling new magnetic phases of undoped and doped manganites},\
  }\href {https://doi.org/10.1103/PhysRevLett.90.247203} {\bibfield  {journal}
  {\bibinfo  {journal} {Phys. Rev. Lett.}\ }\textbf {\bibinfo {volume} {90}},\
  \bibinfo {pages} {247203} (\bibinfo {year} {2003})}\BibitemShut {NoStop}%
\bibitem [{\citenamefont {Salafranca}\ and\ \citenamefont
  {Brey}(2006)}]{salafranca06}%
  \BibitemOpen
  \bibfield  {author} {\bibinfo {author} {\bibfnamefont {J.}~\bibnamefont
  {Salafranca}}\ and\ \bibinfo {author} {\bibfnamefont {L.}~\bibnamefont
  {Brey}},\ }\bibfield  {title} {\bibinfo {title} {Phase diagram and
  incommensurate phases in undoped manganites},\ }\href
  {https://doi.org/10.1103/PhysRevB.73.024422} {\bibfield  {journal} {\bibinfo
  {journal} {Phys. Rev. B}\ }\textbf {\bibinfo {volume} {73}},\ \bibinfo
  {pages} {024422} (\bibinfo {year} {2006})}\BibitemShut {NoStop}%
\bibitem [{\citenamefont {Nanda}\ and\ \citenamefont
  {Satpathy}(2010)}]{nanda10}%
  \BibitemOpen
  \bibfield  {author} {\bibinfo {author} {\bibfnamefont {B.~R.~K.}\
  \bibnamefont {Nanda}}\ and\ \bibinfo {author} {\bibfnamefont
  {S.}~\bibnamefont {Satpathy}},\ }\bibfield  {title} {\bibinfo {title}
  {Magnetic and orbital order in ${\text{lamno}}_{3}$ under uniaxial strain: A
  model study},\ }\href {https://doi.org/10.1103/PhysRevB.81.174423} {\bibfield
   {journal} {\bibinfo  {journal} {Phys. Rev. B}\ }\textbf {\bibinfo {volume}
  {81}},\ \bibinfo {pages} {174423} (\bibinfo {year} {2010})}\BibitemShut
  {NoStop}%
\bibitem [{\citenamefont {Allen}\ and\ \citenamefont {Cahn}(1972)}]{Allen1972}%
  \BibitemOpen
  \bibfield  {author} {\bibinfo {author} {\bibfnamefont {S.}~\bibnamefont
  {Allen}}\ and\ \bibinfo {author} {\bibfnamefont {J.}~\bibnamefont {Cahn}},\
  }\bibfield  {title} {\bibinfo {title} {Ground state structures in ordered
  binary alloys with second neighbor interactions},\ }\href
  {https://doi.org/https://doi.org/10.1016/0001-6160(72)90037-5} {\bibfield
  {journal} {\bibinfo  {journal} {Acta Metallurgica}\ }\textbf {\bibinfo
  {volume} {20}},\ \bibinfo {pages} {423} (\bibinfo {year} {1972})}\BibitemShut
  {NoStop}%
\bibitem [{\citenamefont {Bray}(1994)}]{Bray1994}%
  \BibitemOpen
  \bibfield  {author} {\bibinfo {author} {\bibfnamefont {A.~J.}\ \bibnamefont
  {Bray}},\ }\bibfield  {title} {\bibinfo {title} {Theory of phase-ordering
  kinetics},\ }\href {https://doi.org/10.1080/00018739400101505} {\bibfield
  {journal} {\bibinfo  {journal} {Advances in Physics}\ }\textbf {\bibinfo
  {volume} {43}},\ \bibinfo {pages} {357} (\bibinfo {year} {1994})}\BibitemShut
  {NoStop}%
\bibitem [{\citenamefont {Onuki}(2002)}]{Onuki2002}%
  \BibitemOpen
  \bibfield  {author} {\bibinfo {author} {\bibfnamefont {A.}~\bibnamefont
  {Onuki}},\ }\href@noop {} {\emph {\bibinfo {title} {Phase Transition
  Dynamics}}}\ (\bibinfo  {publisher} {Cambridge University Press},\ \bibinfo
  {year} {2002})\BibitemShut {NoStop}%
\bibitem [{\citenamefont {Puri}\ and\ \citenamefont
  {Wadhawan}(2009)}]{Puri2009}%
  \BibitemOpen
  \bibfield  {author} {\bibinfo {author} {\bibfnamefont {S.}~\bibnamefont
  {Puri}}\ and\ \bibinfo {author} {\bibfnamefont {V.}~\bibnamefont
  {Wadhawan}},\ }\href@noop {} {\emph {\bibinfo {title} {Kinetics of phase
  transitions}}}\ (\bibinfo  {publisher} {CRC press},\ \bibinfo {year}
  {2009})\BibitemShut {NoStop}%
\bibitem [{\citenamefont {Marx}\ and\ \citenamefont {Hutter}(2009)}]{marx09}%
  \BibitemOpen
  \bibfield  {author} {\bibinfo {author} {\bibfnamefont {D.}~\bibnamefont
  {Marx}}\ and\ \bibinfo {author} {\bibfnamefont {J.}~\bibnamefont {Hutter}},\
  }\href@noop {} {\emph {\bibinfo {title} {Ab initio molecular dynamics: basic
  theory and advanced methods}}}\ (\bibinfo  {publisher} {Cambridge University
  Press},\ \bibinfo {year} {2009})\BibitemShut {NoStop}%
\bibitem [{\citenamefont {Kohn}(1996)}]{kohn96}%
  \BibitemOpen
  \bibfield  {author} {\bibinfo {author} {\bibfnamefont {W.}~\bibnamefont
  {Kohn}},\ }\bibfield  {title} {\bibinfo {title} {Density functional and
  density matrix method scaling linearly with the number of atoms},\ }\href
  {https://doi.org/10.1103/PhysRevLett.76.3168} {\bibfield  {journal} {\bibinfo
   {journal} {Phys. Rev. Lett.}\ }\textbf {\bibinfo {volume} {76}},\ \bibinfo
  {pages} {3168} (\bibinfo {year} {1996})}\BibitemShut {NoStop}%
\bibitem [{\citenamefont {Prodan}\ and\ \citenamefont {Kohn}(2005)}]{prodan05}%
  \BibitemOpen
  \bibfield  {author} {\bibinfo {author} {\bibfnamefont {E.}~\bibnamefont
  {Prodan}}\ and\ \bibinfo {author} {\bibfnamefont {W.}~\bibnamefont {Kohn}},\
  }\bibfield  {title} {\bibinfo {title} {Nearsightedness of electronic
  matter},\ }\href {https://doi.org/10.1073/pnas.0505436102} {\bibfield
  {journal} {\bibinfo  {journal} {Proceedings of the National Academy of
  Sciences}\ }\textbf {\bibinfo {volume} {102}},\ \bibinfo {pages} {11635}
  (\bibinfo {year} {2005})}\BibitemShut {NoStop}%
\bibitem [{\citenamefont {Li}\ \emph {et~al.}(2015)\citenamefont {Li},
  \citenamefont {Kermode},\ and\ \citenamefont {De~Vita}}]{li15}%
  \BibitemOpen
  \bibfield  {author} {\bibinfo {author} {\bibfnamefont {Z.}~\bibnamefont
  {Li}}, \bibinfo {author} {\bibfnamefont {J.~R.}\ \bibnamefont {Kermode}},\
  and\ \bibinfo {author} {\bibfnamefont {A.}~\bibnamefont {De~Vita}},\
  }\bibfield  {title} {\bibinfo {title} {Molecular dynamics with on-the-fly
  machine learning of quantum-mechanical forces},\ }\href
  {https://doi.org/10.1103/PhysRevLett.114.096405} {\bibfield  {journal}
  {\bibinfo  {journal} {Phys. Rev. Lett.}\ }\textbf {\bibinfo {volume} {114}},\
  \bibinfo {pages} {096405} (\bibinfo {year} {2015})}\BibitemShut {NoStop}%
\bibitem [{\citenamefont {Shapeev}(2016)}]{shapeev16}%
  \BibitemOpen
  \bibfield  {author} {\bibinfo {author} {\bibfnamefont {A.~V.}\ \bibnamefont
  {Shapeev}},\ }\bibfield  {title} {\bibinfo {title} {Moment tensor potentials:
  A class of systematically improvable interatomic potentials},\ }\href
  {https://doi.org/10.1137/15M1054183} {\bibfield  {journal} {\bibinfo
  {journal} {Multiscale Modeling \& Simulation}\ }\textbf {\bibinfo {volume}
  {14}},\ \bibinfo {pages} {1153} (\bibinfo {year} {2016})}\BibitemShut
  {NoStop}%
\bibitem [{\citenamefont {Behler}(2016)}]{behler16}%
  \BibitemOpen
  \bibfield  {author} {\bibinfo {author} {\bibfnamefont {J.}~\bibnamefont
  {Behler}},\ }\bibfield  {title} {\bibinfo {title} {{Perspective: Machine
  learning potentials for atomistic simulations}},\ }\href
  {https://doi.org/10.1063/1.4966192} {\bibfield  {journal} {\bibinfo
  {journal} {The Journal of Chemical Physics}\ }\textbf {\bibinfo {volume}
  {145}},\ \bibinfo {pages} {170901} (\bibinfo {year} {2016})}\BibitemShut
  {NoStop}%
\bibitem [{\citenamefont {Botu}\ \emph {et~al.}(2017)\citenamefont {Botu},
  \citenamefont {Batra}, \citenamefont {Chapman},\ and\ \citenamefont
  {Ramprasad}}]{botu17}%
  \BibitemOpen
  \bibfield  {author} {\bibinfo {author} {\bibfnamefont {V.}~\bibnamefont
  {Botu}}, \bibinfo {author} {\bibfnamefont {R.}~\bibnamefont {Batra}},
  \bibinfo {author} {\bibfnamefont {J.}~\bibnamefont {Chapman}},\ and\ \bibinfo
  {author} {\bibfnamefont {R.}~\bibnamefont {Ramprasad}},\ }\bibfield  {title}
  {\bibinfo {title} {Machine learning force fields: Construction, validation,
  and outlook},\ }\href {https://doi.org/10.1021/acs.jpcc.6b10908} {\bibfield
  {journal} {\bibinfo  {journal} {The Journal of Physical Chemistry C}\
  }\textbf {\bibinfo {volume} {121}},\ \bibinfo {pages} {511} (\bibinfo {year}
  {2017})}\BibitemShut {NoStop}%
\bibitem [{\citenamefont {Smith}\ \emph {et~al.}(2017)\citenamefont {Smith},
  \citenamefont {Isayev},\ and\ \citenamefont {Roitberg}}]{smith17}%
  \BibitemOpen
  \bibfield  {author} {\bibinfo {author} {\bibfnamefont {J.~S.}\ \bibnamefont
  {Smith}}, \bibinfo {author} {\bibfnamefont {O.}~\bibnamefont {Isayev}},\ and\
  \bibinfo {author} {\bibfnamefont {A.~E.}\ \bibnamefont {Roitberg}},\
  }\bibfield  {title} {\bibinfo {title} {Ani-1: an extensible neural network
  potential with dft accuracy at force field computational cost},\ }\href
  {https://doi.org/10.1039/C6SC05720A} {\bibfield  {journal} {\bibinfo
  {journal} {Chem. Sci.}\ }\textbf {\bibinfo {volume} {8}},\ \bibinfo {pages}
  {3192} (\bibinfo {year} {2017})}\BibitemShut {NoStop}%
\bibitem [{\citenamefont {Chmiela}\ \emph {et~al.}(2017)\citenamefont
  {Chmiela}, \citenamefont {Tkatchenko}, \citenamefont {Sauceda}, \citenamefont
  {Poltavsky}, \citenamefont {Schütt},\ and\ \citenamefont
  {Müller}}]{chmiela17}%
  \BibitemOpen
  \bibfield  {author} {\bibinfo {author} {\bibfnamefont {S.}~\bibnamefont
  {Chmiela}}, \bibinfo {author} {\bibfnamefont {A.}~\bibnamefont {Tkatchenko}},
  \bibinfo {author} {\bibfnamefont {H.~E.}\ \bibnamefont {Sauceda}}, \bibinfo
  {author} {\bibfnamefont {I.}~\bibnamefont {Poltavsky}}, \bibinfo {author}
  {\bibfnamefont {K.~T.}\ \bibnamefont {Schütt}},\ and\ \bibinfo {author}
  {\bibfnamefont {K.-R.}\ \bibnamefont {Müller}},\ }\bibfield  {title}
  {\bibinfo {title} {Machine learning of accurate energy-conserving molecular
  force fields},\ }\href {https://doi.org/10.1126/sciadv.1603015} {\bibfield
  {journal} {\bibinfo  {journal} {Science Advances}\ }\textbf {\bibinfo
  {volume} {3}},\ \bibinfo {pages} {e1603015} (\bibinfo {year}
  {2017})}\BibitemShut {NoStop}%
\bibitem [{\citenamefont {Zhang}\ \emph {et~al.}(2018)\citenamefont {Zhang},
  \citenamefont {Han}, \citenamefont {Wang}, \citenamefont {Car},\ and\
  \citenamefont {E}}]{zhang18}%
  \BibitemOpen
  \bibfield  {author} {\bibinfo {author} {\bibfnamefont {L.}~\bibnamefont
  {Zhang}}, \bibinfo {author} {\bibfnamefont {J.}~\bibnamefont {Han}}, \bibinfo
  {author} {\bibfnamefont {H.}~\bibnamefont {Wang}}, \bibinfo {author}
  {\bibfnamefont {R.}~\bibnamefont {Car}},\ and\ \bibinfo {author}
  {\bibfnamefont {W.}~\bibnamefont {E}},\ }\bibfield  {title} {\bibinfo {title}
  {Deep potential molecular dynamics: A scalable model with the accuracy of
  quantum mechanics},\ }\href {https://doi.org/10.1103/PhysRevLett.120.143001}
  {\bibfield  {journal} {\bibinfo  {journal} {Phys. Rev. Lett.}\ }\textbf
  {\bibinfo {volume} {120}},\ \bibinfo {pages} {143001} (\bibinfo {year}
  {2018})}\BibitemShut {NoStop}%
\bibitem [{\citenamefont {Chmiela}\ \emph {et~al.}(2018)\citenamefont
  {Chmiela}, \citenamefont {Sauceda}, \citenamefont {M{\"u}ller},\ and\
  \citenamefont {Tkatchenko}}]{chmiela18}%
  \BibitemOpen
  \bibfield  {author} {\bibinfo {author} {\bibfnamefont {S.}~\bibnamefont
  {Chmiela}}, \bibinfo {author} {\bibfnamefont {H.~E.}\ \bibnamefont
  {Sauceda}}, \bibinfo {author} {\bibfnamefont {K.-R.}\ \bibnamefont
  {M{\"u}ller}},\ and\ \bibinfo {author} {\bibfnamefont {A.}~\bibnamefont
  {Tkatchenko}},\ }\bibfield  {title} {\bibinfo {title} {Towards exact
  molecular dynamics simulations with machine-learned force fields},\ }\href
  {https://doi.org/10.1038/s41467-018-06169-2} {\bibfield  {journal} {\bibinfo
  {journal} {Nature Communications}\ }\textbf {\bibinfo {volume} {9}},\
  \bibinfo {pages} {3887} (\bibinfo {year} {2018})}\BibitemShut {NoStop}%
\bibitem [{\citenamefont {Deringer}\ \emph {et~al.}(2019)\citenamefont
  {Deringer}, \citenamefont {Caro},\ and\ \citenamefont {Csanyi}}]{deringer19}%
  \BibitemOpen
  \bibfield  {author} {\bibinfo {author} {\bibfnamefont {V.~L.}\ \bibnamefont
  {Deringer}}, \bibinfo {author} {\bibfnamefont {M.~A.}\ \bibnamefont {Caro}},\
  and\ \bibinfo {author} {\bibfnamefont {G.}~\bibnamefont {Csanyi}},\
  }\bibfield  {title} {\bibinfo {title} {Machine learning interatomic
  potentials as emerging tools for materials science},\ }\href
  {https://doi.org/https://doi.org/10.1002/adma.201902765} {\bibfield
  {journal} {\bibinfo  {journal} {Advanced Materials}\ }\textbf {\bibinfo
  {volume} {31}},\ \bibinfo {pages} {1902765} (\bibinfo {year}
  {2019})}\BibitemShut {NoStop}%
\bibitem [{\citenamefont {McGibbon}\ \emph {et~al.}(2017)\citenamefont
  {McGibbon}, \citenamefont {Taube}, \citenamefont {Donchev}, \citenamefont
  {Siva}, \citenamefont {Hernandez}, \citenamefont {Hargus}, \citenamefont
  {Law}, \citenamefont {Klepeis},\ and\ \citenamefont {Shaw}}]{mcgibbon17}%
  \BibitemOpen
  \bibfield  {author} {\bibinfo {author} {\bibfnamefont {R.~T.}\ \bibnamefont
  {McGibbon}}, \bibinfo {author} {\bibfnamefont {A.~G.}\ \bibnamefont {Taube}},
  \bibinfo {author} {\bibfnamefont {A.~G.}\ \bibnamefont {Donchev}}, \bibinfo
  {author} {\bibfnamefont {K.}~\bibnamefont {Siva}}, \bibinfo {author}
  {\bibfnamefont {F.}~\bibnamefont {Hernandez}}, \bibinfo {author}
  {\bibfnamefont {C.}~\bibnamefont {Hargus}}, \bibinfo {author} {\bibfnamefont
  {K.-H.}\ \bibnamefont {Law}}, \bibinfo {author} {\bibfnamefont {J.~L.}\
  \bibnamefont {Klepeis}},\ and\ \bibinfo {author} {\bibfnamefont {D.~E.}\
  \bibnamefont {Shaw}},\ }\bibfield  {title} {\bibinfo {title} {Improving the
  accuracy of m\"oller-plesset perturbation theory with neural networks},\
  }\href {https://doi.org/10.1063/1.4986081} {\bibfield  {journal} {\bibinfo
  {journal} {The Journal of Chemical Physics}\ }\textbf {\bibinfo {volume}
  {147}},\ \bibinfo {pages} {161725} (\bibinfo {year} {2017})}\BibitemShut
  {NoStop}%
\bibitem [{\citenamefont {Suwa}\ \emph {et~al.}(2019)\citenamefont {Suwa},
  \citenamefont {Smith}, \citenamefont {Lubbers}, \citenamefont {Batista},
  \citenamefont {Chern},\ and\ \citenamefont {Barros}}]{suwa19}%
  \BibitemOpen
  \bibfield  {author} {\bibinfo {author} {\bibfnamefont {H.}~\bibnamefont
  {Suwa}}, \bibinfo {author} {\bibfnamefont {J.~S.}\ \bibnamefont {Smith}},
  \bibinfo {author} {\bibfnamefont {N.}~\bibnamefont {Lubbers}}, \bibinfo
  {author} {\bibfnamefont {C.~D.}\ \bibnamefont {Batista}}, \bibinfo {author}
  {\bibfnamefont {G.-W.}\ \bibnamefont {Chern}},\ and\ \bibinfo {author}
  {\bibfnamefont {K.}~\bibnamefont {Barros}},\ }\bibfield  {title} {\bibinfo
  {title} {Machine learning for molecular dynamics with strongly correlated
  electrons},\ }\href {https://doi.org/10.1103/PhysRevB.99.161107} {\bibfield
  {journal} {\bibinfo  {journal} {Phys. Rev. B}\ }\textbf {\bibinfo {volume}
  {99}},\ \bibinfo {pages} {161107} (\bibinfo {year} {2019})}\BibitemShut
  {NoStop}%
\bibitem [{\citenamefont {Sauceda}\ \emph {et~al.}(2020)\citenamefont
  {Sauceda}, \citenamefont {Gastegger}, \citenamefont {Chmiela}, \citenamefont
  {Müller},\ and\ \citenamefont {Tkatchenko}}]{sauceda20}%
  \BibitemOpen
  \bibfield  {author} {\bibinfo {author} {\bibfnamefont {H.~E.}\ \bibnamefont
  {Sauceda}}, \bibinfo {author} {\bibfnamefont {M.}~\bibnamefont {Gastegger}},
  \bibinfo {author} {\bibfnamefont {S.}~\bibnamefont {Chmiela}}, \bibinfo
  {author} {\bibfnamefont {K.-R.}\ \bibnamefont {Müller}},\ and\ \bibinfo
  {author} {\bibfnamefont {A.}~\bibnamefont {Tkatchenko}},\ }\bibfield  {title}
  {\bibinfo {title} {{Molecular force fields with gradient-domain machine
  learning (GDML): Comparison and synergies with classical force fields}},\
  }\href {https://doi.org/10.1063/5.0023005} {\bibfield  {journal} {\bibinfo
  {journal} {The Journal of Chemical Physics}\ }\textbf {\bibinfo {volume}
  {153}},\ \bibinfo {pages} {124109} (\bibinfo {year} {2020})}\BibitemShut
  {NoStop}%
\bibitem [{\citenamefont {Zhang}\ \emph {et~al.}(2020)\citenamefont {Zhang},
  \citenamefont {Saha},\ and\ \citenamefont {Chern}}]{zhang20}%
  \BibitemOpen
  \bibfield  {author} {\bibinfo {author} {\bibfnamefont {P.}~\bibnamefont
  {Zhang}}, \bibinfo {author} {\bibfnamefont {P.}~\bibnamefont {Saha}},\ and\
  \bibinfo {author} {\bibfnamefont {G.-W.}\ \bibnamefont {Chern}},\ }\href@noop
  {} {\bibinfo {title} {Machine learning dynamics of phase separation in
  correlated electron magnets}} (\bibinfo {year} {2020}),\ \Eprint
  {https://arxiv.org/abs/2006.04205} {arXiv:2006.04205 [cond-mat.str-el]}
  \BibitemShut {NoStop}%
\bibitem [{\citenamefont {Zhang}\ and\ \citenamefont {Chern}(2021)}]{zhang21}%
  \BibitemOpen
  \bibfield  {author} {\bibinfo {author} {\bibfnamefont {P.}~\bibnamefont
  {Zhang}}\ and\ \bibinfo {author} {\bibfnamefont {G.-W.}\ \bibnamefont
  {Chern}},\ }\bibfield  {title} {\bibinfo {title} {Arrested phase separation
  in double-exchange models: Large-scale simulation enabled by machine
  learning},\ }\href {https://doi.org/10.1103/PhysRevLett.127.146401}
  {\bibfield  {journal} {\bibinfo  {journal} {Phys. Rev. Lett.}\ }\textbf
  {\bibinfo {volume} {127}},\ \bibinfo {pages} {146401} (\bibinfo {year}
  {2021})}\BibitemShut {NoStop}%
\bibitem [{\citenamefont {Zhang}\ \emph
  {et~al.}(2022{\natexlab{a}})\citenamefont {Zhang}, \citenamefont {Zhang},\
  and\ \citenamefont {Chern}}]{zhang22}%
  \BibitemOpen
  \bibfield  {author} {\bibinfo {author} {\bibfnamefont {P.}~\bibnamefont
  {Zhang}}, \bibinfo {author} {\bibfnamefont {S.}~\bibnamefont {Zhang}},\ and\
  \bibinfo {author} {\bibfnamefont {G.-W.}\ \bibnamefont {Chern}},\ }\href@noop
  {} {\bibinfo {title} {Descriptors for machine learning model of generalized
  force field in condensed matter systems}} (\bibinfo {year}
  {2022}{\natexlab{a}}),\ \Eprint {https://arxiv.org/abs/2201.00798}
  {arXiv:2201.00798 [cond-mat.str-el]} \BibitemShut {NoStop}%
\bibitem [{\citenamefont {Zhang}\ \emph
  {et~al.}(2022{\natexlab{b}})\citenamefont {Zhang}, \citenamefont {Zhang},\
  and\ \citenamefont {Chern}}]{zhang22b}%
  \BibitemOpen
  \bibfield  {author} {\bibinfo {author} {\bibfnamefont {S.}~\bibnamefont
  {Zhang}}, \bibinfo {author} {\bibfnamefont {P.}~\bibnamefont {Zhang}},\ and\
  \bibinfo {author} {\bibfnamefont {G.-W.}\ \bibnamefont {Chern}},\ }\bibfield
  {title} {\bibinfo {title} {Anomalous phase separation in a correlated
  electron system: Machine-learning enabled large-scale kinetic monte carlo
  simulations},\ }\href {https://doi.org/10.1073/pnas.2119957119} {\bibfield
  {journal} {\bibinfo  {journal} {Proceedings of the National Academy of
  Sciences}\ }\textbf {\bibinfo {volume} {119}},\ \bibinfo {pages}
  {e2119957119} (\bibinfo {year} {2022}{\natexlab{b}})}\BibitemShut {NoStop}%
\bibitem [{\citenamefont {Zhang}\ and\ \citenamefont {Chern}(2023)}]{zhang23}%
  \BibitemOpen
  \bibfield  {author} {\bibinfo {author} {\bibfnamefont {P.}~\bibnamefont
  {Zhang}}\ and\ \bibinfo {author} {\bibfnamefont {G.-W.}\ \bibnamefont
  {Chern}},\ }\bibfield  {title} {\bibinfo {title} {Machine learning
  nonequilibrium electron forces for spin dynamics of itinerant magnets},\
  }\href {https://doi.org/10.1038/s41524-023-00990-0} {\bibfield  {journal}
  {\bibinfo  {journal} {npj Computational Materials}\ }\textbf {\bibinfo
  {volume} {9}},\ \bibinfo {pages} {32} (\bibinfo {year} {2023})}\BibitemShut
  {NoStop}%
\bibitem [{\citenamefont {Cheng}\ \emph
  {et~al.}(2023{\natexlab{a}})\citenamefont {Cheng}, \citenamefont {Zhang},\
  and\ \citenamefont {Chern}}]{cheng23}%
  \BibitemOpen
  \bibfield  {author} {\bibinfo {author} {\bibfnamefont {C.}~\bibnamefont
  {Cheng}}, \bibinfo {author} {\bibfnamefont {S.}~\bibnamefont {Zhang}},\ and\
  \bibinfo {author} {\bibfnamefont {G.-W.}\ \bibnamefont {Chern}},\ }\bibfield
  {title} {\bibinfo {title} {Machine learning for phase ordering dynamics of
  charge density waves},\ }\href {https://doi.org/10.1103/PhysRevB.108.014301}
  {\bibfield  {journal} {\bibinfo  {journal} {Phys. Rev. B}\ }\textbf {\bibinfo
  {volume} {108}},\ \bibinfo {pages} {014301} (\bibinfo {year}
  {2023}{\natexlab{a}})}\BibitemShut {NoStop}%
\bibitem [{\citenamefont {Cheng}\ \emph
  {et~al.}(2023{\natexlab{b}})\citenamefont {Cheng}, \citenamefont {Zhang},
  \citenamefont {Nguyen}, \citenamefont {Azarfar}, \citenamefont {Chern},\ and\
  \citenamefont {Baek}}]{cheng23b}%
  \BibitemOpen
  \bibfield  {author} {\bibinfo {author} {\bibfnamefont {X.}~\bibnamefont
  {Cheng}}, \bibinfo {author} {\bibfnamefont {S.}~\bibnamefont {Zhang}},
  \bibinfo {author} {\bibfnamefont {P.~C.~H.}\ \bibnamefont {Nguyen}}, \bibinfo
  {author} {\bibfnamefont {S.}~\bibnamefont {Azarfar}}, \bibinfo {author}
  {\bibfnamefont {G.-W.}\ \bibnamefont {Chern}},\ and\ \bibinfo {author}
  {\bibfnamefont {S.~S.}\ \bibnamefont {Baek}},\ }\bibfield  {title} {\bibinfo
  {title} {Convolutional neural networks for large-scale dynamical modeling of
  itinerant magnets},\ }\href
  {https://doi.org/10.1103/PhysRevResearch.5.033188} {\bibfield  {journal}
  {\bibinfo  {journal} {Phys. Rev. Res.}\ }\textbf {\bibinfo {volume} {5}},\
  \bibinfo {pages} {033188} (\bibinfo {year} {2023}{\natexlab{b}})}\BibitemShut
  {NoStop}%
\bibitem [{\citenamefont {Cybenko}(1989)}]{cybenko89}%
  \BibitemOpen
  \bibfield  {author} {\bibinfo {author} {\bibfnamefont {G.}~\bibnamefont
  {Cybenko}},\ }\bibfield  {title} {\bibinfo {title} {Approximation by
  superpositions of a sigmoidal function},\ }\href
  {https://doi.org/10.1007/BF02551274} {\bibfield  {journal} {\bibinfo
  {journal} {Mathematics of Control, Signals and Systems}\ }\textbf {\bibinfo
  {volume} {2}},\ \bibinfo {pages} {303} (\bibinfo {year} {1989})}\BibitemShut
  {NoStop}%
\bibitem [{\citenamefont {Hornik}\ \emph {et~al.}(1989)\citenamefont {Hornik},
  \citenamefont {Stinchcombe},\ and\ \citenamefont {White}}]{hornik89}%
  \BibitemOpen
  \bibfield  {author} {\bibinfo {author} {\bibfnamefont {K.}~\bibnamefont
  {Hornik}}, \bibinfo {author} {\bibfnamefont {M.}~\bibnamefont
  {Stinchcombe}},\ and\ \bibinfo {author} {\bibfnamefont {H.}~\bibnamefont
  {White}},\ }\bibfield  {title} {\bibinfo {title} {Multilayer feedforward
  networks are universal approximators},\ }\href
  {https://doi.org/https://doi.org/10.1016/0893-6080(89)90020-8} {\bibfield
  {journal} {\bibinfo  {journal} {Neural Networks}\ }\textbf {\bibinfo {volume}
  {2}},\ \bibinfo {pages} {359} (\bibinfo {year} {1989})}\BibitemShut {NoStop}%
\bibitem [{\citenamefont {Paszke}\ \emph {et~al.}(2017)\citenamefont {Paszke},
  \citenamefont {Gross}, \citenamefont {Chintala}, \citenamefont {Chanan},
  \citenamefont {Yang}, \citenamefont {DeVito}, \citenamefont {Lin},
  \citenamefont {Desmaison}, \citenamefont {Antiga},\ and\ \citenamefont
  {Lerer}}]{paszke17}%
  \BibitemOpen
  \bibfield  {author} {\bibinfo {author} {\bibfnamefont {A.}~\bibnamefont
  {Paszke}}, \bibinfo {author} {\bibfnamefont {S.}~\bibnamefont {Gross}},
  \bibinfo {author} {\bibfnamefont {S.}~\bibnamefont {Chintala}}, \bibinfo
  {author} {\bibfnamefont {G.}~\bibnamefont {Chanan}}, \bibinfo {author}
  {\bibfnamefont {E.}~\bibnamefont {Yang}}, \bibinfo {author} {\bibfnamefont
  {Z.}~\bibnamefont {DeVito}}, \bibinfo {author} {\bibfnamefont
  {Z.}~\bibnamefont {Lin}}, \bibinfo {author} {\bibfnamefont {A.}~\bibnamefont
  {Desmaison}}, \bibinfo {author} {\bibfnamefont {L.}~\bibnamefont {Antiga}},\
  and\ \bibinfo {author} {\bibfnamefont {A.}~\bibnamefont {Lerer}},\ }\bibfield
   {title} {\bibinfo {title} {Automatic differentiation in pytorch},\ }in\
  \href {https://openreview.net/forum?id=BJJsrmfCZ} {\emph {\bibinfo
  {booktitle} {NIPS 2017 Workshop on Autodiff}}}\ (\bibinfo {year}
  {2017})\BibitemShut {NoStop}%
\bibitem [{\citenamefont {Baydin}\ \emph {et~al.}(2018)\citenamefont {Baydin},
  \citenamefont {Pearlmutter}, \citenamefont {Radul},\ and\ \citenamefont
  {Siskind}}]{baydin18}%
  \BibitemOpen
  \bibfield  {author} {\bibinfo {author} {\bibfnamefont {A.~G.}\ \bibnamefont
  {Baydin}}, \bibinfo {author} {\bibfnamefont {B.~A.}\ \bibnamefont
  {Pearlmutter}}, \bibinfo {author} {\bibfnamefont {A.~A.}\ \bibnamefont
  {Radul}},\ and\ \bibinfo {author} {\bibfnamefont {J.~M.}\ \bibnamefont
  {Siskind}},\ }\bibfield  {title} {\bibinfo {title} {Automatic differentiation
  in machine learning: a survey},\ }\href
  {http://jmlr.org/papers/v18/17-468.html} {\bibfield  {journal} {\bibinfo
  {journal} {Journal of Machine Learning Research}\ }\textbf {\bibinfo {volume}
  {18}},\ \bibinfo {pages} {1} (\bibinfo {year} {2018})}\BibitemShut {NoStop}%
\bibitem [{\citenamefont {Behler}(2011)}]{behler11}%
  \BibitemOpen
  \bibfield  {author} {\bibinfo {author} {\bibfnamefont {J.}~\bibnamefont
  {Behler}},\ }\bibfield  {title} {\bibinfo {title} {{Atom-centered symmetry
  functions for constructing high-dimensional neural network potentials}},\
  }\href {https://doi.org/10.1063/1.3553717} {\bibfield  {journal} {\bibinfo
  {journal} {The Journal of Chemical Physics}\ }\textbf {\bibinfo {volume}
  {134}},\ \bibinfo {pages} {074106} (\bibinfo {year} {2011})}\BibitemShut
  {NoStop}%
\bibitem [{\citenamefont {Ghiringhelli}\ \emph {et~al.}(2015)\citenamefont
  {Ghiringhelli}, \citenamefont {Vybiral}, \citenamefont {Levchenko},
  \citenamefont {Draxl},\ and\ \citenamefont {Scheffler}}]{ghiringhelli15}%
  \BibitemOpen
  \bibfield  {author} {\bibinfo {author} {\bibfnamefont {L.~M.}\ \bibnamefont
  {Ghiringhelli}}, \bibinfo {author} {\bibfnamefont {J.}~\bibnamefont
  {Vybiral}}, \bibinfo {author} {\bibfnamefont {S.~V.}\ \bibnamefont
  {Levchenko}}, \bibinfo {author} {\bibfnamefont {C.}~\bibnamefont {Draxl}},\
  and\ \bibinfo {author} {\bibfnamefont {M.}~\bibnamefont {Scheffler}},\
  }\bibfield  {title} {\bibinfo {title} {Big data of materials science:
  Critical role of the descriptor},\ }\href
  {https://doi.org/10.1103/PhysRevLett.114.105503} {\bibfield  {journal}
  {\bibinfo  {journal} {Phys. Rev. Lett.}\ }\textbf {\bibinfo {volume} {114}},\
  \bibinfo {pages} {105503} (\bibinfo {year} {2015})}\BibitemShut {NoStop}%
\bibitem [{\citenamefont {Bart\'ok}\ \emph {et~al.}(2013)\citenamefont
  {Bart\'ok}, \citenamefont {Kondor},\ and\ \citenamefont
  {Cs\'anyi}}]{bartok13}%
  \BibitemOpen
  \bibfield  {author} {\bibinfo {author} {\bibfnamefont {A.~P.}\ \bibnamefont
  {Bart\'ok}}, \bibinfo {author} {\bibfnamefont {R.}~\bibnamefont {Kondor}},\
  and\ \bibinfo {author} {\bibfnamefont {G.}~\bibnamefont {Cs\'anyi}},\
  }\bibfield  {title} {\bibinfo {title} {On representing chemical
  environments},\ }\href {https://doi.org/10.1103/PhysRevB.87.184115}
  {\bibfield  {journal} {\bibinfo  {journal} {Phys. Rev. B}\ }\textbf {\bibinfo
  {volume} {87}},\ \bibinfo {pages} {184115} (\bibinfo {year}
  {2013})}\BibitemShut {NoStop}%
\bibitem [{\citenamefont {Drautz}(2019)}]{drautz19}%
  \BibitemOpen
  \bibfield  {author} {\bibinfo {author} {\bibfnamefont {R.}~\bibnamefont
  {Drautz}},\ }\bibfield  {title} {\bibinfo {title} {Atomic cluster expansion
  for accurate and transferable interatomic potentials},\ }\href
  {https://doi.org/10.1103/PhysRevB.99.014104} {\bibfield  {journal} {\bibinfo
  {journal} {Phys. Rev. B}\ }\textbf {\bibinfo {volume} {99}},\ \bibinfo
  {pages} {014104} (\bibinfo {year} {2019})}\BibitemShut {NoStop}%
\bibitem [{\citenamefont {Himanen}\ \emph {et~al.}(2020)\citenamefont
  {Himanen}, \citenamefont {Jäger}, \citenamefont {Morooka}, \citenamefont
  {{Federici Canova}}, \citenamefont {Ranawat}, \citenamefont {Gao},
  \citenamefont {Rinke},\ and\ \citenamefont {Foster}}]{himanen20}%
  \BibitemOpen
  \bibfield  {author} {\bibinfo {author} {\bibfnamefont {L.}~\bibnamefont
  {Himanen}}, \bibinfo {author} {\bibfnamefont {M.~O.}\ \bibnamefont {Jäger}},
  \bibinfo {author} {\bibfnamefont {E.~V.}\ \bibnamefont {Morooka}}, \bibinfo
  {author} {\bibfnamefont {F.}~\bibnamefont {{Federici Canova}}}, \bibinfo
  {author} {\bibfnamefont {Y.~S.}\ \bibnamefont {Ranawat}}, \bibinfo {author}
  {\bibfnamefont {D.~Z.}\ \bibnamefont {Gao}}, \bibinfo {author} {\bibfnamefont
  {P.}~\bibnamefont {Rinke}},\ and\ \bibinfo {author} {\bibfnamefont {A.~S.}\
  \bibnamefont {Foster}},\ }\bibfield  {title} {\bibinfo {title} {Dscribe:
  Library of descriptors for machine learning in materials science},\ }\href
  {https://doi.org/https://doi.org/10.1016/j.cpc.2019.106949} {\bibfield
  {journal} {\bibinfo  {journal} {Computer Physics Communications}\ }\textbf
  {\bibinfo {volume} {247}},\ \bibinfo {pages} {106949} (\bibinfo {year}
  {2020})}\BibitemShut {NoStop}%
\bibitem [{\citenamefont {Huo}\ and\ \citenamefont {Rupp}(2022)}]{huo22}%
  \BibitemOpen
  \bibfield  {author} {\bibinfo {author} {\bibfnamefont {H.}~\bibnamefont
  {Huo}}\ and\ \bibinfo {author} {\bibfnamefont {M.}~\bibnamefont {Rupp}},\
  }\bibfield  {title} {\bibinfo {title} {Unified representation of molecules
  and crystals for machine learning},\ }\href
  {https://doi.org/10.1088/2632-2153/aca005} {\bibfield  {journal} {\bibinfo
  {journal} {Machine Learning: Science and Technology}\ }\textbf {\bibinfo
  {volume} {3}},\ \bibinfo {pages} {045017} (\bibinfo {year}
  {2022})}\BibitemShut {NoStop}%
\bibitem [{\citenamefont {Kondor}(2007)}]{kondor07}%
  \BibitemOpen
  \bibfield  {author} {\bibinfo {author} {\bibfnamefont {R.}~\bibnamefont
  {Kondor}},\ }\href@noop {} {\bibinfo {title} {A novel set of rotationally and
  translationally invariant features for images based on the non-commutative
  bispectrum}} (\bibinfo {year} {2007}),\ \Eprint
  {https://arxiv.org/abs/cs/0701127} {arXiv:cs/0701127 [cs.CV]} \BibitemShut
  {NoStop}%
\bibitem [{\citenamefont {Ma}\ \emph {et~al.}(2019)\citenamefont {Ma},
  \citenamefont {Zhang}, \citenamefont {Tan}, \citenamefont {Ghosh},\ and\
  \citenamefont {Chern}}]{Ma19}%
  \BibitemOpen
  \bibfield  {author} {\bibinfo {author} {\bibfnamefont {J.}~\bibnamefont
  {Ma}}, \bibinfo {author} {\bibfnamefont {P.}~\bibnamefont {Zhang}}, \bibinfo
  {author} {\bibfnamefont {Y.}~\bibnamefont {Tan}}, \bibinfo {author}
  {\bibfnamefont {A.~W.}\ \bibnamefont {Ghosh}},\ and\ \bibinfo {author}
  {\bibfnamefont {G.-W.}\ \bibnamefont {Chern}},\ }\bibfield  {title} {\bibinfo
  {title} {Machine learning electron correlation in a disordered medium},\
  }\href {https://doi.org/10.1103/PhysRevB.99.085118} {\bibfield  {journal}
  {\bibinfo  {journal} {Phys. Rev. B}\ }\textbf {\bibinfo {volume} {99}},\
  \bibinfo {pages} {085118} (\bibinfo {year} {2019})}\BibitemShut {NoStop}%
\bibitem [{\citenamefont {Liu}\ \emph {et~al.}(2022)\citenamefont {Liu},
  \citenamefont {Zhang}, \citenamefont {Zhang}, \citenamefont {Lee},\ and\
  \citenamefont {Chern}}]{Liu22}%
  \BibitemOpen
  \bibfield  {author} {\bibinfo {author} {\bibfnamefont {Y.-H.}\ \bibnamefont
  {Liu}}, \bibinfo {author} {\bibfnamefont {S.}~\bibnamefont {Zhang}}, \bibinfo
  {author} {\bibfnamefont {P.}~\bibnamefont {Zhang}}, \bibinfo {author}
  {\bibfnamefont {T.-K.}\ \bibnamefont {Lee}},\ and\ \bibinfo {author}
  {\bibfnamefont {G.-W.}\ \bibnamefont {Chern}},\ }\bibfield  {title} {\bibinfo
  {title} {Machine learning predictions for local electronic properties of
  disordered correlated electron systems},\ }\href
  {https://doi.org/10.1103/PhysRevB.106.035131} {\bibfield  {journal} {\bibinfo
   {journal} {Phys. Rev. B}\ }\textbf {\bibinfo {volume} {106}},\ \bibinfo
  {pages} {035131} (\bibinfo {year} {2022})}\BibitemShut {NoStop}%
\bibitem [{\citenamefont {Tian}\ \emph {et~al.}(2023)\citenamefont {Tian},
  \citenamefont {Zhang},\ and\ \citenamefont {Chern}}]{Tian23}%
  \BibitemOpen
  \bibfield  {author} {\bibinfo {author} {\bibfnamefont {Z.}~\bibnamefont
  {Tian}}, \bibinfo {author} {\bibfnamefont {S.}~\bibnamefont {Zhang}},\ and\
  \bibinfo {author} {\bibfnamefont {G.-W.}\ \bibnamefont {Chern}},\ }\bibfield
  {title} {\bibinfo {title} {Machine learning for structure-property mapping of
  {Ising} models: Scalability and limitations},\ }\href
  {https://doi.org/10.1103/PhysRevE.108.065304} {\bibfield  {journal} {\bibinfo
   {journal} {Phys. Rev. E}\ }\textbf {\bibinfo {volume} {108}},\ \bibinfo
  {pages} {065304} (\bibinfo {year} {2023})}\BibitemShut {NoStop}%
\bibitem [{\citenamefont {Hamermesh}(1962)}]{hamermesh62}%
  \BibitemOpen
  \bibfield  {author} {\bibinfo {author} {\bibfnamefont {M.}~\bibnamefont
  {Hamermesh}},\ }\href@noop {} {\emph {\bibinfo {title} {Group Theory and Its
  Application to Physical Problems}}}\ (\bibinfo  {publisher} {Dover},\
  \bibinfo {address} {New York},\ \bibinfo {year} {1962})\BibitemShut {NoStop}%
\bibitem [{\citenamefont {Paszke}\ \emph {et~al.}(2019)\citenamefont {Paszke},
  \citenamefont {Gross}, \citenamefont {Massa}, \citenamefont {Lerer},
  \citenamefont {Bradbury}, \citenamefont {Chanan}, \citenamefont {Killeen},
  \citenamefont {Lin}, \citenamefont {Gimelshein}, \citenamefont {Antiga},
  \citenamefont {Desmaison}, \citenamefont {Kopf}, \citenamefont {Yang},
  \citenamefont {DeVito}, \citenamefont {Raison}, \citenamefont {Tejani},
  \citenamefont {Chilamkurthy}, \citenamefont {Steiner}, \citenamefont {Fang},
  \citenamefont {Bai},\ and\ \citenamefont {Chintala}}]{paszke19}%
  \BibitemOpen
  \bibfield  {author} {\bibinfo {author} {\bibfnamefont {A.}~\bibnamefont
  {Paszke}}, \bibinfo {author} {\bibfnamefont {S.}~\bibnamefont {Gross}},
  \bibinfo {author} {\bibfnamefont {F.}~\bibnamefont {Massa}}, \bibinfo
  {author} {\bibfnamefont {A.}~\bibnamefont {Lerer}}, \bibinfo {author}
  {\bibfnamefont {J.}~\bibnamefont {Bradbury}}, \bibinfo {author}
  {\bibfnamefont {G.}~\bibnamefont {Chanan}}, \bibinfo {author} {\bibfnamefont
  {T.}~\bibnamefont {Killeen}}, \bibinfo {author} {\bibfnamefont
  {Z.}~\bibnamefont {Lin}}, \bibinfo {author} {\bibfnamefont {N.}~\bibnamefont
  {Gimelshein}}, \bibinfo {author} {\bibfnamefont {L.}~\bibnamefont {Antiga}},
  \bibinfo {author} {\bibfnamefont {A.}~\bibnamefont {Desmaison}}, \bibinfo
  {author} {\bibfnamefont {A.}~\bibnamefont {Kopf}}, \bibinfo {author}
  {\bibfnamefont {E.}~\bibnamefont {Yang}}, \bibinfo {author} {\bibfnamefont
  {Z.}~\bibnamefont {DeVito}}, \bibinfo {author} {\bibfnamefont
  {M.}~\bibnamefont {Raison}}, \bibinfo {author} {\bibfnamefont
  {A.}~\bibnamefont {Tejani}}, \bibinfo {author} {\bibfnamefont
  {S.}~\bibnamefont {Chilamkurthy}}, \bibinfo {author} {\bibfnamefont
  {B.}~\bibnamefont {Steiner}}, \bibinfo {author} {\bibfnamefont
  {L.}~\bibnamefont {Fang}}, \bibinfo {author} {\bibfnamefont {J.}~\bibnamefont
  {Bai}},\ and\ \bibinfo {author} {\bibfnamefont {S.}~\bibnamefont
  {Chintala}},\ }\bibfield  {title} {\bibinfo {title} {Pytorch: An imperative
  style, high-performance deep learning library},\ }in\ \href
  {https://proceedings.neurips.cc/paper_files/paper/2019/file/bdbca288fee7f92f2bfa9f7012727740-Paper.pdf}
  {\emph {\bibinfo {booktitle} {Advances in Neural Information Processing
  Systems}}},\ Vol.~\bibinfo {volume} {32},\ \bibinfo {editor} {edited by\
  \bibinfo {editor} {\bibfnamefont {H.}~\bibnamefont {Wallach}}, \bibinfo
  {editor} {\bibfnamefont {H.}~\bibnamefont {Larochelle}}, \bibinfo {editor}
  {\bibfnamefont {A.}~\bibnamefont {Beygelzimer}}, \bibinfo {editor}
  {\bibfnamefont {F.}~\bibnamefont {d'Alch\'{e} Buc}}, \bibinfo {editor}
  {\bibfnamefont {E.}~\bibnamefont {Fox}},\ and\ \bibinfo {editor}
  {\bibfnamefont {R.}~\bibnamefont {Garnett}}}\ (\bibinfo  {publisher} {Curran
  Associates, Inc.},\ \bibinfo {year} {2019})\BibitemShut {NoStop}%
\bibitem [{\citenamefont {Nair}\ and\ \citenamefont {Hinton}(2010)}]{nair10}%
  \BibitemOpen
  \bibfield  {author} {\bibinfo {author} {\bibfnamefont {V.}~\bibnamefont
  {Nair}}\ and\ \bibinfo {author} {\bibfnamefont {G.~E.}\ \bibnamefont
  {Hinton}},\ }\bibfield  {title} {\bibinfo {title} {Rectified linear units
  improve restricted boltzmann machines},\ }in\ \href@noop {} {\emph {\bibinfo
  {booktitle} {Proceedings of the 27th International Conference on
  International Conference on Machine Learning}}},\ \bibinfo {series and
  number} {ICML'10}\ (\bibinfo  {publisher} {Omnipress},\ \bibinfo {address}
  {Madison, WI, USA},\ \bibinfo {year} {2010})\ p.\ \bibinfo {pages}
  {807–814}\BibitemShut {NoStop}%
\bibitem [{\citenamefont {Barron}(2017)}]{barron17}%
  \BibitemOpen
  \bibfield  {author} {\bibinfo {author} {\bibfnamefont {J.~T.}\ \bibnamefont
  {Barron}},\ }\href@noop {} {\bibinfo {title} {Continuously differentiable
  exponential linear units}} (\bibinfo {year} {2017}),\ \Eprint
  {https://arxiv.org/abs/1704.07483} {arXiv:1704.07483 [cs.LG]} \BibitemShut
  {NoStop}%
\bibitem [{\citenamefont {Kingma}\ and\ \citenamefont {Ba}(2017)}]{kingma17}%
  \BibitemOpen
  \bibfield  {author} {\bibinfo {author} {\bibfnamefont {D.~P.}\ \bibnamefont
  {Kingma}}\ and\ \bibinfo {author} {\bibfnamefont {J.}~\bibnamefont {Ba}},\
  }\href@noop {} {\bibinfo {title} {Adam: A method for stochastic
  optimization}} (\bibinfo {year} {2017}),\ \Eprint
  {https://arxiv.org/abs/1412.6980} {arXiv:1412.6980 [cs.LG]} \BibitemShut
  {NoStop}%
\bibitem [{\citenamefont {Noack}\ \emph {et~al.}(1991)\citenamefont {Noack},
  \citenamefont {Scalapino},\ and\ \citenamefont {Scalettar}}]{noack91}%
  \BibitemOpen
  \bibfield  {author} {\bibinfo {author} {\bibfnamefont {R.~M.}\ \bibnamefont
  {Noack}}, \bibinfo {author} {\bibfnamefont {D.~J.}\ \bibnamefont
  {Scalapino}},\ and\ \bibinfo {author} {\bibfnamefont {R.~T.}\ \bibnamefont
  {Scalettar}},\ }\bibfield  {title} {\bibinfo {title} {Charge-density-wave and
  pairing susceptibilities in a two-dimensional electron-phonon model},\ }\href
  {https://doi.org/10.1103/PhysRevLett.66.778} {\bibfield  {journal} {\bibinfo
  {journal} {Phys. Rev. Lett.}\ }\textbf {\bibinfo {volume} {66}},\ \bibinfo
  {pages} {778} (\bibinfo {year} {1991})}\BibitemShut {NoStop}%
\bibitem [{\citenamefont {Hohenadler}\ and\ \citenamefont
  {Batrouni}(2019)}]{hohenadler19}%
  \BibitemOpen
  \bibfield  {author} {\bibinfo {author} {\bibfnamefont {M.}~\bibnamefont
  {Hohenadler}}\ and\ \bibinfo {author} {\bibfnamefont {G.~G.}\ \bibnamefont
  {Batrouni}},\ }\bibfield  {title} {\bibinfo {title} {Dominant charge density
  wave correlations in the holstein model on the half-filled square lattice},\
  }\href {https://doi.org/10.1103/PhysRevB.100.165114} {\bibfield  {journal}
  {\bibinfo  {journal} {Phys. Rev. B}\ }\textbf {\bibinfo {volume} {100}},\
  \bibinfo {pages} {165114} (\bibinfo {year} {2019})}\BibitemShut {NoStop}%
\bibitem [{\citenamefont {Esterlis}\ \emph {et~al.}(2019)\citenamefont
  {Esterlis}, \citenamefont {Kivelson},\ and\ \citenamefont
  {Scalapino}}]{esterlis19}%
  \BibitemOpen
  \bibfield  {author} {\bibinfo {author} {\bibfnamefont {I.}~\bibnamefont
  {Esterlis}}, \bibinfo {author} {\bibfnamefont {S.~A.}\ \bibnamefont
  {Kivelson}},\ and\ \bibinfo {author} {\bibfnamefont {D.~J.}\ \bibnamefont
  {Scalapino}},\ }\bibfield  {title} {\bibinfo {title} {Pseudogap crossover in
  the electron-phonon system},\ }\href
  {https://doi.org/10.1103/PhysRevB.99.174516} {\bibfield  {journal} {\bibinfo
  {journal} {Phys. Rev. B}\ }\textbf {\bibinfo {volume} {99}},\ \bibinfo
  {pages} {174516} (\bibinfo {year} {2019})}\BibitemShut {NoStop}%
\bibitem [{\citenamefont {Wei\ss{}e}\ \emph {et~al.}(2006)\citenamefont
  {Wei\ss{}e}, \citenamefont {Wellein}, \citenamefont {Alvermann},\ and\
  \citenamefont {Fehske}}]{weisse06}%
  \BibitemOpen
  \bibfield  {author} {\bibinfo {author} {\bibfnamefont {A.}~\bibnamefont
  {Wei\ss{}e}}, \bibinfo {author} {\bibfnamefont {G.}~\bibnamefont {Wellein}},
  \bibinfo {author} {\bibfnamefont {A.}~\bibnamefont {Alvermann}},\ and\
  \bibinfo {author} {\bibfnamefont {H.}~\bibnamefont {Fehske}},\ }\bibfield
  {title} {\bibinfo {title} {The kernel polynomial method},\ }\href
  {https://doi.org/10.1103/RevModPhys.78.275} {\bibfield  {journal} {\bibinfo
  {journal} {Rev. Mod. Phys.}\ }\textbf {\bibinfo {volume} {78}},\ \bibinfo
  {pages} {275} (\bibinfo {year} {2006})}\BibitemShut {NoStop}%
\bibitem [{\citenamefont {Barros}\ and\ \citenamefont {Kato}(2013)}]{barros13}%
  \BibitemOpen
  \bibfield  {author} {\bibinfo {author} {\bibfnamefont {K.}~\bibnamefont
  {Barros}}\ and\ \bibinfo {author} {\bibfnamefont {Y.}~\bibnamefont {Kato}},\
  }\bibfield  {title} {\bibinfo {title} {Efficient langevin simulation of
  coupled classical fields and fermions},\ }\href
  {https://doi.org/10.1103/PhysRevB.88.235101} {\bibfield  {journal} {\bibinfo
  {journal} {Phys. Rev. B}\ }\textbf {\bibinfo {volume} {88}},\ \bibinfo
  {pages} {235101} (\bibinfo {year} {2013})}\BibitemShut {NoStop}%
\bibitem [{\citenamefont {Wang}\ \emph {et~al.}(2018)\citenamefont {Wang},
  \citenamefont {Chern}, \citenamefont {Batista},\ and\ \citenamefont
  {Barros}}]{wang18}%
  \BibitemOpen
  \bibfield  {author} {\bibinfo {author} {\bibfnamefont {Z.}~\bibnamefont
  {Wang}}, \bibinfo {author} {\bibfnamefont {G.-W.}\ \bibnamefont {Chern}},
  \bibinfo {author} {\bibfnamefont {C.~D.}\ \bibnamefont {Batista}},\ and\
  \bibinfo {author} {\bibfnamefont {K.}~\bibnamefont {Barros}},\ }\bibfield
  {title} {\bibinfo {title} {{Gradient-based stochastic estimation of the
  density matrix}},\ }\href {https://doi.org/10.1063/1.5017741} {\bibfield
  {journal} {\bibinfo  {journal} {The Journal of Chemical Physics}\ }\textbf
  {\bibinfo {volume} {148}},\ \bibinfo {pages} {094107} (\bibinfo {year}
  {2018})}\BibitemShut {NoStop}%
\end{thebibliography}%

\end{document}